\documentclass[11pt]{article} 
\date{\today}

\usepackage{geometry}
\usepackage{amsmath,amsfonts,amssymb}  
\usepackage{bbold}

\usepackage{graphicx}       
\usepackage{subcaption}      
\usepackage{float}           

\usepackage[natbibapa]{apacite} 
\usepackage[
    colorlinks=true,
    linkcolor=blue,
    citecolor=blue,
    urlcolor=blue
]{hyperref}

\usepackage{comment}         
\usepackage[normalem]{ulem}  

\usepackage{authblk} 
\usepackage[table,xcdraw]{xcolor} 

\usepackage{makecell}
\usepackage[export]{adjustbox} 

\usepackage{booktabs}

\title{Revisiting Equity–Bond Diversification under Tail Risk and Structural Market Change: Evidence from Global ETFs and Model-Implied Option Prices}
\author[1]{Farzana Afroz}
\author[2]{Abootaleb Shirvani}
\author[1]{Nicholas Appiah}
\author[1]{Svetlozar T. Rachev}
\author[3]{Frank J. Fabozzi}

\affil[1]{\small Department of Mathematics \& Statistics, Texas Tech University, email: fafroz@ttu.edu, niappiah@ttu.edu, zari.rachev@ttu.edu}
\affil[2]{\small Department of Mathematical Science, Kean University, email: ashirvan@kean.edu}
\affil[3]{\small Carey Business School, Johns Hopkins University, email: fabozzi321@aol.com}

\begin{document}
\maketitle

\begin{abstract}
This study examines thirty equity--bond and multi-asset exchange-traded funds (ETFs) listed in the United States, Canada, and Australia using daily U.S.-dollar returns over 2020--2024. A rolling-window framework compares minimum-variance and Conditional Value-at-Risk (CVaR) portfolio optimization under long-only and long--short constraints against an equally weighted benchmark. Performance is assessed using risk-adjusted measures, downside risk, extreme-tail estimation, and conditional volatility models. Optimized portfolios exhibit lower empirical VaR and CVaR than the benchmark, with long--short strategies generally providing greater downside-risk reduction. However, the equally weighted benchmark achieves the highest reported Sharpe, Sortino, and Rachev ratios. GARCH and FIGARCH comparisons favor GARCH based on information criteria. A historically estimated Normal Inverse Gaussian model further generates European option prices and model-implied volatility surfaces under a martingale-corrected pricing specification. The findings highlight trade-offs between risk reduction, risk-adjusted performance, and practical implementation costs.
\end{abstract}


\section{Introduction}
Balanced equity--bond and multi-asset portfolios have long occupied a central position in strategic asset allocation and institutional portfolio management. The classical justification for combining equities and fixed-income securities rests on imperfect correlations between growth-sensitive and defensive asset classes, thereby allowing investors to improve risk-adjusted returns through diversification \citep{markowitz1952,elton1997}. Multi-asset allocation strategies extend this diversification approach by incorporating additional asset classes with potentially different risk and return characteristics.

Over the last two decades, exchange-traded funds (ETFs) have become important vehicles for implementing such allocation strategies because of their transparency, liquidity, relatively low costs, and broad market accessibility \citep{madhavan2012,bendavid2018etfvolatility}.

Recent market developments, however, have raised important questions regarding the robustness of traditional diversification strategies. The period 2020--2024 included the COVID-19 crisis, substantial inflationary pressures, aggressive monetary tightening, and changes in global interest-rate expectations. These developments highlight the importance of evaluating portfolio strategies not only in terms of overall volatility and risk-adjusted returns, but also in terms of their exposure to downside risk and extreme market losses.

Traditional portfolio optimization methods, particularly the mean--variance framework, primarily characterize risk through return variance. Although such approaches provide a well-established foundation for portfolio allocation, variance treats positive and negative deviations from expected returns symmetrically. Consequently, portfolios constructed using variance-based optimization may not adequately account for asymmetric losses or extreme downside outcomes. This limitation is particularly relevant when evaluating equity--bond and multi-asset portfolios, where diversification across asset classes does not necessarily eliminate exposure to substantial losses during periods of market stress.

These considerations motivate the use of downside-risk-based portfolio optimization methods. In particular, Conditional Value-at-Risk (CVaR) provides a coherent risk measure that evaluates expected losses beyond a specified VaR threshold \citep{rockafellar_uryasev_2000,pflug2000}. Unlike variance, CVaR directly focuses on adverse outcomes in the tail of the loss distribution. Comparing minimum-variance and CVaR-based portfolio strategies therefore allows us to examine how alternative risk objectives influence portfolio allocation, downside exposure, and out-of-sample performance.

Beyond portfolio optimization, distributional characteristics such as skewness and heavy tails provide additional information about portfolio risk that may not be fully captured by variance-based measures. Examining these features can therefore contribute to a broader assessment of downside and extreme-tail risk across alternative portfolio strategies.

Another important challenge concerns the distributional assumptions underlying classical portfolio theory. Empirical asset returns frequently exhibit skewness, excess kurtosis, and persistent volatility dynamics inconsistent with Gaussian models \citep{mandelbrot1963,cont2001}.

L\'evy processes provide a flexible framework for modeling financial returns because they can accommodate jumps, heavy tails, and asymmetric return distributions while maintaining analytical tractability. Among these models, the Normal Inverse Gaussian (NIG) process introduced by \citet{barndorffnielsen1997} provides a particularly flexible specification for capturing distributional characteristics observed in financial return data.

This paper develops a unified empirical framework for analyzing equity--bond and multi-asset ETFs over the period 2020--2024, characterized by substantial market fluctuations and changing economic conditions. Our contribution is fourfold.

First, we provide a cross-country empirical analysis of equity--bond and multi-asset ETFs spanning the United States, Canada, and Australia. Second, we compare variance-based and CVaR-based portfolio optimization methods under both long-only and long--short constraints, examining downside risk, extreme-tail exposure, portfolio exposure, turnover, transaction-cost sensitivity, and out-of-sample portfolio performance. Third, we examine volatility persistence in the portfolio return series using GARCH and FIGARCH specifications. Fourth, we develop a model-based option-pricing analysis under the NIG L\'evy framework to complement the historical portfolio and risk analysis.

The empirical analysis considers portfolios constructed from thirty equity--bond and multi-asset ETFs representing a broad range of asset allocation strategies, including conservative income-oriented and growth-oriented portfolios. Portfolio performance is evaluated using a 504-trading-day rolling estimation window and subsequent out-of-sample daily simple returns. The evaluation incorporates multiple risk and performance measures, including Sharpe ratios, Sortino ratios, Rachev ratios, Value-at-Risk, Conditional Value-at-Risk, and extreme-tail estimators \citep{Lee2026TaiwanETFs}. Portfolio exposure, turnover, and sensitivity to proportional transaction costs are also examined to assess the implications of portfolio rebalancing.

In addition, volatility dynamics are examined using GARCH and FIGARCH models, while model-implied option prices and corresponding implied-volatility surfaces are generated under the NIG specification using Fourier-transform methods.

The empirical results show that the optimized portfolios exhibit lower out-of-sample daily volatility, maximum drawdown, VaR, and CVaR than the equally weighted benchmark. The long--short strategies consistently achieve lower values of these risk measures than their corresponding long-only portfolios. However, these reductions in realized risk do not consistently translate into higher Sharpe, Sortino, or Rachev ratios. The transaction-cost sensitivity analysis further shows that the performance of long--short strategies can be substantially affected by portfolio turnover, particularly under higher assumed transaction-cost rates.

The analysis also identifies persistent volatility dynamics, with the GARCH specification providing an adequate representation of the observed volatility clustering. Furthermore, the fitted NIG model captures important distributional characteristics of portfolio returns, including skewness and heavy-tailed behavior, and provides the basis for the model-implied option-pricing analysis.

The remainder of the paper is organized as follows. Section \ref{sec:2} describes the data, return construction procedures, and descriptive characteristics of the ETF universe. Section \ref{sec:3} presents the portfolio optimization framework. Section \ref{sec:portfolio-perfomance} examines empirical portfolio risk-return performance and extreme-tail behavior using Hill tail-index estimation. Section \ref{sec:long-range} examines volatility persistence using GARCH and FIGARCH models. Section \ref{sec:option-pricing} develops the model-based option-pricing analysis under the NIG L\'evy framework. Section \ref{conclusion} concludes.

\section{Data and Summary Statistics}\label{sec:2}
This section describes the dataset, return construction methodology, and principal empirical characteristics of the balanced and multi-asset ETF universe analyzed in this study. The objective is to establish a consistent cross-country framework for evaluating portfolio diversification and downside-risk behavior under changing market conditions.
\subsection{Data Description}
The empirical analysis is based on a sample of thirty equity–bond and multi-asset exchange-traded funds listed in three developed financial markets: the United States, Canada, and Australia. The sample consists of sixteen U.S. ETFs, ten Canadian ETFs, and four Australian ETFs. These funds collectively represent a broad spectrum of strategic allocation profiles, ranging from conservative income-oriented portfolios with relatively high bond exposure to aggressive growth-oriented portfolios dominated by equities.

Balanced ETFs provide diversified exposure to equity and fixed-income markets within a single tradable instrument and therefore constitute an economically meaningful laboratory for studying portfolio diversification under realistic investment conditions. Although several ETFs in the sample contain modest allocations to alternative assets such as real estate investment trusts, preferred securities, commodities, or tactical allocation overlays, the dominant sources of systematic risk remain the underlying equity and fixed-income exposures. Consistent with both academic literature and industry classification standards, the ETFs are therefore categorized according to their primary equity–bond allocation structure.

Daily adjusted closing prices are obtained from Yahoo Finance over the period January 2, 2020 through December 31, 2024. This sample period intentionally captures several economically distinct market environments, including the COVID-19 market collapse, the subsequent recovery phase, the inflation-driven monetary tightening cycle of 2022, and the later stabilization period. These changing market conditions provide a useful setting for evaluating diversification benefits and comparing alternative portfolio optimization frameworks.

In addition to the ETF universe, the Dow Jones Industrial Average (DJIA) is included as a benchmark representing broad U.S. equity market performance. The DJIA serves as a useful reference point for comparing the behavior of balanced portfolios with concentrated large-cap equity exposure during periods of market turbulence and recovery.

Daily ETF closing prices for funds listed in the United States, Canada, and Australia were expressed in U.S. dollars before calculating returns. Canadian-listed ETF prices were converted using the DEXCAUS daily exchange-rate series (Canadian dollars per U.S. dollar), while Australian-listed ETF prices were converted using the DEXUSAL series (U.S. dollars per Australian dollar), both obtained from the Federal Reserve Economic Data (FRED) database maintained by the Federal Reserve Bank of St. Louis \citep{FRED_DEXCAUS,FRED_DEXUSAL}.
ETF closing prices and exchange-rate observations were aligned to the available U.S. trading dates. For each date, the corresponding closing prices reported by Yahoo Finance for the U.S., Canadian, and Australian markets were matched according to their respective trading dates. Thus, observations associated with the same date represent closing prices from the respective local exchanges rather than prices recorded simultaneously. 

Differences in market closing times were not explicitly adjusted for, as the analysis was conducted using daily observations rather than intraday synchronized prices. When an exchange-rate observation was unavailable on a corresponding ETF observation date, the most recent preceding available exchange rate was used.
Under the rolling-window framework, portfolio weights were estimated using observations through date $t$ and applied to returns on date $t+1$. Although the exchanges operate in different time zones, the Australian and Canadian closing prices associated with date $t$ would ordinarily have been available by the close of the U.S. market on that date. Thus, calendar-date alignment of the ETF closing prices does not, by itself, require future ETF price information when weights are formed after the U.S. market close. However, the analysis does not explicitly verify the publication timing of the corresponding daily exchange rates or account for the timing of trade execution across exchanges.

Differences in international trading calendars resulted in missing ETF price observations on some U.S. trading dates. These missing values were imputed using a forward-filling procedure, whereby each missing ETF price was replaced with the most recent available price from a preceding trading date. In total, $374$ of $37,740$ ETF price observations ($0.99\%$) were imputed, occurring across $46$ of the $1,258$ dates in the sample. Because forward filling relies exclusively on previously available price observations, the imputation procedure does not introduce future price information. However, this approach may generate zero returns on dates when an ETF did not trade and may affect the measured return dynamics.

Accordingly, the reported out-of-sample results should be interpreted as a daily-frequency portfolio backtest based on date-aligned closing prices rather than as the performance of a fully synchronized, executable cross-market trading strategy. The analysis does not explicitly account for cross-market execution timing. Portfolio turnover and gross and short exposures are examined separately in Section~\ref{porfolio-exposure}, while the effects of proportional transaction costs are evaluated through a sensitivity analysis in Section~\ref{transaction_cost_sensitivity}. However, short-selling costs, financing costs, and other leverage-related implementation considerations are not explicitly incorporated.

\subsection{Return Construction and Risk-Free Rate}\label{return-construction}

Daily simple returns are computed for each ETF according to
\begin{equation*}
\label{eq:1}
R_{k,t} = \frac{P_{k,t}}{P_{k,t-1}}-1,
\end{equation*}
where $P_{k,t}$ denotes the adjusted closing price of ETF $k$ on date $t$, expressed in U.S. dollars. Daily simple returns are used in rolling portfolio optimization, out-of-sample performance evaluation, and historical tail-risk estimation. Daily logarithmic returns are used separately for the full-sample efficient-frontier analysis. 

To construct excess returns for portfolio performance evaluation, we use the 3-month U.S. Treasury yield as the short-term risk-free-rate proxy. Daily Treasury yields are aligned with the corresponding portfolio evaluation dates. When a Treasury yield is unavailable on a particular date, the most recent preceding available observation is used. Because the empirical analysis is conducted at the daily frequency, the annualized yields, expressed in percentages, are converted into daily simple equivalents under the assumption of 252 trading days per year:
\begin{equation*}
\label{eq:2}
R_{f,t}^{\text{daily}} =
\left(1+\frac{y_{f,t}}{100}\right)^{1/252}-1,
\end{equation*}
where $y_{f,t}$ denotes the annualized 3-month Treasury yield observed or carried forward to date $t$.

The resulting time-varying daily risk-free rate is used to calculate portfolio excess returns and the associated risk-adjusted performance measures. In addition, the 10-year U.S. Treasury yield is considered as an alternative benchmark in the efficient-frontier analysis, allowing comparisons under different Treasury reference-yield assumptions.

The use of date-aligned Treasury yields reflects changes in interest rates over the 2020--2024 sample period, particularly during the monetary tightening cycle. This return construction framework maintains consistency between the portfolio optimization and subsequent out-of-sample performance and risk evaluation.

\subsection{Descriptive Statistics and Stylized Facts}

Table~\ref{tab:balanced_etf_summary} summarizes the principal characteristics of the ETF sample, including ticker symbol, country, investment strategy, and strategic allocation or target exposure. The sample consists of
30 balanced and multi-asset ETFs from the United States, Canada, and Australia over the 2020--2024 study period. The funds were identified based on their balanced or multi-asset investment characteristics and the availability of historical data over the study period. The sample is not intended to represent an exhaustive universe of balanced and multi-asset ETFs.

The investment strategies represented in the sample range from conventional
equity--bond allocation portfolios to broader multi-asset strategies. Many
funds maintain strategic equity--fixed-income allocations, ranging from
conservative allocations with relatively low equity exposure to growth-oriented
allocations with substantially higher equity exposure. Other funds supplement
equity and fixed-income exposures with real estate, preferred securities, commodities, alternative strategies, tactical allocations, or derivative positions. Accordingly, these funds are more appropriately characterized as multi-asset rather than conventional balanced ETFs.

Despite this heterogeneity, equity and fixed-income exposures constitute the core of the investment universe, while the broader multi-asset funds allow the
analysis to examine equity--bond diversification in the presence of additional asset-class and strategy exposures. Strategic allocations and target exposures reported in Table~\ref{tab:balanced_etf_summary} are based on issuer
documentation applicable at or near the beginning of the sample where a fixed strategic allocation is available. Material strategy changes during the sample period and funds employing derivative-based exposures are identified in the table notes.

\begin{table}[H]
\centering
\caption{Balanced and Multi-Asset ETF Sample and Strategic Allocations.}
\label{tab:balanced_etf_summary}

\renewcommand{\arraystretch}{1.}
\resizebox{1\textwidth}{!}{%
\begin{tabular}{|l|l|l|l|}
\hline

\textbf{Ticker} &
\textbf{ETF Name} &
\textbf{Strategy / Allocation Type} &
\textbf{Allocation / Exposure} \\
\hline

\multicolumn{4}{|c|}{\textbf{United States}} \\
\hline

AOA & iShares Core Aggressive Allocation ETF
& Aggressive Allocation
& 80\% Equity / 20\% Fixed Income \\

AOR & iShares Core Growth Allocation ETF
& Growth Allocation
& 60\% Equity / 40\% Fixed Income \\

AOM & iShares Core Moderate Allocation ETF
& Moderate Allocation
& 40\% Equity / 60\% Fixed Income \\

AOK & iShares Core Conservative Allocation ETF
& Conservative Allocation
& 30\% Equity / 70\% Fixed Income \\

GAL & SPDR SSGA Global Allocation ETF
& Global Multi-Asset Allocation
& Dynamic \\

MDIV & First Trust Multi-Asset Diversified Income Index Fund
& Multi-Asset Income
& Dynamic \\

IYLD & iShares Morningstar Multi-Asset Income ETF
& Multi-Asset Income
& Dynamic \\

CVY & Invesco Zacks Multi-Asset Income ETF
& Multi-Asset Income
& Dynamic \\

RLY & SPDR SSGA Multi-Asset Real Return ETF
& Multi-Asset Real Return
& Dynamic \\

PWS & Pacer WealthShield ETF
& Tactical Allocation
& Dynamic \\

GAA & Cambria Global Asset Allocation ETF
& Global Multi-Asset (ETF-of-ETFs)
& 45\% Equity / 45\% Fixed Income / 10\% Other \\

TRTY & Cambria Trinity ETF
& Global Multi-Asset
& 25\% Equity / 25\% Fixed Income / 50\% Other \\

OCIO & ClearShares OCIO ETF
& Tactical Multi-Asset Allocation
& Dynamic \\

GDMA & Gadsden Dynamic Multi-Asset ETF
& Dynamic Multi-Asset
& Dynamic \\

NTSX & WisdomTree U.S. Efficient Core Fund
& Capital-Efficient Equity--Bond
& 90\% Equity + 60\% Treasury Futures \\

RPAR & RPAR Risk Parity ETF
& Risk-Parity Multi-Asset
& 25\% Equity / 70\% Fixed Income / 25\% Other \\

\hline
\multicolumn{4}{|c|}{\textbf{Canada}} \\
\hline

VBAL.TO & Vanguard Balanced ETF Portfolio
& Balanced Allocation
& 60\% Equity / 40\% Fixed Income \\

XBAL.TO & iShares Core Balanced ETF Portfolio
& Balanced Allocation
& 60\% Equity / 40\% Fixed Income \\

ZBAL.TO & BMO Balanced ETF
& Balanced Allocation
& 60\% Equity / 40\% Fixed Income \\

VGRO.TO & Vanguard Growth ETF Portfolio
& Growth Allocation
& 80\% Equity / 20\% Fixed Income \\

XGRO.TO & iShares Core Growth ETF Portfolio
& Growth Allocation
& 80\% Equity / 20\% Fixed Income \\

ZGRO.TO & BMO Growth ETF
& Growth Allocation
& 80\% Equity / 20\% Fixed Income \\

VCNS.TO & Vanguard Conservative ETF Portfolio
& Conservative Allocation
& 40\% Equity / 60\% Fixed Income \\

VCIP.TO & Vanguard Conservative Income ETF Portfolio
& Conservative Income Allocation
& 20\% Equity / 80\% Fixed Income \\

ZCON.TO & BMO Conservative ETF
& Conservative Allocation
& 40\% Equity / 60\% Fixed Income \\

XINC.TO & iShares Core Income Balanced ETF Portfolio
& Income Allocation
& 20\% Equity / 80\% Fixed Income \\

\hline
\multicolumn{4}{|c|}{\textbf{Australia}} \\
\hline

VDBA.AX & Vanguard Diversified Balanced Index ETF
& Diversified Balanced
& 50\% Equity / 50\% Fixed Income \\

VDGR.AX & Vanguard Diversified Growth Index ETF
& Diversified Growth
& 70\% Equity / 30\% Fixed Income \\

DBBF.AX & BetaShares Diversified Balanced ETF
& Diversified Balanced
& 47.5\% Equity / 40\% Fixed Income / 12.5\% Other \\

DGGF.AX & BetaShares Diversified Growth ETF
& Diversified Growth
& 66.5\% Equity / 25\% Fixed Income / 8.5\% Other \\

\hline
\end{tabular}%
}

\vspace{1mm}

\begin{minipage}{\textwidth}
\footnotesize
\textit{Notes:} Strategic allocations and target exposures are based on issuer documentation applicable at or near the beginning of the sample
period where a fixed allocation is available. ``Dynamic'' indicates that a fixed equity--fixed-income--other allocation does not adequately characterize the fund's strategy. ``Other'' aggregates exposures outside the conventional
equity and fixed-income categories, including, where applicable, commodities, gold, real estate or property, preferred securities, MLPs, cash, trend-following strategies, and alternative strategies. The composition of ``Other'' therefore
varies across funds. NTSX and RPAR use Treasury futures, so their target exposures may exceed 100\% of net assets. DBBF and DGGF underwent material
strategy changes effective 15 December 2020; the table reports their pre-change strategic allocations.
\end{minipage}

\end{table}

\begin{figure}[H]
\centering

\begin{subfigure}{0.38\textwidth}
    \centering
    \includegraphics[width=\linewidth]{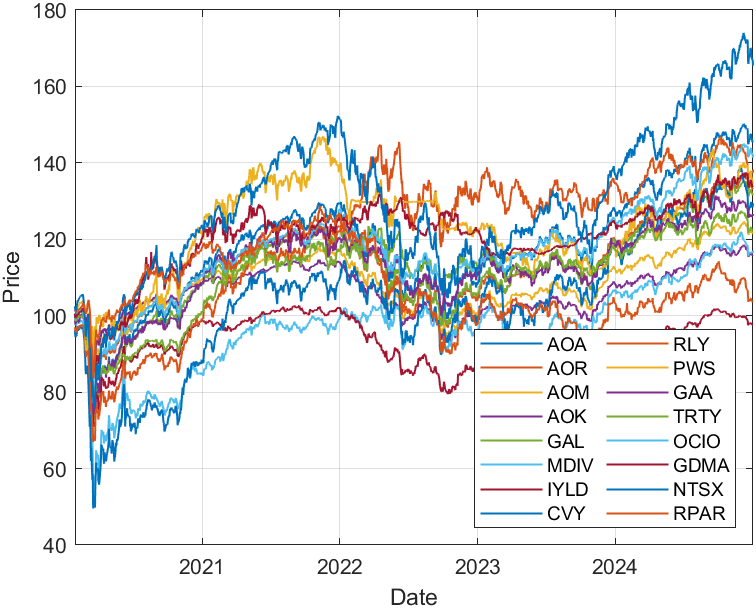}
    \caption{American}
\end{subfigure}
\begin{subfigure}{0.38\textwidth}
    \centering
    \includegraphics[width=\linewidth]{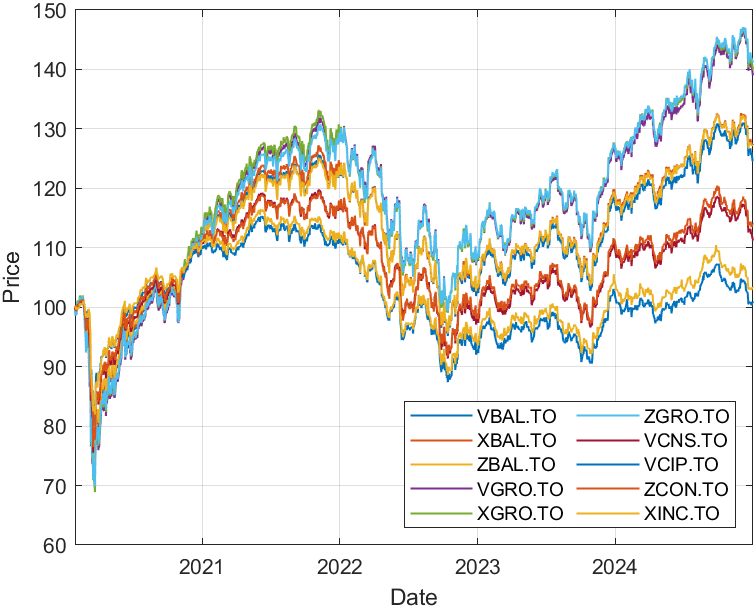}
    \caption{Canadian}
\end{subfigure}

\medskip

\begin{subfigure}{0.38\textwidth}
    \centering
    \includegraphics[width=\linewidth]{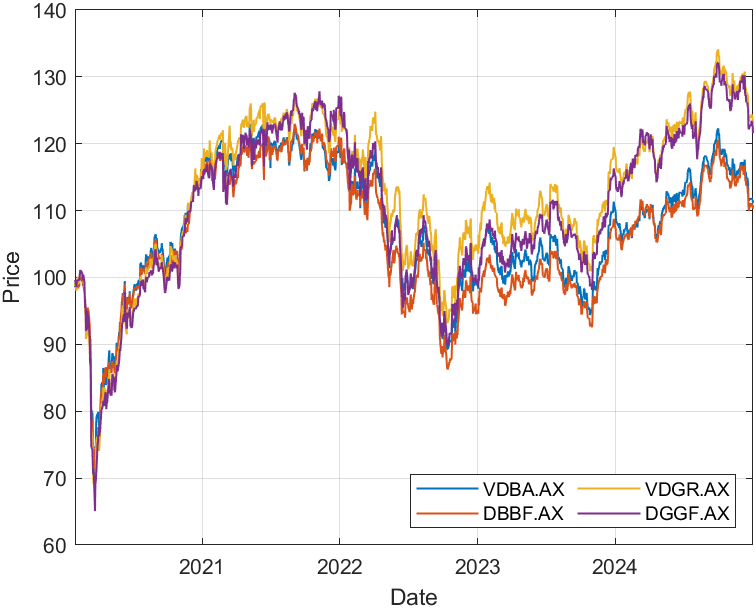}
    \caption{Australian}
\end{subfigure}

\caption{Comparative growth trajectories of selected Equity–Bond and multi-asset ETFs, normalized to \$100 on January 2, 2020, to facilitate direct performance comparisons.}
\label{fig:image_grid}
\end{figure}

\begin{figure}[H]
\centering

\begin{subfigure}{0.4\textwidth}
    \centering
    \includegraphics[width=\linewidth]{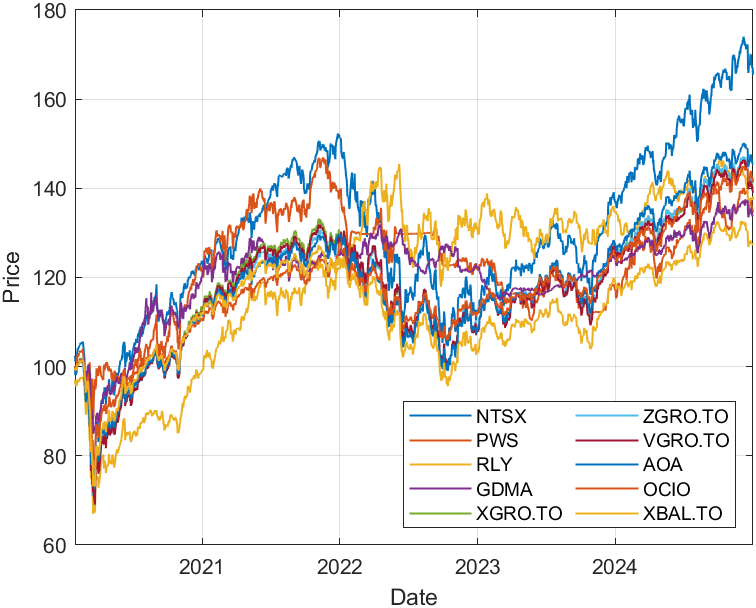}
    \caption{Top 10 ETFs}
\end{subfigure}
\begin{subfigure}{0.4\textwidth}
    \centering
    \includegraphics[width=\linewidth]{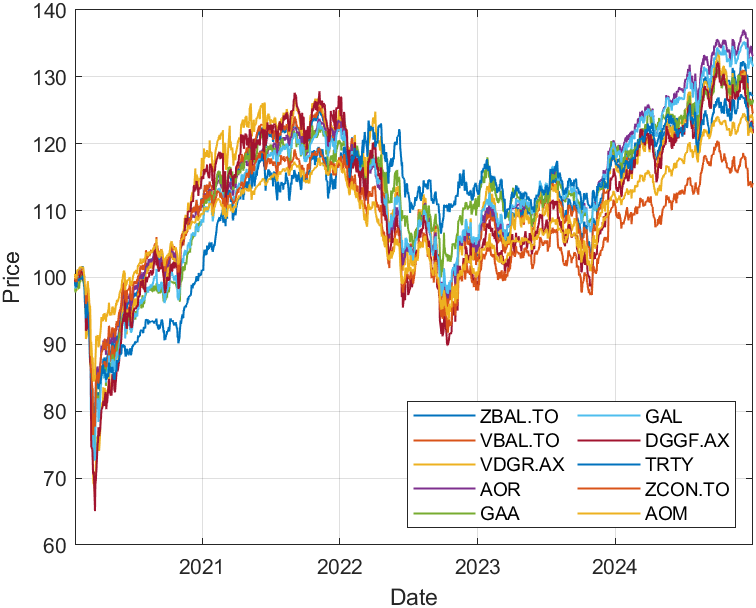}
    \caption{Middle 10 ETFs}
\end{subfigure}

\medskip

\begin{subfigure}{0.4\textwidth}
    \centering
    \includegraphics[width=\linewidth]{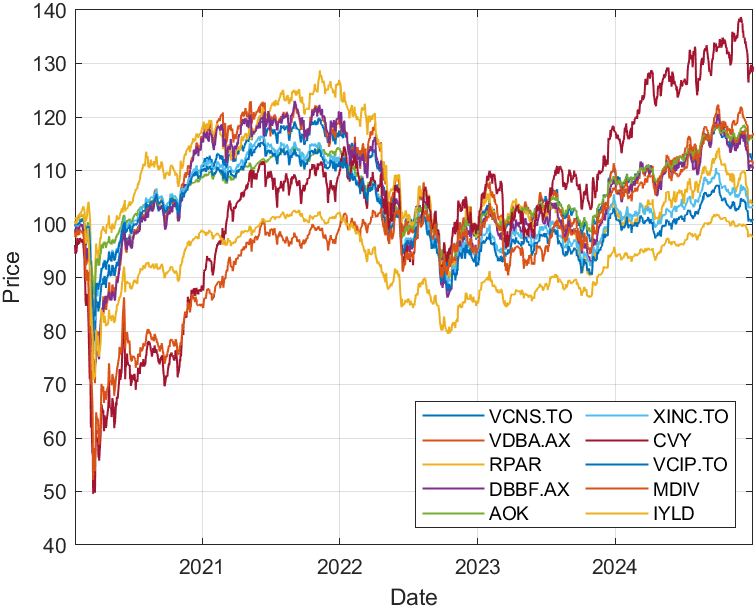}
    \caption{Bottom 10 ETFs}
\end{subfigure}

\caption{Cross-sectional classification of Equity–Bond ETFs by mean normalized price over the sample period (2020–2024).}
\label{fig:image_grid_mean}
\end{figure}

Figure \ref{fig:image_grid} presents normalized cumulative price dynamics for the ETF universe over the sample period. The three regional groups display noticeable differences in their cumulative price trajectories. U.S. ETFs exhibit the greatest dispersion across the sample, reflecting substantial differences in the performance paths of the individual funds. Canadian ETFs are more tightly clustered, particularly among funds with similar allocation profiles, although their cumulative performance begins to diverge more clearly toward the end of the sample. Australian ETFs also remain relatively clustered, with differences in cumulative performance becoming more pronounced over time.
Across all three regions, a sharp decline is visible in early 2020, followed by a strong recovery through 2021. The trajectories subsequently weaken during 2022, with many ETFs experiencing substantial declines or relatively flat performance. From 2023 onward, the funds generally recover, although the magnitude of the recovery differs considerably across ETFs and regions. By the end of 2024, the U.S. ETF universe displays the widest range of cumulative outcomes, while the Canadian and Australian ETFs remain comparatively more concentrated.
Figure \ref{fig:image_grid_mean} further groups the ETFs according to their mean normalized performance over the sample period. The high-, middle-, and low-mean groups display clear differences in their cumulative price trajectories, highlighting substantial cross-sectional variation within the ETF universe. The high-mean group contains several of the strongest cumulative growth paths toward the end of the sample, while the middle-mean group exhibits more concentrated performance trajectories. The low-mean group generally shows weaker cumulative growth, although considerable variation remains within the group.
Despite these differences, the three groups share several common patterns over time. A pronounced decline occurs in early 2020, followed by a broad recovery through 2021. Performance weakens again during 2022, after which most ETFs recover to varying degrees through 2023 and 2024. The variation both across and within the mean-performance groups motivates the subsequent portfolio analysis, where differences in return, volatility, and downside risk are examined more systematically.

\section{Portfolio Optimization Framework}\label{sec:3}
\begin{figure}[H]
\centering
\includegraphics[width=0.51\textwidth]{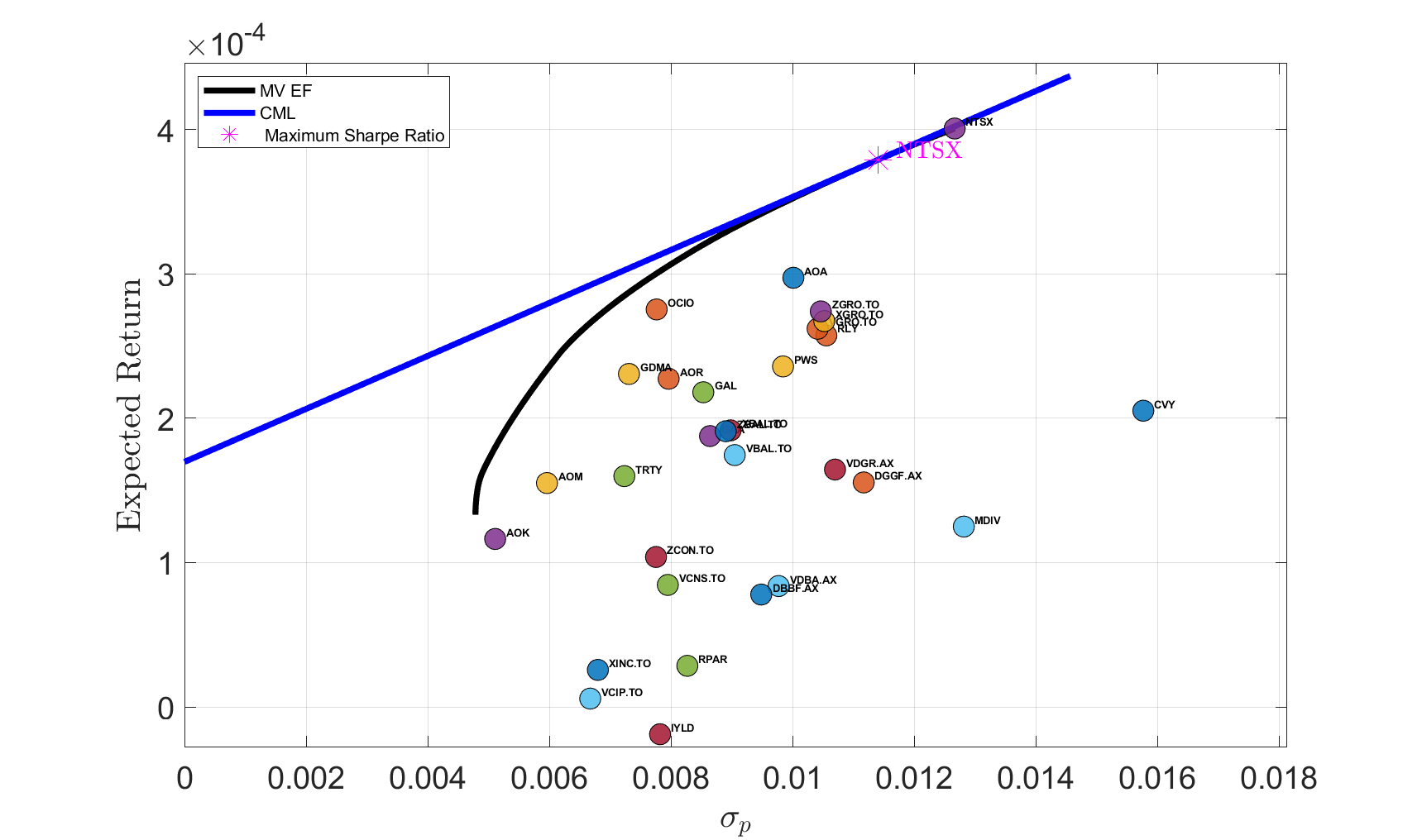}%
\includegraphics[width=0.50\textwidth]{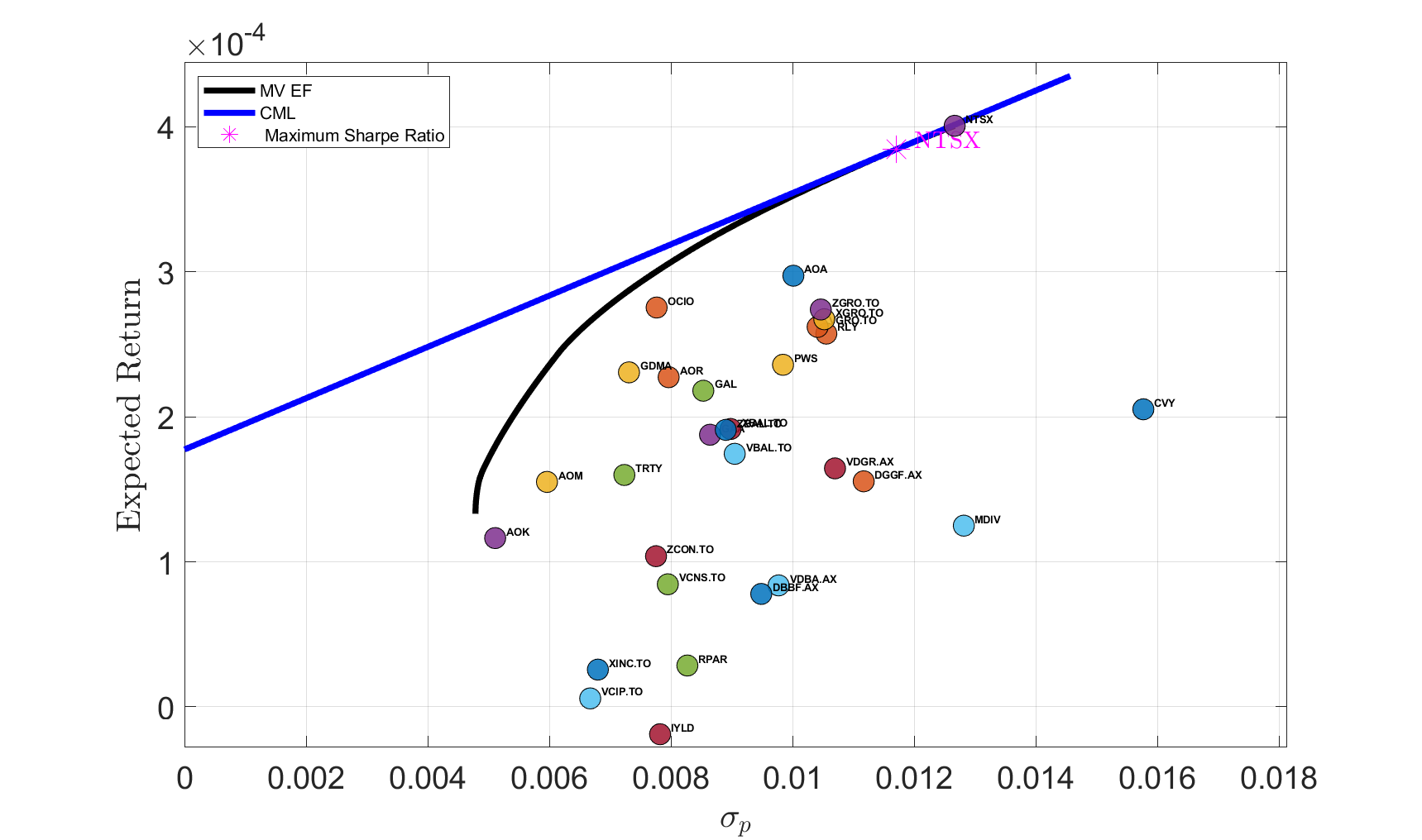}
\caption{Mean--variance efficient frontiers based on historical returns using (left) the 3-month U.S. Treasury yield and (right) the 10-year U.S. Treasury yield as alternative risk-free benchmarks.}
\label{fig:markowitz_compare}
\end{figure}

\begin{figure}[H]
\centering
\includegraphics[width=0.50\textwidth]{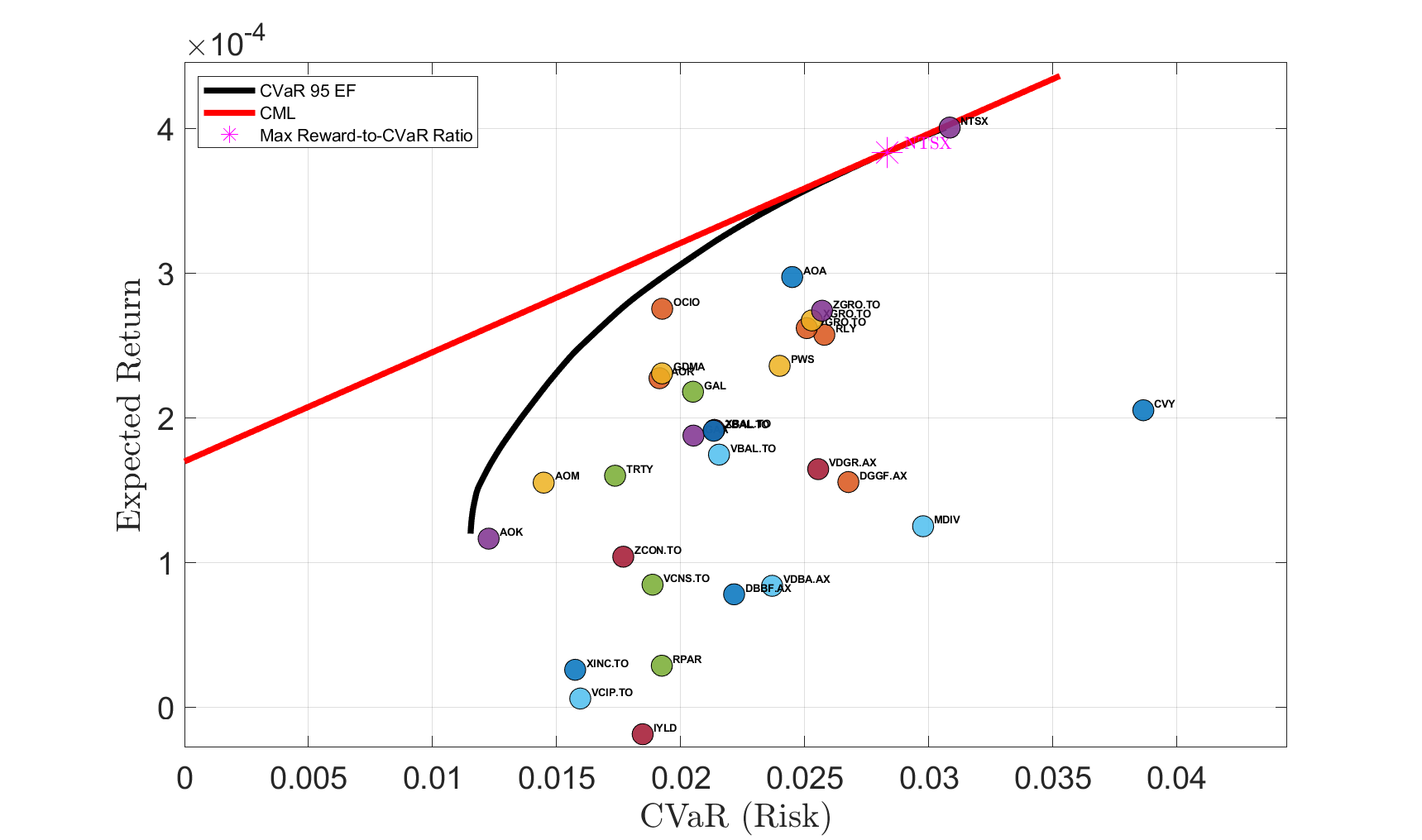}%
\includegraphics[width=0.50\textwidth]{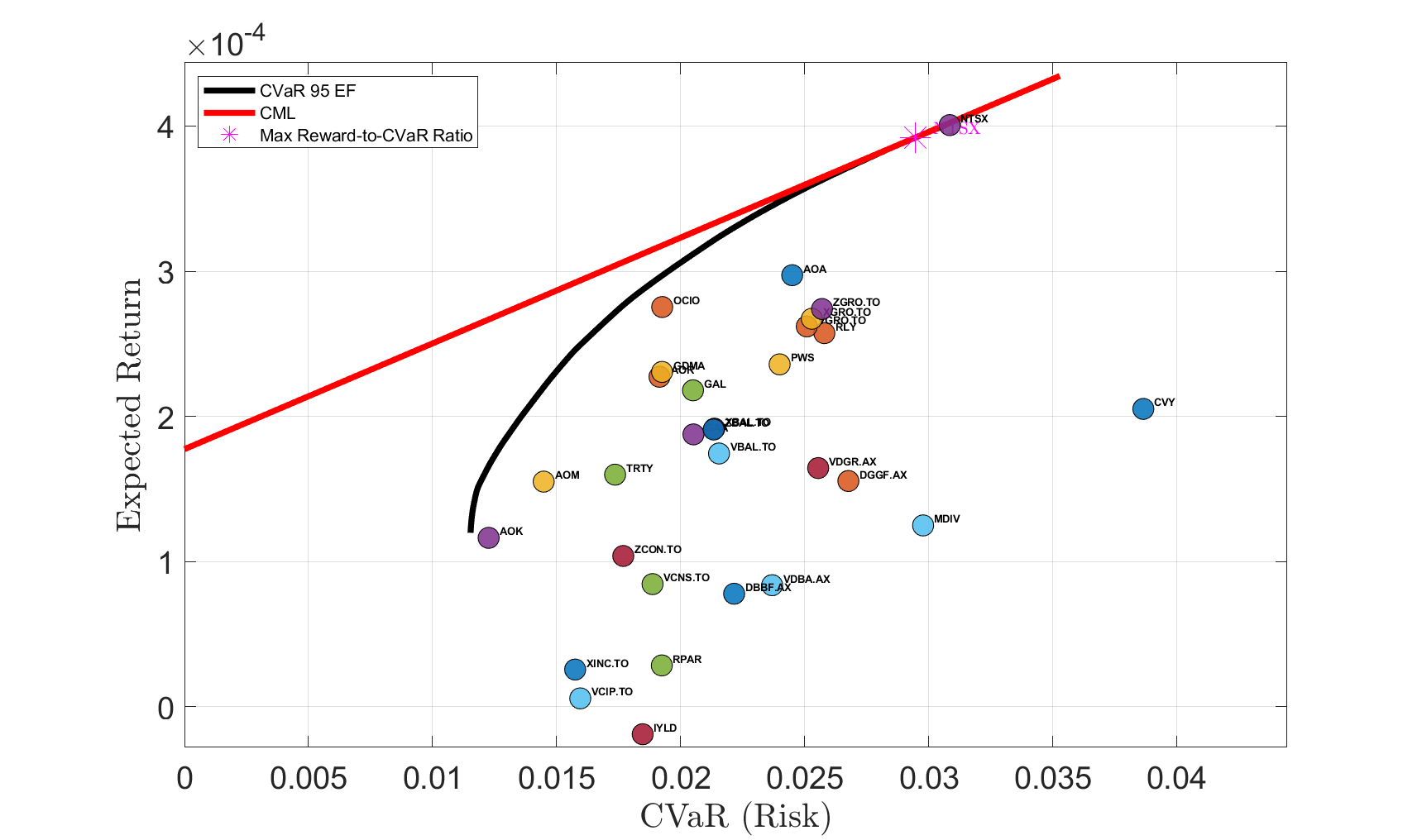}
\caption{CVaR (95\%) efficient frontiers based on historical returns using (left) the 3-month U.S. Treasury yield and (right) the 10-year U.S. Treasury yield as alternative risk-free benchmarks.}
\label{fig:cvar_compare}
\end{figure}

This section develops the portfolio optimization framework used to evaluate alternative allocation strategies under changing market conditions. The objective is to examine how different notions of portfolio risk behave in the presence of heavy-tailed return distributions, unstable correlation structures, and persistent volatility dynamics. Particular emphasis is placed on the comparison between variance-based and downside-risk-based portfolio construction methods.
To account for the evolving nature of financial markets, we employ a rolling-window estimation framework. Portfolio weights are estimated using a fixed window of 504 trading days, corresponding approximately to two years of historical observations. For each out-of-sample return date $t$, target weights $w_{p,k,t}$ are estimated using the preceding 504 daily return observations, indexed from $t-504$ through $t-1$, and are applied to the simple ETF returns realized on date $t$. The estimation window subsequently advances by one trading day and the procedure is repeated throughout the sample period. This rolling procedure updates allocations as new observations become available and allows the estimated portfolio weights to respond to changes in historical volatility, correlations, and downside-risk characteristics.
Figures~\ref{fig:markowitz_compare} and \ref{fig:cvar_compare} present the full-sample (in-sample) efficient frontiers obtained under the mean--variance and CVaR optimization frameworks using the 3-month U.S. Treasury yield as a short-term risk-free-rate proxy and the 10-year U.S. Treasury yield as an alternative reference yield.

The full-sample efficient frontiers are constructed using daily logarithmic returns calculated from the forward-filled ETF prices expressed in U.S. dollars. These in-sample frontier calculations are distinct from the rolling portfolio optimizations and out-of-sample wealth calculations, which use daily simple returns. For the frontier comparisons, the annualized Treasury yields are converted to daily log-equivalent rates using $\log(1+y/100)/252$, where $y$ denotes the annualized yield in percent. The reference rate for each comparison is based on the Treasury yield observed on December 31, 2024. These reference rates are used to identify the maximum reward-to-risk portfolios along the estimated frontiers.

Under both frameworks, the individual ETFs occupy a wide range of risk--return positions relative to the corresponding efficient frontiers. The use of the two Treasury benchmarks also allows the sensitivity of the optimal portfolio allocation to the choice of reference yield to be examined. The mean--variance framework measures portfolio risk through return variance, whereas the CVaR framework focuses on losses in the lower tail of the return distribution. The two approaches therefore evaluate portfolio risk from different perspectives.

Under the mean--variance framework, the maximum-Sharpe portfolio is strongly concentrated in NTSX under both risk-free-rate specifications. Using the 3-month Treasury yield, the optimal portfolio consists of 86.882\% NTSX and 13.118\% PWS. When the 10-year Treasury yield is used, the allocation changes to 90.161\% NTSX and 9.839\% PWS. Thus, NTSX remains the dominant component under both specifications, while the allocation to PWS decreases when the 10-year Treasury yield is used as the alternative reference yield.

A similar pattern is observed under the 95\% CVaR framework. Using the 3-month Treasury yield, the maximum reward-to-CVaR portfolio assigns 89.672\% of the portfolio weight to NTSX and 10.328\% to PWS. With the 10-year Treasury yield, the optimal portfolio consists of 94.836\% NTSX and 5.164\% PWS. The CVaR results therefore also place the dominant portfolio weight on NTSX, although the allocation changes with the choice of risk-free benchmark.

Taken together, the four specifications show a consistent concentration in NTSX, with a smaller allocation to PWS appearing in all four optimal portfolios. The precise weights vary across the mean--variance and CVaR frameworks and between the two Treasury benchmarks, indicating that both the definition of portfolio risk and the choice of reference yield affect the resulting optimal allocation. The concentration in NTSX should also be interpreted in light of the structure of the fund itself. NTSX combines U.S. equity exposure with U.S. Treasury futures exposure and therefore incorporates both equity and fixed-income components within a single ETF. Consequently, the large weight assigned to NTSX does not represent exposure to a single underlying asset class; rather, the optimization procedures place substantial weight on an ETF that itself combines equity and Treasury exposures.

\medskip

Having examined the full-sample efficient frontiers under the mean--variance and CVaR frameworks, we next turn to the broader empirical comparison of the portfolio strategies considered in this study. We consider three portfolio construction approaches:
\begin{itemize}
    \item mean--variance optimization,
    \item Conditional Value-at-Risk optimization,
    \item the equally weighted portfolio (EWP) as a benchmark specification.
\end{itemize}
The empirical analysis evaluates both long-only and long--short portfolio configurations in order to examine how portfolio constraints affect diversification and downside-risk management. The portfolio strategies considered throughout the paper are defined as follows:
\begin{itemize}
\item LO-MVP: Long-only minimum-variance portfolio,
\item LO-C95: Long-only CVaR portfolio at the 95\% confidence level,
\item LO-C99: Long-only CVaR portfolio at the 99\% confidence level,
\item LS-MVP: long--short minimum-variance portfolio,
\item LS-C95: long--short CVaR portfolio at the 95\% confidence level,
\item LS-C99: long--short CVaR portfolio at the 99\% confidence level.
\end{itemize}

\subsection{Mean--Variance Optimization}

Let $r_t \in \mathbb{R}^N$ denote the vector of asset returns at time $t$, and let $\Sigma$ represent the sample covariance matrix estimated within the rolling estimation window. The minimum-variance portfolio (MVP) considered in the empirical analysis is obtained by solving
\[
\min_{w} \quad w^\top \Sigma w
\quad \text{subject to} \quad
\mathbf{1}^\top w = 1,
\]
where $w \in \mathbb{R}^N$ denotes the portfolio weight vector.

For the long-only specification, the portfolio weights satisfy
$0 \leq w_i \leq 1$, whereas for the long--short specification, individual
asset weights are bounded by $-0.3 \leq w_i \leq 1.3$. These bounds permit
short positions in the long--short portfolios while restricting the position taken in each individual asset. No explicit turnover constraint is imposed between consecutive portfolio rebalancing dates. The main reported portfolio performance is evaluated before transaction costs; a separate sensitivity analysis examines proportional transaction costs.

The mean--variance framework introduced by \citet{markowitz1952} remains one of the foundational models in modern portfolio theory because it provides a tractable mechanism for balancing expected return against total portfolio variance. For the minimum-variance strategy studied here, the optimization objective depends on the estimated covariance matrix and treats positive and negative return deviations symmetrically. It therefore does not directly target extreme downside losses.

\subsection{Conditional Value-at-Risk Optimization}

To address the limitations of variance-based risk measures, we consider Conditional Value-at-Risk (CVaR), which explicitly focuses on downside losses beyond the Value-at-Risk threshold. Let portfolio loss be defined as
\[
\ell_t(w) = -w^\top r_t.
\]
Following \citet{rockafellar_uryasev_2000}, the empirical CVaR representation is based on the auxiliary objective function
\begin{equation*}
F_\alpha(w,\nu)
=
\nu + \frac{1}{(1-\alpha)T}
\sum_{t=1}^{T}
\max\bigl(0,\,\ell_t(w)-\nu\bigr),
\end{equation*}
where $\nu$ is an auxiliary threshold variable and $\alpha$ is the confidence level. For a given portfolio $w$, empirical CVaR is obtained by minimizing this objective over $\nu$:
\begin{equation*}
\mathrm{CVaR}_\alpha(w)
=
\min_{\nu\in\mathbb{R}} F_\alpha(w,\nu).
\end{equation*}
The corresponding portfolio optimization problem is
\begin{equation*}
\begin{aligned}
\min_{w,\nu}\quad&
\nu + \frac{1}{(1-\alpha)T}
\sum_{t=1}^{T}
\max\bigl(0,\,\ell_t(w)-\nu\bigr)\\
\text{subject to}\quad&
\mathbf{1}^\top w=1,\\
& w\in\mathcal{W},
\end{aligned}
\end{equation*}
where the feasible weight set is defined as
\[
\mathcal{W}=
\begin{cases}
\{w:0\leq w_i\leq 1\}, & \text{long-only},\\
\{w:-0.3\leq w_i\leq 1.3\}, & \text{long--short},
\end{cases}
\qquad i=1,\ldots,N.
\]

Unlike variance minimization, CVaR optimization explicitly targets losses in the lower tail of the empirical return distribution. With the stated linear portfolio constraints, the sample-based problem is convex and admits a linear-programming formulation \citep{lindquist2022}.

\subsection{Equally Weighted Portfolio}
As a benchmark specification, we consider the equally weighted portfolio (EWP), defined by
\[
w_i = \frac{1}{N}, \quad i = 1,\dots,N.
\]
The equally weighted portfolio remains an important benchmark in empirical asset allocation studies because of its robustness to parameter estimation error and its relative insensitivity to model misspecification. Although simplistic from an optimization perspective, the equally weighted approach often performs competitively in practice, particularly when covariance and expected-return estimates are unstable.

The EWP therefore serves as a useful reference point for evaluating whether optimization-based strategies generate economically meaningful improvements in risk-adjusted performance.

\subsection{Portfolio Constraints}
The empirical analysis considers both long-only and long--short portfolio specifications.
In the long-only framework, portfolio weights satisfy
\[
w_i \ge 0,
\quad
\sum_{i=1}^{N} w_i = 1.
\]
Under the long--short framework, negative portfolio weights are permitted, thereby allowing the investor to exploit relative-value opportunities across assets and hedge undesirable exposures more effectively. Following the constrained framework used in the empirical portfolio literature \citep{lindquist2022}, portfolio weights satisfy $w_i \in [-0.3,1.3], \quad i=1,\dots,N$, which restricts individual asset positions while allowing substantial aggregate gross exposure. 

Long--short allocations permit the optimizer to take negative positions in selected ETFs, although these positions can also increase aggregate exposure, trading activity, and implementation costs.

\subsection{Portfolio Wealth Construction}
Portfolio performance is evaluated using daily simple ETF returns (Section~\ref{return-construction}) and portfolio weights obtained from the preceding 504 trading days under the optimization frameworks described above. For each out-of-sample evaluation date $t$, the daily simple return of portfolio $p$ is calculated as
\begin{equation*}
R_{p,t} = \sum_{k=1}^{N} w_{p,k,t}R_{k,t},
\end{equation*}
where $w_{p,k,t}$ denotes the target weight assigned to ETF $k$ using information available through date $t-1$, $R_{k,t}$ is the corresponding daily simple ETF return, and $N$ is the number of ETFs.
The cumulative wealth of portfolio $p$ is constructed recursively as
\begin{equation*}
V_{p,t} = V_{p,t-1}(1+R_{p,t}),
\end{equation*}
with initial wealth $V_{p,0}=\$100$. Equivalently,
\begin{equation*}
V_{p,t} = V_{p,0}\prod_{s=1}^{t}(1+R_{p,s}).
\end{equation*}
The resulting wealth series represents the compounded out-of-sample portfolio performance before transaction costs and other implementation expenses.

\subsection{Portfolio Exposure and Turnover}\label{porfolio-exposure}

To examine the exposure characteristics and trading activity of the optimized portfolios, we calculate gross exposure, short exposure, and daily portfolio turnover over the rolling out-of-sample evaluation period. These measures provide additional information about the portfolio allocations and their potential implementation requirements under the long-only and long--short investment constraints.

Gross exposure is defined as
\begin{equation*}
G_{p,t} = \sum_{k=1}^{N}|w_{p,k,t}|,
\end{equation*}
where $w_{p,k,t}$ denotes the target weight assigned to ETF $k$ in portfolio $p$ at rebalancing date $t$, and $N=30$ is the number of ETFs. Short exposure is calculated as
\begin{equation*}
S_{p,t} = \sum_{k=1}^{N}\max(-w_{p,k,t},0).
\end{equation*}
Gross exposure measures the total absolute portfolio allocation, while short exposure measures the aggregate magnitude of negative portfolio weights.

Daily portfolio turnover is calculated as
\begin{equation*}
T_{p,t} = \frac{1}{2}\sum_{k=1}^{N}
\left|w_{p,k,t}-\widetilde{w}_{p,k,t}\right|,
\end{equation*}
where $\widetilde{w}_{p,k,t}$ denotes the portfolio weight
immediately before rebalancing, after accounting for changes
in asset values during the preceding holding period.
It is calculated as
\begin{equation*}
\widetilde{w}_{p,k,t}
=
\frac{w_{p,k,t-1}(1+R_{k,t-1})}
{1+R_{p,t-1}},
\end{equation*}
where $R_{k,t-1}$ and $R_{p,t-1}$ denote the simple ETF
and portfolio returns over the preceding holding period. The turnover measure therefore accounts for changes in portfolio weights arising from asset returns rather than measuring only changes in target allocations. Turnover is calculated between consecutive out-of-sample portfolio allocations, excluding the initial portfolio establishment.

Table~\ref{tab:exposure_turnover} summarizes the mean and maximum gross and short exposures, together with mean daily turnover, across the seven portfolio strategies.

\begin{table}[htbp]
\centering
\caption{Gross Exposure, Short Exposure, and Daily Portfolio Turnover}
\label{tab:exposure_turnover}
\small
\begin{tabular}{lccccc}
\hline
Strategy & Mean Gross & Max Gross & Mean Short & Max Short & Mean Daily Turnover \\
\hline
LO-MVP & 1.0000 & 1.0000 & 0.0000 & 0.0000 & 0.47\% \\
LO-C95 & 1.0000 & 1.0000 & 0.0000 & 0.0000 & 0.89\% \\
LO-C99 & 1.0000 & 1.0000 & 0.0000 & 0.0000 & 1.17\% \\
LS-MVP & 4.0854 & 4.6310 & 1.5427 & 1.8155 & 4.61\% \\
LS-C95 & 5.0798 & 5.7384 & 2.0399 & 2.3692 & 12.30\% \\
LS-C99 & 5.3942 & 6.6364 & 2.1971 & 2.8182 & 21.70\% \\
EWP    & 1.0000 & 1.0000 & 0.0000 & 0.0000 & 0.17\% \\
\hline
\end{tabular}
\end{table}

As expected, the long-only portfolios and the equally weighted portfolio maintain gross exposure of 1 and zero short exposure. Their mean daily turnover ranges from 0.17\% for EWP to 1.17\% for LO-C99. In contrast, the long--short portfolios exhibit substantially higher gross and short exposures. Mean gross exposure ranges from 4.0854 for LS-MVP to 5.3942 for LS-C99, while mean short exposure ranges from 1.5427 to 2.1971. The LS-C99 strategy also records the highest maximum gross exposure of 6.6364 and the highest mean daily turnover of 21.70\% among the seven strategies.

Under the long--short optimization framework, individual portfolio weights are constrained by $-0.30\leq w_{p,k,t}\leq 1.30$, with the full-investment condition $\sum_{k=1}^{N}w_{p,k,t}=1$. No separate constraint is imposed on aggregate gross exposure. Consequently, the individual weight bounds do not prevent the portfolios from taking substantial gross and short positions. The higher exposures and turnover of the long--short strategies indicate greater potential implementation requirements, including transaction costs, short-selling costs, and financing considerations. The main portfolio performance results are reported before these implementation costs. Transaction costs are examined separately through a sensitivity analysis, while short-borrowing fees and financing costs are not explicitly modeled.


\section{Portfolio Performance and Risk Evaluation}\label{sec:portfolio-perfomance}

\begin{figure}[H]
\centering
\begin{subfigure}{0.48\textwidth}
    \centering
    \includegraphics[width=\linewidth]{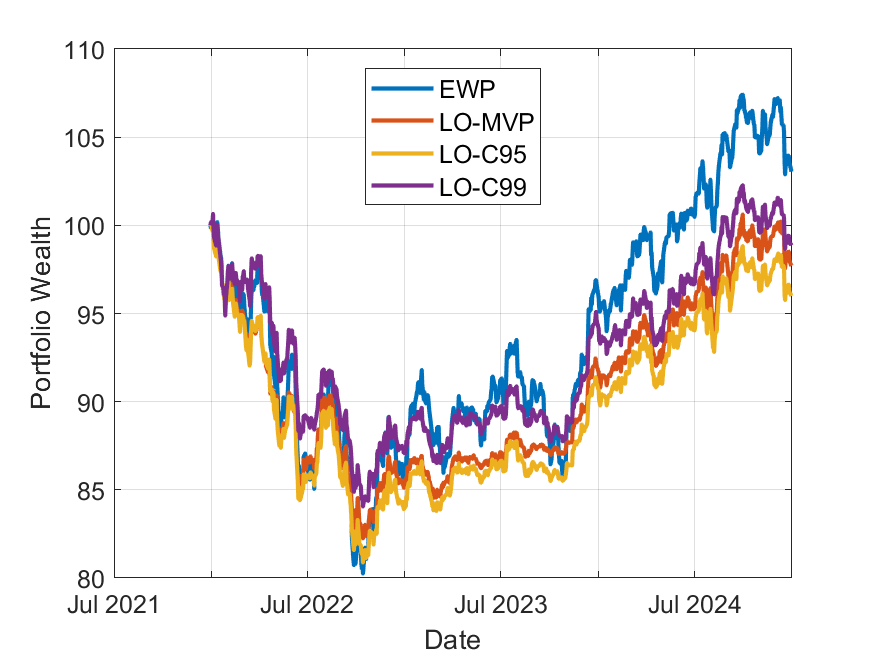}
    \caption{Long-only portfolios}
    \label{fig:long-only}
\end{subfigure}%
\begin{subfigure}{0.48\textwidth}
    \centering
    \includegraphics[width=\linewidth]{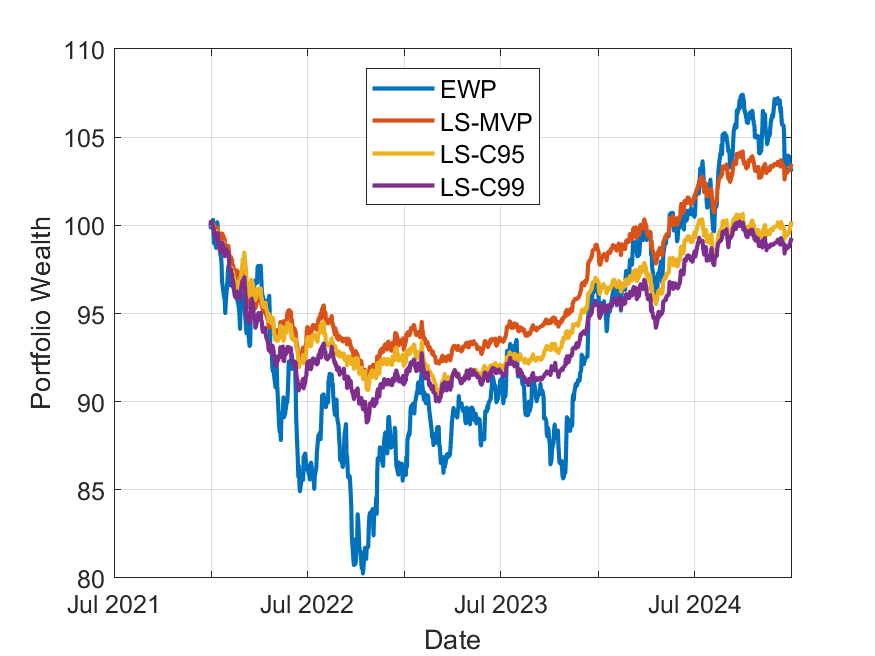}
    \caption{long--short portfolios}
    \label{fig:long--short}
\end{subfigure}

\caption{Cumulative out-of-sample portfolio wealth (normalized to \$100) for long-only and long--short strategies and the equally weighted benchmark, December 31, 2021--December 31, 2024.}
\label{fig:portfolio-comparison}
\end{figure}

Figure~\ref{fig:portfolio-comparison} compares the cumulative performance of long-only and long–short portfolio strategies constructed under the alternative optimization frameworks. Several economically important findings emerge from the empirical analysis.

The cumulative portfolio trajectories reveal clear differences between the long-only and long–short specifications. Within the long-only framework, the optimized portfolios follow relatively similar paths, while the equally weighted portfolio exhibits larger fluctuations over the evaluation period. The EWP experiences a deeper decline, particularly during 2022, followed by a stronger recovery, and finishes above the long-only optimized portfolios by the end of the sample.
The long--short portfolios display a different pattern.
Among these strategies, LS-MVP achieves the highest terminal portfolio wealth, finishing slightly above EWP. In contrast, LS-C95 and LS-C99 finish below EWP.
These results indicate that lower portfolio volatility does not necessarily translate into higher cumulative wealth.

Overall, the cumulative-value paths indicate that both the optimization criterion and portfolio constraints affect the realized portfolio trajectory. However, cumulative performance alone does not provide a complete assessment of portfolio quality. The subsequent analysis therefore evaluates the strategies using complementary measures of risk-adjusted performance, downside risk, and extreme-tail behavior.

\subsection{Portfolio Risk--Return Performance}\label{measures}

To evaluate portfolio performance comprehensively, we consider several complementary risk-adjusted
performance and risk measures, including the Sharpe ratio, Sortino ratio, daily volatility, Rachev ratio, and maximum drawdown. Each metric captures a different dimension of portfolio behavior and therefore provides insight into the interaction between return generation, downside exposure, and tail-risk sensitivity. For the risk-adjusted measures requiring a risk-free rate, the 3-month U.S. Treasury yield is used as the risk-free-rate proxy and converted to an effective daily rate to match the frequency of the portfolio returns.

The Sharpe ratio evaluates excess return relative to total return volatility and remains one of the most widely used measures of mean--variance efficiency \citep{sharpe1966}. However, because the Sharpe ratio relies on variance as the underlying risk measure, it may provide a misleading characterization of performance when returns exhibit asymmetry, excess kurtosis, or substantial tail dependence.

The Sortino ratio partially addresses this limitation by focusing exclusively on downside volatility rather than total return dispersion \citep{sortino_classical1994}. Nevertheless, both Sharpe and Sortino ratios remain fundamentally moment-based performance measures and therefore may still inadequately capture extreme downside exposure under heavy-tailed return distributions.

To incorporate tail asymmetry more explicitly, we additionally consider the Rachev ratio, which compares expected upper-tail gains with expected lower-tail losses \citep{rachev2005}. In this analysis, the Rachev ratio is computed from daily excess returns, defined as the portfolio return minus the corresponding daily risk-free rate, with the upper and lower tails each defined using a 5\% tail probability. The denominator is expressed as the absolute magnitude of the expected lower-tail return. Unlike traditional moment-based performance metrics, the Rachev ratio directly evaluates the balance between favorable and unfavorable tail events and is therefore particularly appropriate in the context of non-Gaussian portfolio returns.
\begin{figure}[H]
    \centering
   \begin{subfigure}[b]{0.43\textwidth}
        \includegraphics[width=\textwidth]{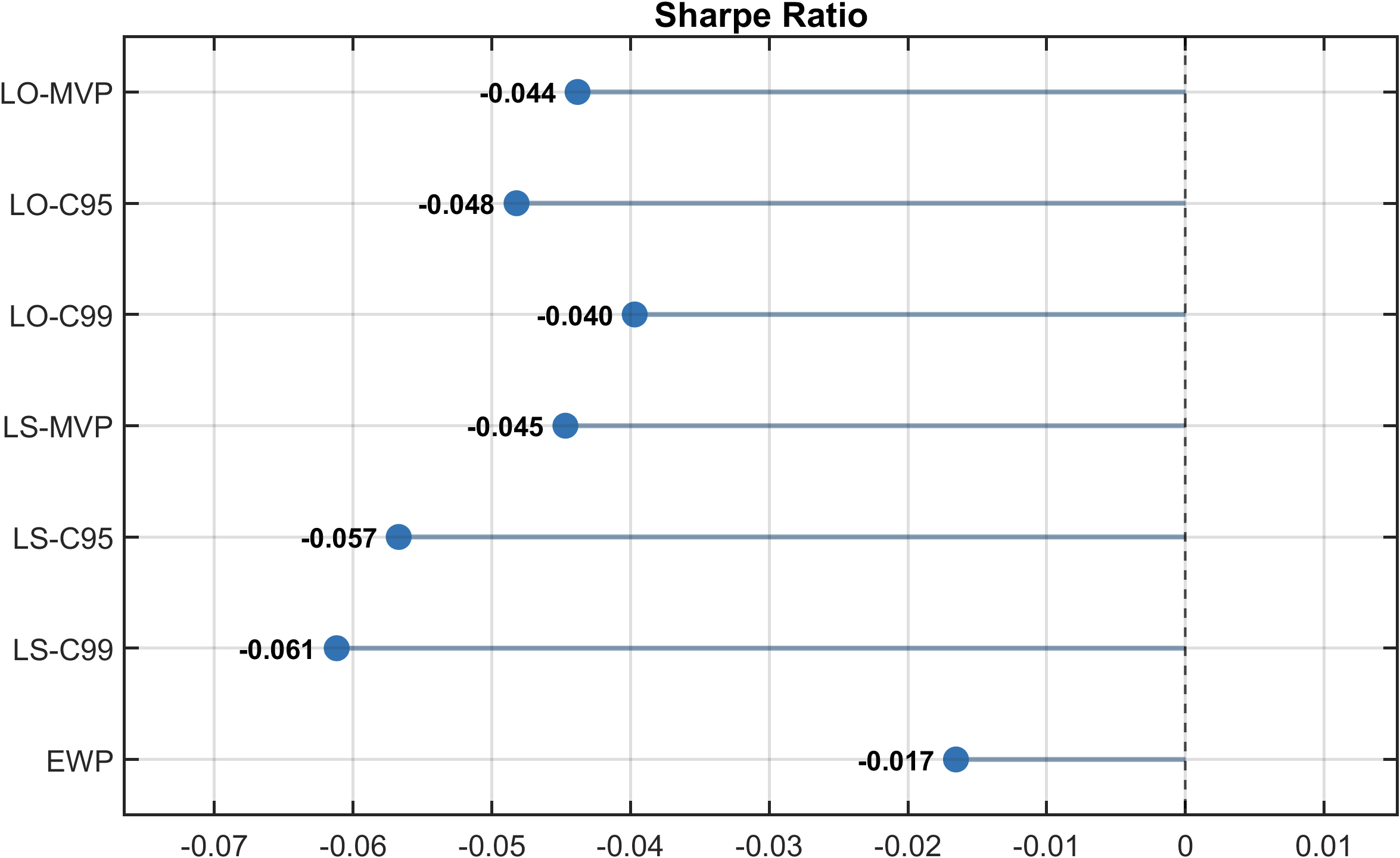}
        \caption{}
        \label{fig:sharpe}
    \end{subfigure}
     \hspace{0.02\textwidth}
     \vspace{0.02\textwidth}
    \begin{subfigure}[b]{0.43\textwidth}
        \includegraphics[width=\textwidth]{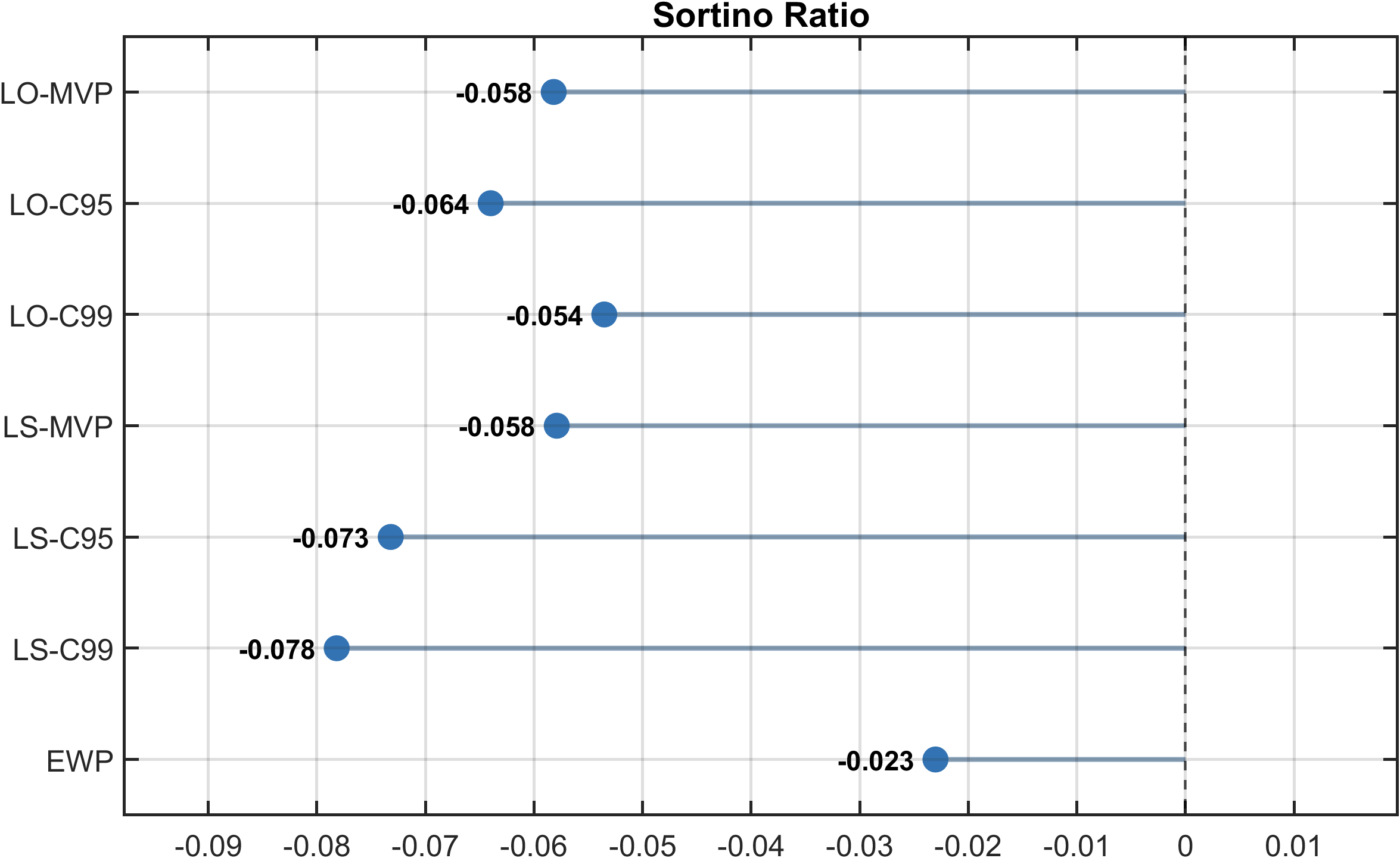}
        \caption{}
        \label{fig:sortino}
    \end{subfigure}
     \hspace{0.02\textwidth}
      \vspace{0.02\textwidth}
    \begin{subfigure}[b]{0.43\textwidth}
        \includegraphics[width=\textwidth]{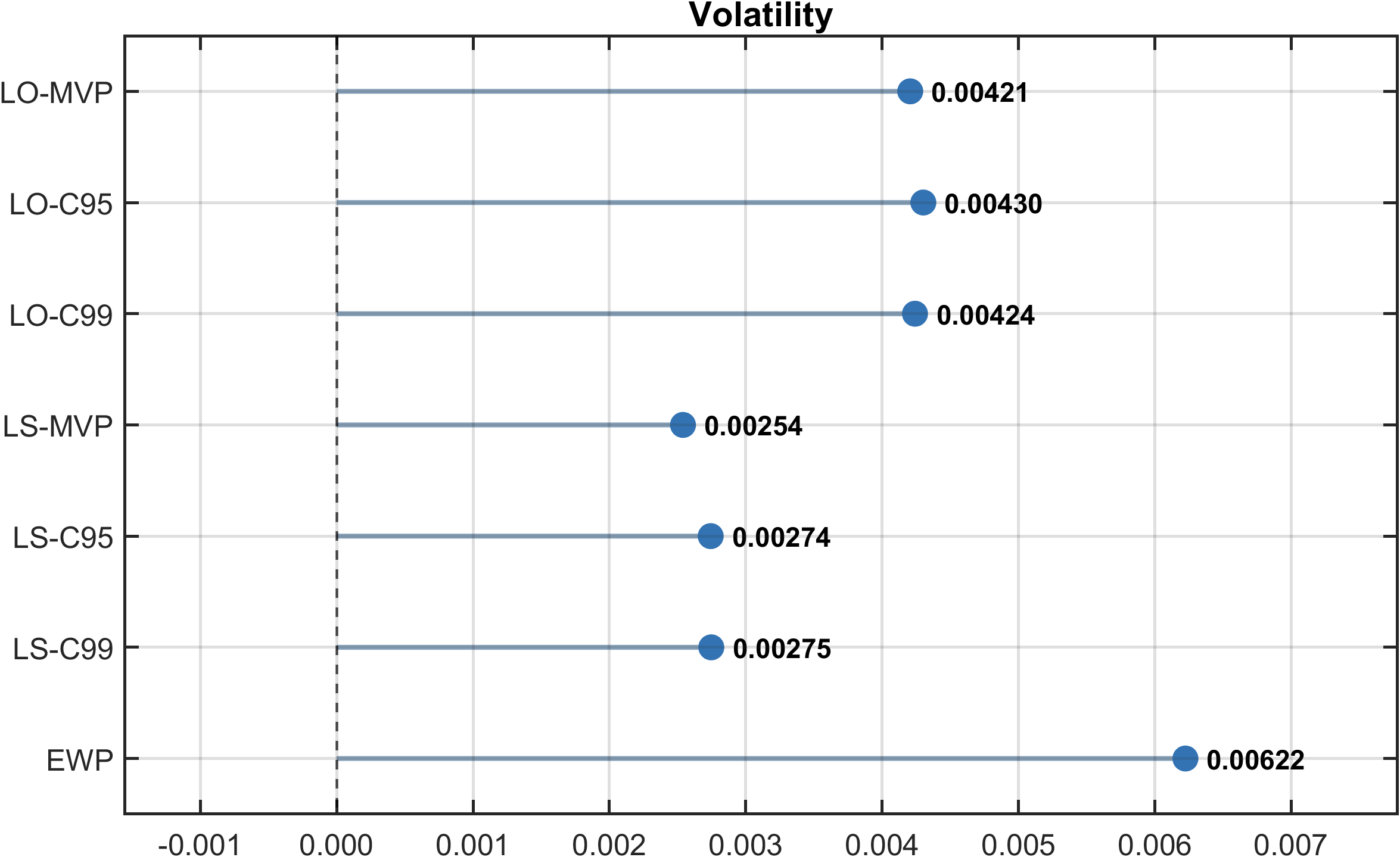}
        \caption{}
        \label{fig:volatility}
    \end{subfigure}
     \hspace{0.02\textwidth}
      \vspace{0.02\textwidth}
    \begin{subfigure}[b]{0.43\textwidth}
        \includegraphics[width=\textwidth]{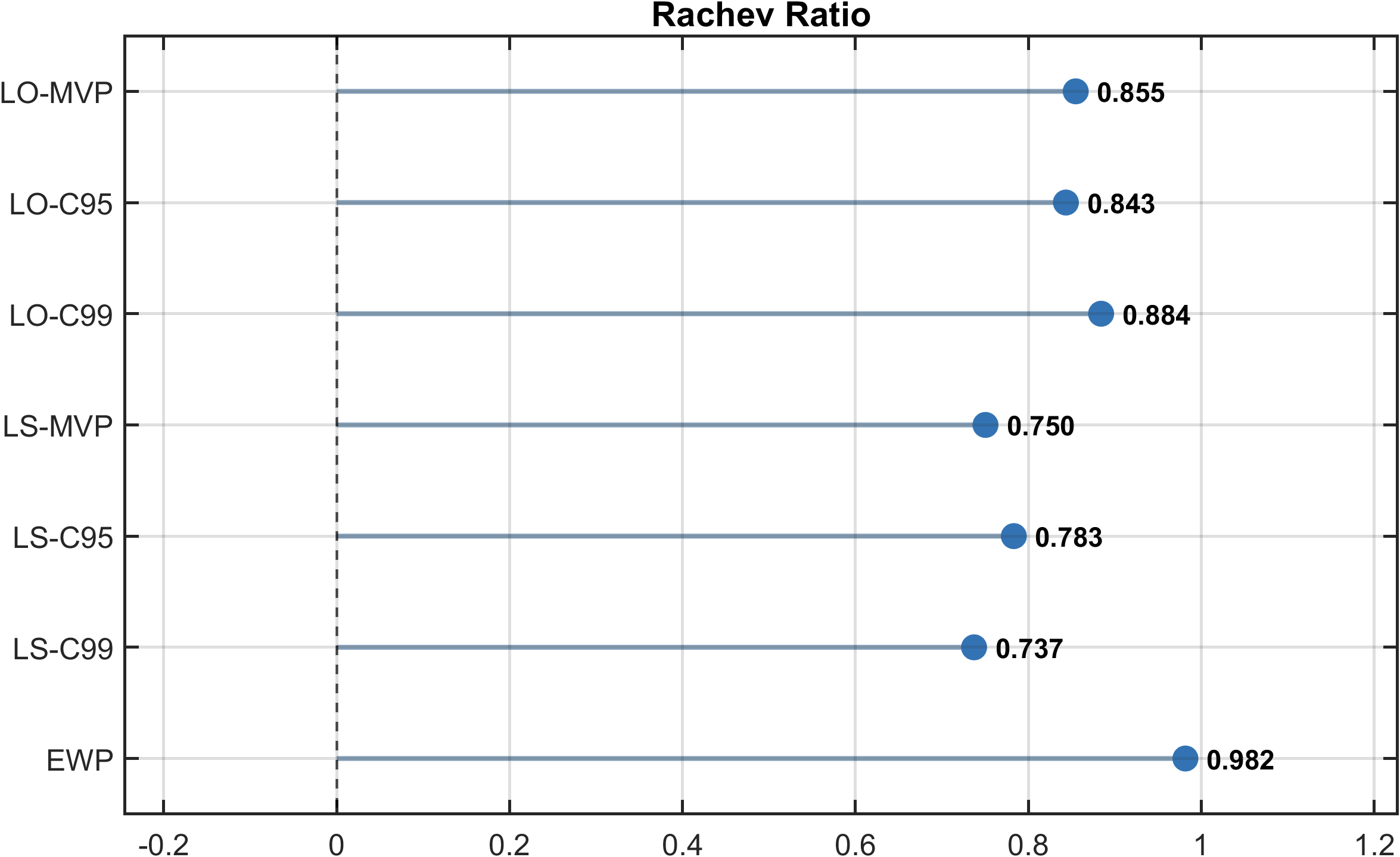}
        \caption{}
        \label{fig:rachev}
    \end{subfigure}
     \hspace{0.02\textwidth}
    \begin{subfigure}[b]{0.45\textwidth}
        \includegraphics[width=\textwidth]{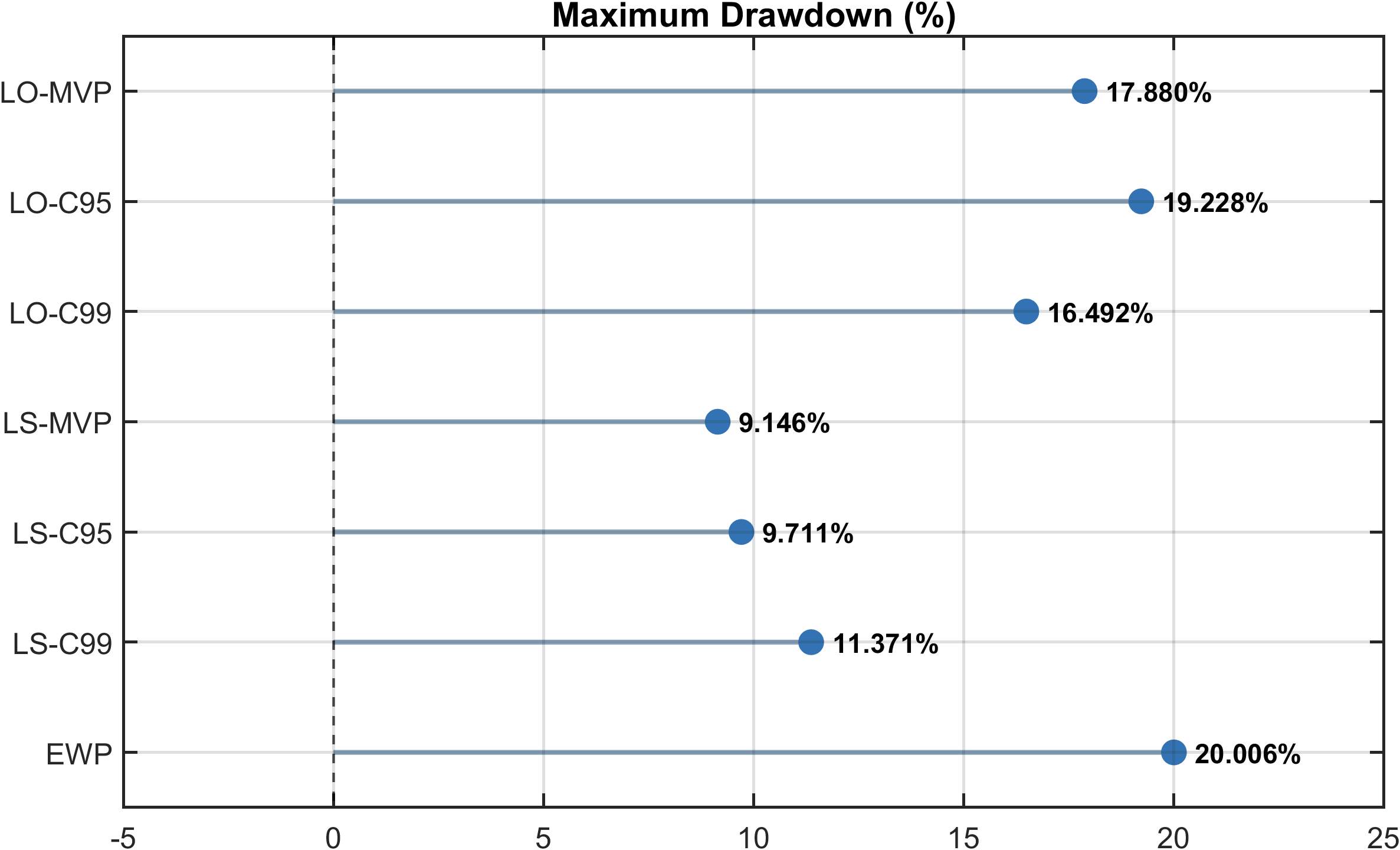}
        \caption{}
        \label{fig:maxdrawdown}
    \end{subfigure}
    \caption{Comparative performance and risk measures
across portfolios: (a) Sharpe ratio, (b) Sortino ratio,
(c) daily volatility, (d) Rachev ratio, and (e) maximum drawdown.}
    \label{fig:performance_metrics_fig}
\end{figure}


\begin{table}[H]
\centering
\caption{Performance and risk metrics across portfolio strategies for the out-of-sample daily returns.}
\label{tab:performance_metrics_table}
\resizebox{\textwidth}{!}{
\setlength{\arrayrulewidth}{0.7pt}
\renewcommand{\arraystretch}{1.5}
\begin{tabular}{|l|c|c|c|c|c|}
\hline
\textbf{Portfolio} & \textbf{Sharpe Ratio} & \textbf{Sortino Ratio} & \textbf{Daily Volatility} & \textbf{Rachev Ratio} & \textbf{Max Drawdown (\%)} \\
\hline
EWP-Return
& -0.0165 & -0.0230 & 0.0062243 & 0.9817 & 20.0059 \\
\hline
LS-$C95$-Return
& -0.0567 & -0.0732 & 0.0027428 & 0.7832 & 9.7110 \\
\hline
LS-$C99$-Return
& -0.0612 & -0.0782 & 0.0027471 & 0.7372 & 11.3713 \\
\hline
LS-MVP-Return
& -0.0447 & -0.0579 & 0.0025395 & 0.7504 & 9.1461 \\
\hline
LO-$C95$-Return
& -0.0482 & -0.0640 & 0.0043021 & 0.8435 & 19.2280 \\
\hline
LO-$C99$-Return
& -0.0397 & -0.0535 & 0.0042411 & 0.8842 & 16.4921 \\
\hline
LO-MVP-Return
& -0.0438 & -0.0582 & 0.0042060 & 0.8548 & 17.8797 \\
\hline
\end{tabular}}
\end{table}

Figure~\ref{fig:performance_metrics_fig} and
Table~\ref{tab:performance_metrics_table} summarize the
empirical performance and risk measures across the portfolio
strategies based on 754 out-of-sample daily observations.
Several important findings emerge from the analysis.

First, the long--short portfolios exhibit substantially lower
daily volatility than their corresponding long-only
portfolios. Among the minimum-variance strategies, LS-MVP
records the lowest daily volatility of 0.00254, compared
with 0.00421 for LO-MVP. Similarly, LS-C95 and LS-C99
record daily volatilities of 0.00274 and 0.00275,
respectively, compared with 0.00430 and 0.00424 for
their long-only counterparts. All six optimized portfolios
also exhibit lower daily volatility than EWP, which
records the highest value of 0.00622.

Second, the optimized portfolios exhibit smaller maximum drawdowns than EWP. LS-MVP records the lowest maximum drawdown of 9.15\%, followed by LS-C95 at 9.71\% and LS-C99 at 11.37\%. These values are substantially below the maximum drawdown of 20.01\% recorded by EWP. The long-only portfolios also maintain lower maximum drawdowns than EWP, ranging from 16.49\% for LO-C99
to 19.23\% for LO-C95. These findings indicate that the optimized portfolios experienced smaller peak-to-trough wealth declines over the out-of-sample
evaluation period.

Third, the Sharpe and Sortino ratios present a different
ranking of portfolio performance. EWP records the highest Sharpe ratio of -0.0165 and Sortino ratio of -0.0230, while all optimized portfolios also exhibit negative
values. All reported ratios are negative, indicating negative average excess returns relative to the time-varying daily risk-free-rate proxy derived from the three-month Treasury yield. Importantly, because the Sharpe ratios are negative, a larger volatility
denominator can produce a less negative ratio. Consequently, EWP's higher Sharpe ratio should not be interpreted independently as evidence of lower portfolio risk, particularly given its higher daily volatility and maximum drawdown.

Fourth, the Rachev ratio provides additional information about the relative magnitudes of favorable and adverse tail excess returns. EWP records the highest Rachev
ratio of 0.9817, followed by LO-C99 at 0.8842 and LO-MVP at 0.8548. Among the long--short portfolios, LS-C95 records the highest Rachev ratio of 0.7832,
compared with 0.7504 for LS-MVP and 0.7372 for LS-C99. All reported Rachev ratios are below one, indicating that the average magnitude of adverse-tail
excess returns exceeds that of favorable-tail excess returns under the tail probabilities considered. Thus, although the optimized portfolios exhibit lower
absolute risk as measured by volatility and maximum drawdown, they do not outperform EWP in terms of the Rachev ratio.

Finally, the comparison between long-only and long--short strategies reveals a distinction between risk reduction and risk-adjusted performance. The long--short portfolios consistently achieve lower daily volatility and maximum drawdown than their corresponding long-only portfolios. However, their Sharpe, Sortino, and Rachev ratios do not consistently improve. In particular, the long-only CVaR-based strategies exhibit less negative Sharpe and Sortino ratios and higher Rachev ratios than their corresponding long--short strategies. For the minimum-variance strategies, LS-MVP records a slightly less negative Sortino ratio than LO-MVP, while LO-MVP records a slightly less negative Sharpe ratio and a higher
Rachev ratio.

Taken together, the results indicate that portfolio rankings depend on the risk and performance measures considered. Although EWP achieves the highest Sharpe, Sortino, and Rachev ratios, it also exhibits the highest daily volatility and maximum drawdown. In contrast,
the optimized portfolios, particularly the long--short
strategies, provide substantial reductions in these absolute risk measures. Among the long--short strategies, LS-MVP records the lowest daily volatility and maximum drawdown, while LS-C95 achieves lower values of these
measures than LS-C99. These findings highlight the importance of evaluating both risk-adjusted performance and absolute risk when comparing balanced equity--bond portfolio strategies.

\subsection{Portfolio Risk Assessment: Value-at-Risk and Conditional Value-at-Risk}\label{subsec:3.7}

To further evaluate portfolio robustness under adverse market conditions, we examine both Value-at-Risk (VaR) and Conditional Value-at-Risk (CVaR) across all portfolio strategies.\footnote{See Appendix~\ref{porfolio-definition}}

For a given confidence level, Value-at-Risk measures the threshold loss exceeded with a specified probability. Although VaR remains widely used in institutional risk management, it provides only a quantile-based measure of downside exposure and therefore does not capture the magnitude of losses beyond the VaR threshold. CVaR addresses this limitation by measuring the expected loss conditional on the occurrence of extreme downside events \citep{rockafellar_uryasev_2000}. Consequently, CVaR provides a more informative characterization of downside exposure under heavy-tailed and asymmetric return distributions. This distinction is particularly important in the context of balanced equity--bond portfolios because simultaneous declines in both equity and fixed-income markets may generate return asymmetries that are not adequately captured by variance-based or quantile-only risk measures.

Throughout this analysis, VaR and CVaR are reported as positive loss measures computed from the portfolio return series. Thus, larger values indicate greater downside risk.

\begin{figure}[H]
    \centering
    \begin{subfigure}[b]{0.45\textwidth}
        \includegraphics[width=\textwidth]{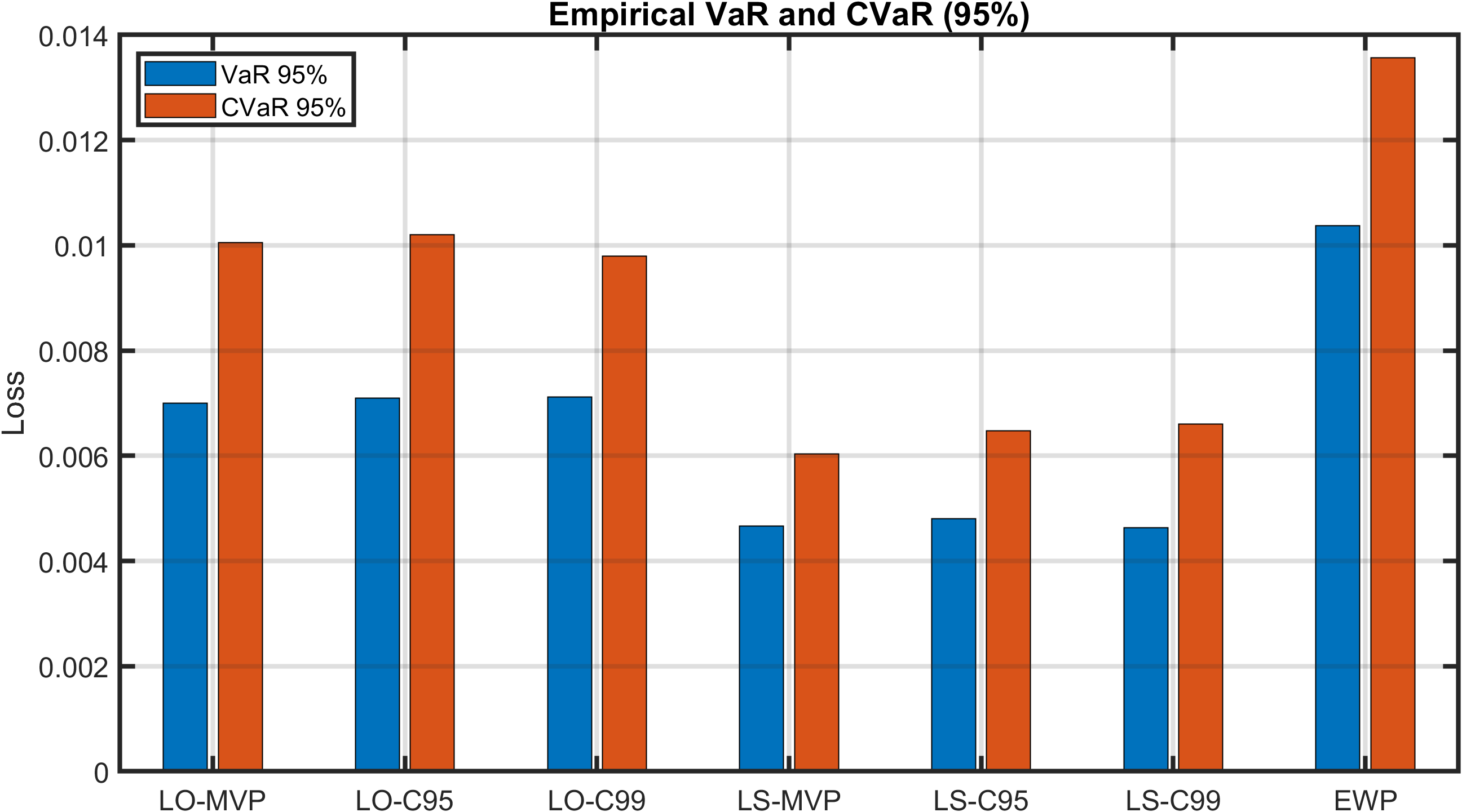}
        \caption{}
        \label{fig:VaR}
    \end{subfigure}
    \hspace{0.04\textwidth}
    \vspace{0.02\textwidth}
    \begin{subfigure}[b]{0.45\textwidth}
        \includegraphics[width=\textwidth]{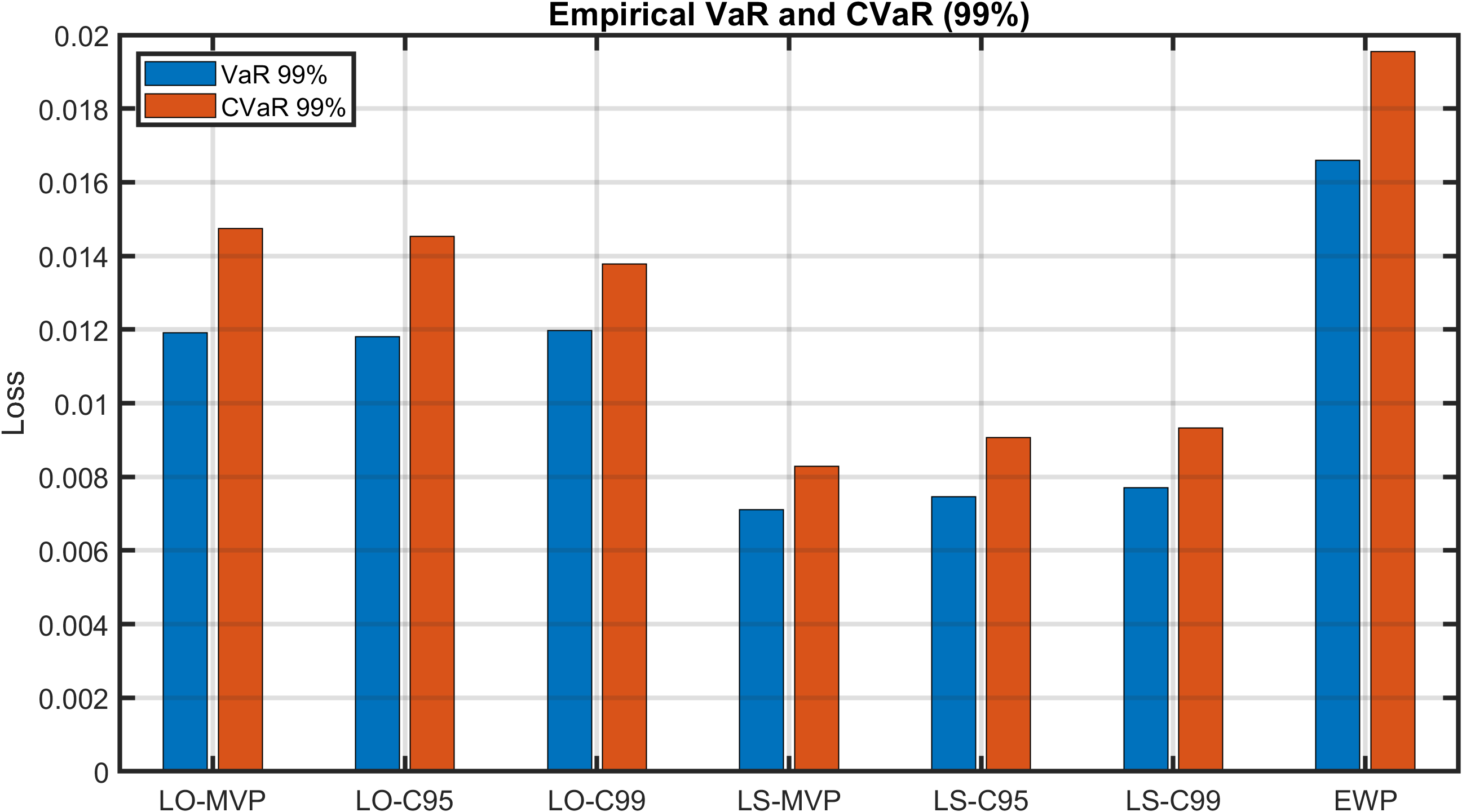}
        \caption{}
        \label{fig:CVaR}
    \end{subfigure}
    \caption{Comparison of historical Value-at-Risk (VaR)
and Conditional Value-at-Risk (CVaR) across portfolio strategies at (a) the 95\% and (b) the 99\%
confidence levels.}
    \label{fig:VaR_CVaR_fig}
\end{figure}


\begin{table}[htb]
\centering
\caption{Out-of-sample VaR and CVaR across portfolio strategies.}
\label{tab:var-cvar-table}
\small
\setlength{\arrayrulewidth}{0.5pt}
\renewcommand{\arraystretch}{1}
\begin{tabular}{|l|c|c|c|c|}
\hline
\textbf{Portfolio} & \textbf{VaR 95\%} & \textbf{VaR 99\%} & \textbf{CVaR 95\%} & \textbf{CVaR 99\%} \\
\hline
EWP-Return       & 0.0104 & 0.0166 & 0.0136 & 0.0195 \\
\hline
LS--C95-Return   & 0.0048 & 0.0075 & 0.0065 & 0.0091 \\
\hline
LS--C99-Return   & 0.0046 & 0.0077 & 0.0066 & 0.0093 \\
\hline
LS-MVP-Return    & 0.0047 & 0.0071 & 0.0060 & 0.0083 \\
\hline
LO--C95-Return   & 0.0071 & 0.0118 & 0.0102 & 0.0145 \\
\hline
LO--C99-Return   & 0.0071 & 0.0120 & 0.0098 & 0.0138 \\
\hline
LO-MVP-Return    & 0.0070 & 0.0119 & 0.0101 & 0.0148 \\
\hline
\end{tabular}
\end{table}

Figure~\ref{fig:VaR_CVaR_fig} and Table~\ref{tab:var-cvar-table} summarize the out-of-sample VaR and CVaR estimates across the
portfolio strategies at the 95\% and 99\% confidence levels. Several important patterns emerge from the analysis.

First, the equally weighted portfolio exhibits the highest
estimated downside-tail risk across all reported measures.
EWP records a 95\% VaR of 0.0104 and a 95\% CVaR of 0.0136, compared with substantially lower values for all six optimized portfolios. At the 99\% confidence
level, EWP also records the highest VaR and CVaR estimates of 0.0166 and 0.0195, respectively.
These results indicate greater realized downside-tail exposure for EWP over the out-of-sample evaluation period.

Second, the optimized long--short portfolios exhibit particularly low VaR and CVaR estimates.
At the 95\% confidence level, LS-C95 records a VaR of 0.0048 and a CVaR of 0.0065, compared with
0.0104 and 0.0136, respectively, for EWP.
Among the long--short strategies, LS-C99 records the lowest 95\% VaR of 0.0046, while LS-MVP records the lowest 95\% CVaR of 0.0060.
LS-MVP also records the lowest 99\% VaR and CVaR of 0.0071 and 0.0083, respectively.
These findings indicate that the minimum-variance and minimum-CVaR strategies are associated with
relatively small realized tail losses in the reported sample.

Third, CVaR exceeds the corresponding VaR estimate
across all portfolio strategies and both confidence levels.
This indicates that the average loss beyond the VaR threshold is greater than the threshold itself,
highlighting the additional information provided by CVaR about the severity of adverse-tail outcomes.

Fourth, the long--short portfolios consistently exhibit lower VaR and CVaR estimates than their corresponding long-only portfolios. For example,
LS-C99 records a 99\% VaR of 0.0077 and a 99\% CVaR of 0.0093, compared with 0.0120 and 0.0138,
respectively, for LO-C99.
Similar differences are observed for the C95 and minimum-variance strategies. Thus, the long--short
portfolio specifications are associated with lower measured downside-tail risk in this sample.
However, these comparisons are based on portfolio returns before transaction costs and do not incorporate short-borrowing or financing costs.

Taken together, the results show that all six optimized portfolios exhibit lower historical VaR
and CVaR estimates than EWP at both confidence levels.
The long--short strategies record the smallest tail-risk estimates, with LS-MVP achieving the
lowest CVaR at both confidence levels.
These findings highlight the importance of considering both VaR and CVaR when evaluating
downside risk, as VaR identifies the estimated loss threshold while CVaR captures the average
severity of losses beyond that threshold.

\subsection{Transaction-Cost Sensitivity Analysis of Portfolio Risk}
\label{transaction_cost_sensitivity}

The preceding analysis indicates that the optimized portfolios, particularly the long--short strategies, generally exhibit lower volatility, maximum drawdown, and historical VaR and CVaR than the EWP. However, these findings are based on portfolio returns before transaction costs. Given the differences in portfolio turnover reported earlier, we examine whether these risk advantages persist after accounting for rebalancing costs.

We consider proportional transaction costs of 0, 5, 10, and 20 basis points per dollar traded, applied to both purchases and sales during portfolio rebalancing. Since daily portfolio turnover is defined as one-half of the sum of absolute changes between target and drift-adjusted portfolio weights, daily net portfolio returns are calculated as
\begin{equation*}
R_{p,t}^{\mathrm{net}}
=
R_{p,t}^{\mathrm{gross}}-2cT_{p,t},
\end{equation*}
where $c$ denotes the proportional transaction-cost rate and $T_{p,t}$ denotes daily portfolio turnover. The original optimized portfolio weights are held fixed across the cost scenarios. Thus, the analysis evaluates the sensitivity of the existing strategies to trading costs without reoptimizing portfolio allocations. For each scenario, we recalculate terminal wealth, daily volatility, maximum drawdown, and historical VaR and CVaR at the 95\% and 99\% confidence levels.

\begin{table}[htbp]
\centering
\caption{Transaction-Cost Sensitivity of Terminal Wealth and daily volatility}
\label{tab:cost_wealth_volatility}
\small
\resizebox{\textwidth}{!}{%
\begin{tabular}{lrrrrrrrr}
\toprule
& \multicolumn{4}{c}{Terminal Wealth}
& \multicolumn{4}{c}{Daily Volatility} \\
\cmidrule(lr){2-5}
\cmidrule(lr){6-9}
Strategy
& 0 bps & 5 bps & 10 bps & 20 bps
& 0 bps & 5 bps & 10 bps & 20 bps \\
\midrule
LO-MVP & 97.702 & 97.359 & 97.016 & 96.335 & 0.0042060 & 0.0042063 & 0.0042067 & 0.0042074 \\
LO-C95 & 95.986 & 95.347 & 94.712 & 93.454 & 0.0043021 & 0.0043044 & 0.0043069 & 0.0043122 \\
LO-C99 & 98.865 & 97.994 & 97.131 & 95.425 & 0.0042411 & 0.0042423 & 0.0042441 & 0.0042498 \\
LS-MVP & 103.490 & 99.963 & 96.552 & 90.076 & 0.0025395 & 0.0025428 & 0.0025468 & 0.0025570 \\
LS-C95 & 100.230 & 91.362 & 83.277 & 69.181 & 0.0027428 & 0.0027495 & 0.0027718 & 0.0028614 \\
LS-C99 & 99.291 & 84.318 & 71.594 & 51.596 & 0.0027471 & 0.0027757 & 0.0028481 & 0.0031105 \\
EWP    & 103.040 & 102.910 & 102.780 & 102.530 & 0.0062243 & 0.0062243 & 0.0062243 & 0.0062243 \\
\bottomrule
\end{tabular}%
}
\par\smallskip
\begin{minipage}{\textwidth}
\footnotesize
\textit{Note:} Terminal wealth is calculated from an initial portfolio value of \$100. Daily volatility is the standard deviation of daily net portfolio returns.
\end{minipage}
\end{table}
\newpage

Table~\ref{tab:cost_wealth_volatility} shows that the effect of transaction costs on daily volatility is relatively small across the seven strategies. At 20 basis points, LS-MVP continues to exhibit the lowest daily volatility (0.0025570), compared with 0.0062243 for EWP. In the zero-cost case, LS-MVP achieves the highest terminal wealth (103.490), slightly exceeding EWP (103.040). However, EWP achieves the highest terminal wealth under all positive transaction-cost scenarios considered (5, 10, and 20 basis points). At 20 basis points, terminal wealth declines to 90.076 for LS-MVP, 69.181 for LS-C95, and 51.596 for LS-C99, while EWP retains terminal wealth of 102.530. These differences are consistent with the substantially higher turnover of the long--short strategies documented earlier.

\begin{table}[htbp]
\centering
\caption{Transaction-Cost Sensitivity of Maximum Drawdown}
\label{tab:cost_mdd}
\begin{tabular}{lrrrr}
\toprule
Strategy & 0 bps & 5 bps & 10 bps & 20 bps \\
\midrule
LO-MVP & 17.880\% & 18.009\% & 18.138\% & 18.395\% \\
LO-C95 & 19.228\% & 19.486\% & 19.744\% & 20.256\% \\
LO-C99 & 16.492\% & 16.909\% & 17.325\% & 18.149\% \\
LS-MVP & 9.146\% & 10.321\% & 11.480\% & 13.999\% \\
LS-C95 & 9.711\% & 14.215\% & 18.497\% & 31.503\% \\
LS-C99 & 11.371\% & 17.452\% & 28.918\% & 48.754\% \\
EWP    & 20.006\% & 20.040\% & 20.075\% & 20.144\% \\
\bottomrule
\end{tabular}
\par\smallskip
\begin{minipage}{0.85\textwidth}
\footnotesize
\textit{Note:} Maximum drawdown is calculated from the cumulative wealth path and expressed as a percentage decline from the preceding peak portfolio wealth.
\end{minipage}
\end{table}

As shown in Table~\ref{tab:cost_mdd}, maximum drawdown generally increases with transaction costs. At zero transaction costs, all six optimized strategies have lower maximum drawdown than EWP. At 20 basis points, LS-MVP records the lowest maximum drawdown of 13.999\%, compared with 20.144\% for EWP. The maximum drawdown of LS-C95 increases from 9.711\% to 31.503\%, while that of LS-C99 rises from 11.371\% to 48.754\%. Both exceed the corresponding EWP value at the highest transaction-cost level. These increases are consistent with the higher turnover of the two long--short CVaR strategies. The results indicate that the maximum-drawdown advantage observed before transaction costs does not persist for every optimized strategy as trading costs increase.

\begin{table}[htbp]
\centering
\caption{Transaction-Cost Sensitivity of Historical VaR and CVaR at the 95\% Confidence Level}
\label{tab:cost_var95}
\small
\resizebox{\textwidth}{!}{%
\begin{tabular}{lrrrrrrrr}
\toprule
& \multicolumn{4}{c}{VaR (95\%)}
& \multicolumn{4}{c}{CVaR (95\%)} \\
\cmidrule(lr){2-5}
\cmidrule(lr){6-9}
Strategy
& 0 bps & 5 bps & 10 bps & 20 bps
& 0 bps & 5 bps & 10 bps & 20 bps \\
\midrule
LO-MVP & 0.0069968 & 0.0070160 & 0.0070351 & 0.0070486 & 0.0100510 & 0.0100600 & 0.0100690 & 0.0100890 \\
LO-C95 & 0.0070944 & 0.0070998 & 0.0071304 & 0.0071851 & 0.0102060 & 0.0102290 & 0.0102520 & 0.0102990 \\
LO-C99 & 0.0071186 & 0.0071216 & 0.0071245 & 0.0071304 & 0.0097987 & 0.0098189 & 0.0098392 & 0.0098796 \\
LS-MVP & 0.0046680 & 0.0047257 & 0.0047725 & 0.0048398 & 0.0060378 & 0.0061203 & 0.0062029 & 0.0063797 \\
LS-C95 & 0.0048068 & 0.0050272 & 0.0052479 & 0.0055651 & 0.0064770 & 0.0066408 & 0.0068294 & 0.0073645 \\
LS-C99 & 0.0046371 & 0.0049672 & 0.0052797 & 0.0062069 & 0.0066053 & 0.0070716 & 0.0075926 & 0.0088237 \\
EWP    & 0.0103760 & 0.0103790 & 0.0103820 & 0.0103880 & 0.0135630 & 0.0135650 & 0.0135670 & 0.0135710 \\
\bottomrule
\end{tabular}%
}
\par\smallskip
\begin{minipage}{\textwidth}
\footnotesize
\textit{Note:} VaR and CVaR are historical daily loss measures expressed in decimal form, with positive values representing losses. Both are calculated from daily net portfolio returns.
\end{minipage}
\end{table}

\begin{table}[htbp]
\centering
\caption{Transaction-Cost Sensitivity of Historical VaR and CVaR at the 99\% Confidence Level}
\label{tab:cost_var99}
\small
\resizebox{\textwidth}{!}{%
\begin{tabular}{lrrrrrrrr}
\toprule
& \multicolumn{4}{c}{VaR (99\%)}
& \multicolumn{4}{c}{CVaR (99\%)} \\
\cmidrule(lr){2-5}
\cmidrule(lr){6-9}
Strategy
& 0 bps & 5 bps & 10 bps & 20 bps
& 0 bps & 5 bps & 10 bps & 20 bps \\
\midrule
LO-MVP & 0.0119080 & 0.0119180 & 0.0119270 & 0.0119460 & 0.0147510 & 0.0147580 & 0.0147650 & 0.0147790 \\
LO-C95 & 0.0118040 & 0.0118870 & 0.0119710 & 0.0121390 & 0.0145230 & 0.0145730 & 0.0146230 & 0.0147220 \\
LO-C99 & 0.0119770 & 0.0119780 & 0.0119800 & 0.0119840 & 0.0137870 & 0.0138760 & 0.0139660 & 0.0141450 \\
LS-MVP & 0.0071121 & 0.0071628 & 0.0072158 & 0.0075808 & 0.0082842 & 0.0083254 & 0.0083666 & 0.0085174 \\
LS-C95 & 0.0074586 & 0.0079499 & 0.0082623 & 0.0084414 & 0.0090717 & 0.0092289 & 0.0094685 & 0.0099478 \\
LS-C99 & 0.0077095 & 0.0080710 & 0.0087783 & 0.0103280 & 0.0093257 & 0.0097625 & 0.0103670 & 0.0123740 \\
EWP    & 0.0165880 & 0.0165920 & 0.0165950 & 0.0166020 & 0.0195470 & 0.0195500 & 0.0195520 & 0.0195570 \\
\bottomrule
\end{tabular}%
}
\par\smallskip
\begin{minipage}{\textwidth}
\footnotesize
\textit{Note:} VaR and CVaR are historical daily loss measures expressed in decimal form, with positive values representing losses. Both are calculated from daily net portfolio returns.
\end{minipage}
\end{table}

Tables~\ref{tab:cost_var95} and~\ref{tab:cost_var99} indicate that the long--short strategies retain lower historical VaR and CVaR than EWP at both confidence levels across all transaction-cost scenarios. Although the estimated tail losses generally increase as transaction costs rise, the relative differences remain substantial. For example, at 20 basis points, LS-MVP has a 99\% VaR of 0.0075808 and CVaR of 0.0085174, compared with 0.0166020 and 0.0195570, respectively, for EWP. These results indicate that the lower historical daily tail-risk estimates of the long--short strategies persist under the transaction-cost assumptions considered.

Taken together, the findings indicate that the effect of transaction costs differs across performance and risk measures. EWP achieves the highest terminal wealth among the seven strategies at each positive transaction-cost level and exhibits relatively small changes in terminal wealth as costs increase. In contrast, the optimized portfolios generally maintain lower daily volatility and historical VaR and CVaR than EWP. Maximum drawdown is more sensitive to transaction costs for the higher-turnover long--short CVaR strategies, with LS-C95 and LS-C99 exceeding EWP at 20 basis points. These findings highlight the trade-off between cumulative investment performance and downside-risk measures when transaction costs are considered.

The analysis retains the original optimized portfolio weights and accounts only for proportional rebalancing costs. Initial portfolio establishment costs, short-borrowing fees, financing costs, market impact, and the practical availability of short positions are not explicitly incorporated. Consequently, the results should be interpreted as a transaction-cost sensitivity assessment rather than a complete evaluation of implementable investment performance.

\subsection{Portfolio Extreme-Tail Risk Estimation}\label{subsec:hill}

While CVaR measures the average loss beyond a specified VaR threshold, it does not fully characterize the behavior of the extreme tail of the loss distribution. To further examine extreme-loss behavior, we employ the Hill estimator \citep{hill1975}, a standard approach for estimating the tail exponent under regularly varying, heavy-tailed distributions.

For each return series, an AR(1)--GARCH(1,1) model with Gaussian innovations is first fitted to account for serial dependence in returns and time-varying volatility. This filtering step aims to reduce the influence of predictable return dynamics and volatility clustering before the extreme-tail analysis is applied, consistent with filtered extreme-value approaches used in the literature \citep{mcneil2000,hill2015}. The standardized residuals obtained from the fitted models are then used for the Hill estimation. Specifically, negative standardized residuals are transformed into positive loss magnitudes, so that increasingly large negative standardized shocks correspond to increasingly large positive losses.

In our analysis, we adopt a Pareto-type tail assumption for these positive loss magnitudes. More specifically, we assume that the extreme upper tail of the loss magnitudes obtained from the negative standardized residuals is regularly varying and can be approximated asymptotically by a power-law relationship of the form
\[
P(L>x)\sim Cx^{-\alpha},
\qquad x\to\infty,
\]
where $L$ denotes the positive magnitude of a negative standardized residual, $C>0$ is a constant, and $\alpha>0$ is the Pareto tail exponent. This assumption concerns only the extreme tail of the transformed residual distribution rather than the entire distribution of portfolio returns. Under this assumption, the probability of observing increasingly large negative standardized shocks decreases according to a power law. A smaller value of $\alpha$ corresponds to slower tail decay and a heavier extreme tail, whereas a larger value indicates faster tail decay and a relatively thinner extreme tail.

The Pareto-type tail assumption provides the basis for applying the Hill estimator in our empirical analysis. By estimating the exponent $\alpha$ from the largest observed loss magnitudes, we examine and compare the extreme left-tail behavior of the standardized innovations across the selected return series.

Let \(L_{(1)} \geq L_{(2)} \geq \cdots \geq L_{(n)}\) denote the ordered positive loss magnitudes obtained from the negative standardized residuals, where $n$ is the number of negative standardized residuals. The Hill estimator of the extreme-value index $\gamma$ is given by
\[
\hat{\gamma}(k)
=
\frac{1}{k}\sum_{i=1}^{k}
\left(
\ln L_{(i)}-\ln L_{(k+1)}
\right),
\]
where $k$ denotes the number of upper-order observations used for estimation, with $1\leq k<n$. The corresponding estimated Pareto tail exponent, referred to here as the Hill tail index, is
\[
\hat{\alpha}(k)=\frac{1}{\hat{\gamma}(k)}.
\]

Under the regularly varying tail assumption, a smaller estimated Hill tail index $\hat{\alpha}$ indicates a heavier extreme-loss tail, whereas a larger $\hat{\alpha}$ indicates a relatively thinner extreme-loss tail. Since the Hill estimator depends on the choice of $k$, we examine its sensitivity to different threshold values. We also construct approximate 95\% Wald-type confidence intervals based on the asymptotic distribution of the Hill estimator \citep{resnick2007}. These intervals are obtained by transforming the approximate confidence bounds for $\gamma$ and do not account for first-stage GARCH estimation uncertainty, residual dependence, or threshold-selection bias.

We apply this procedure to four selected return series: the Dow Jones Industrial Average (DJIA), the equally weighted portfolio (EWP), the long--short minimum-variance portfolio (LS-MVP), and the long--short 95\% CVaR-minimizing portfolio (LS-C95). For comparability, we use a common reference threshold of $k=100$ and examine threshold sensitivity over
$k=40,50,\ldots,200$. This allows us to assess whether the relative ordering of the estimated tail exponents remains consistent across nearby threshold choices. The resulting estimates describe the extreme-tail behavior of the standardized GARCH innovations, rather than the unconditional loss distributions of the portfolios.

\begin{figure}[H]
    \centering
    \begin{subfigure}[b]{0.46\textwidth}
        \includegraphics[width=\textwidth]{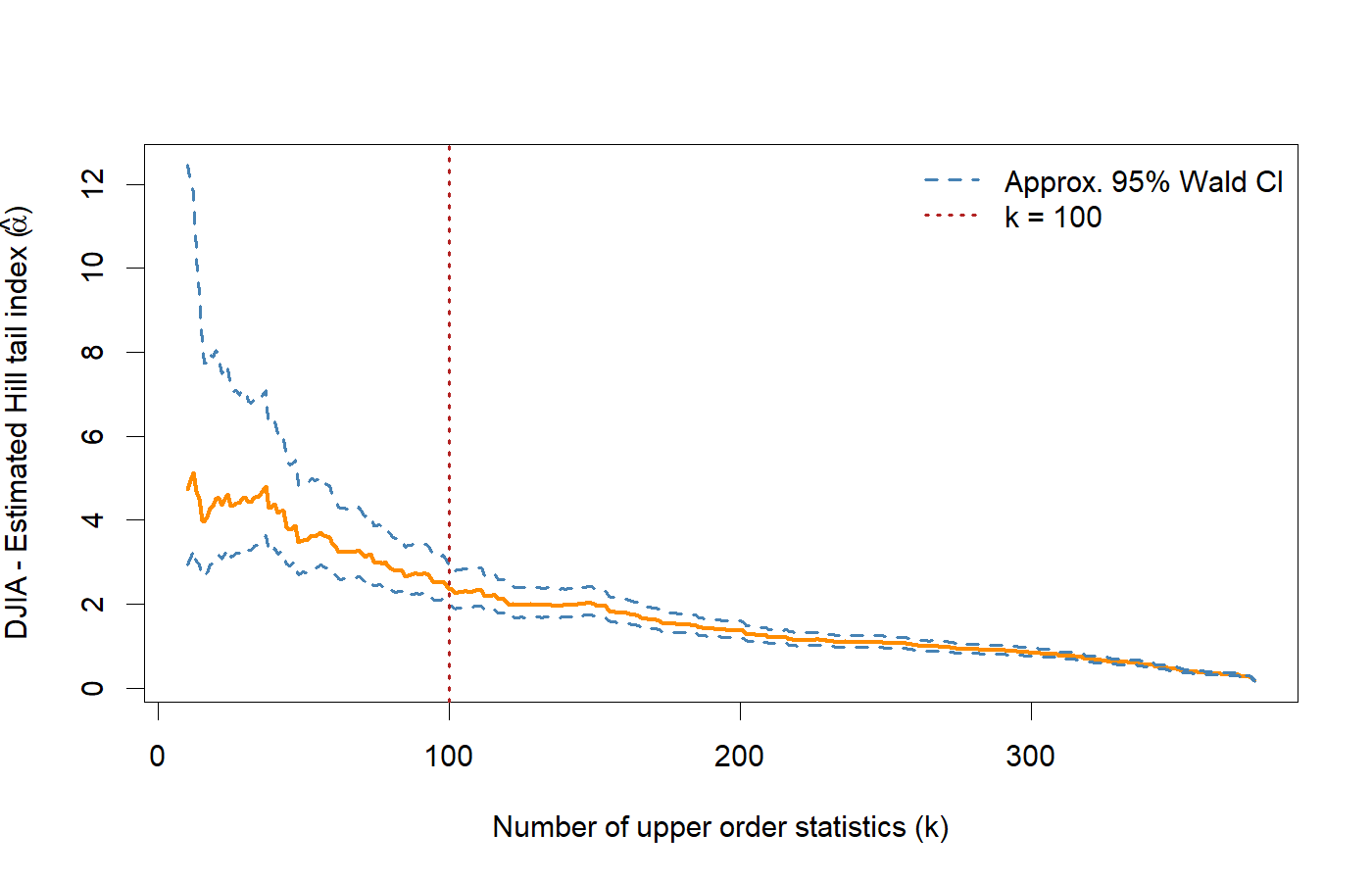}
        \caption{}
        \label{fig:hill_djia}
    \end{subfigure}
    \begin{subfigure}[b]{0.46\textwidth}
        \includegraphics[width=\textwidth]{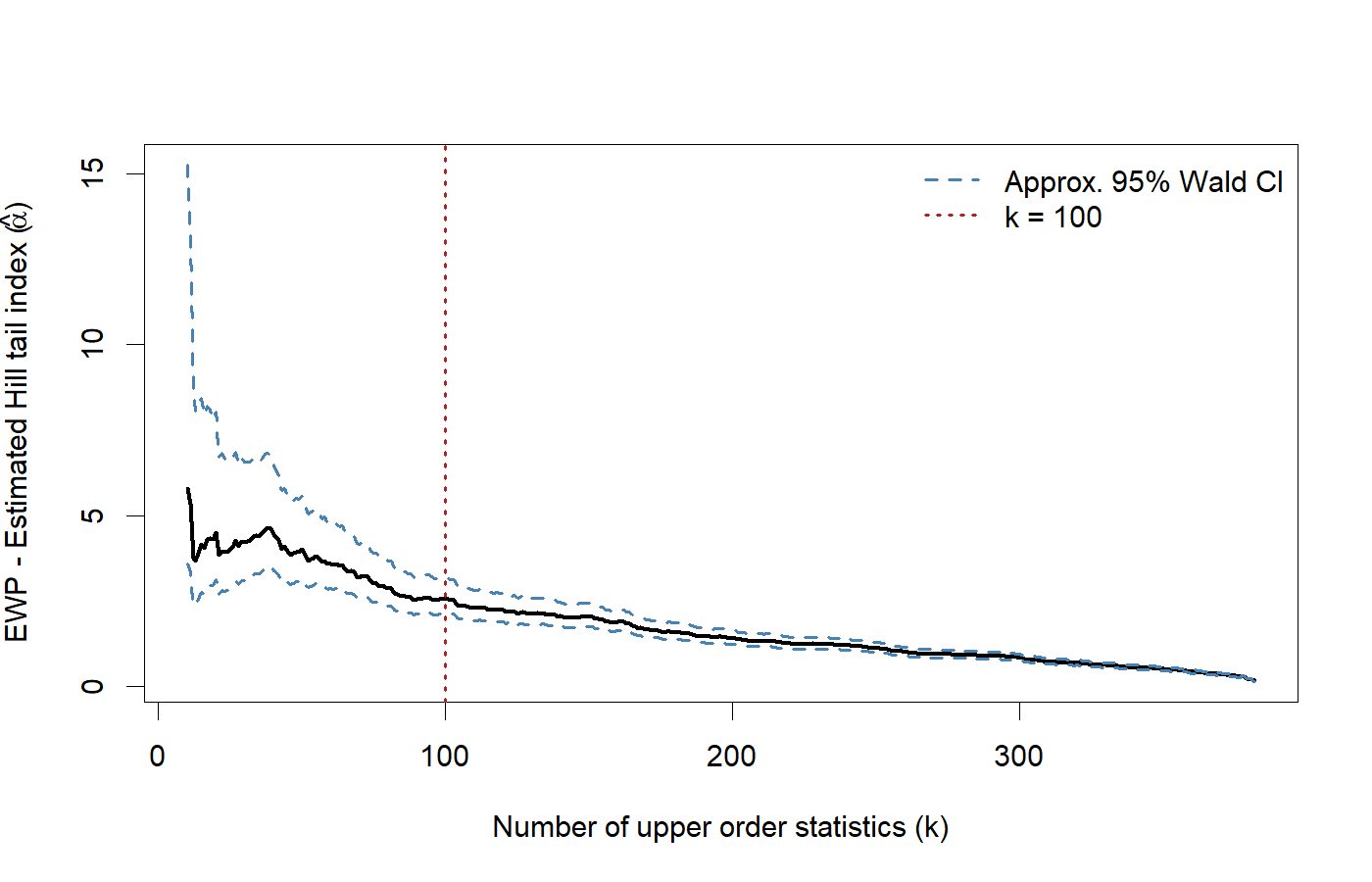}
        \caption{}
        \label{fig:hill_ewp}
    \end{subfigure}
    \begin{subfigure}[b]{0.46\textwidth}
        \includegraphics[width=\textwidth]{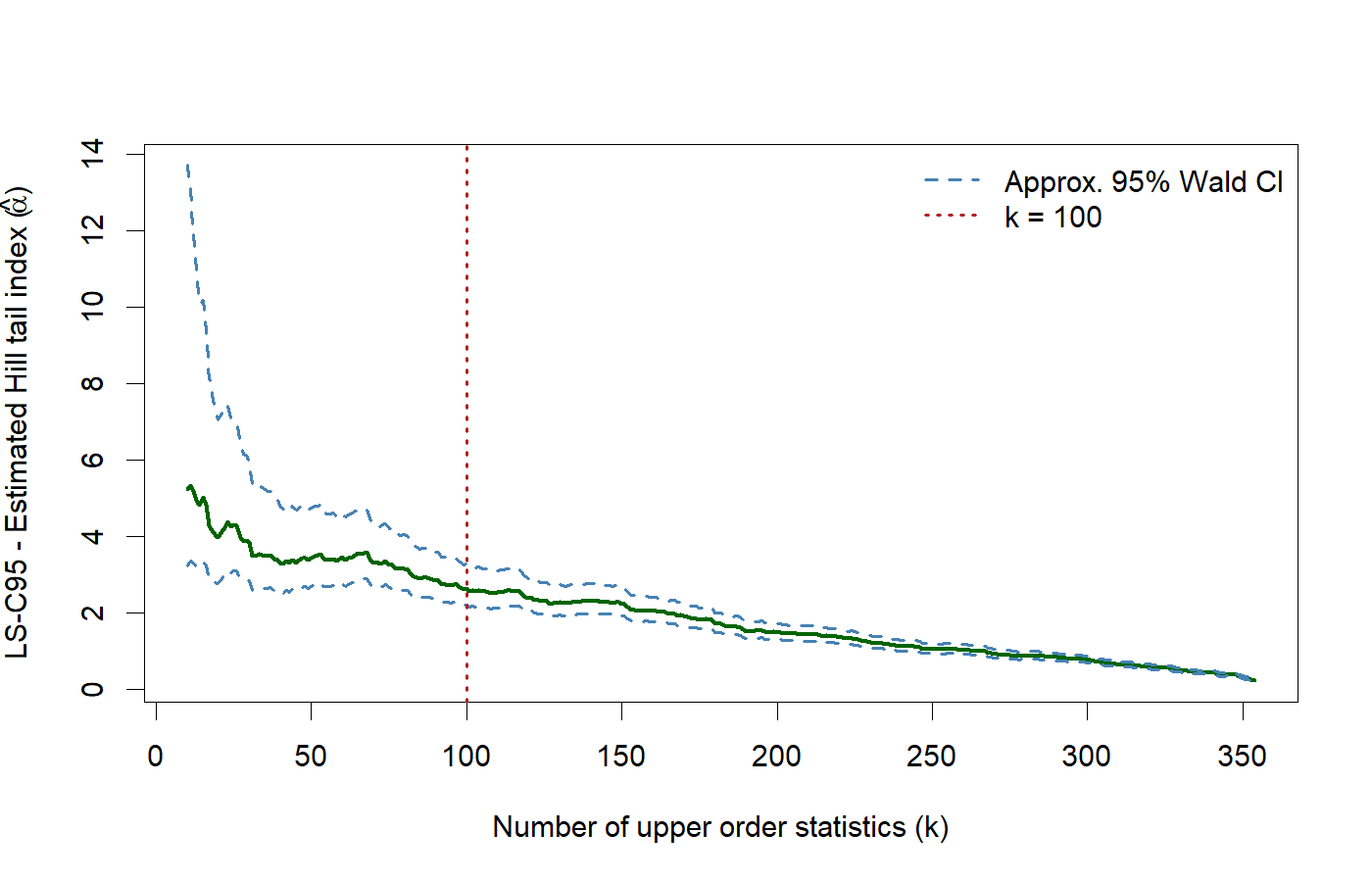}
        \caption{}
        \label{fig:hill_lscvar}
    \end{subfigure}
    \begin{subfigure}[b]{0.46\textwidth}
        \includegraphics[width=\textwidth]{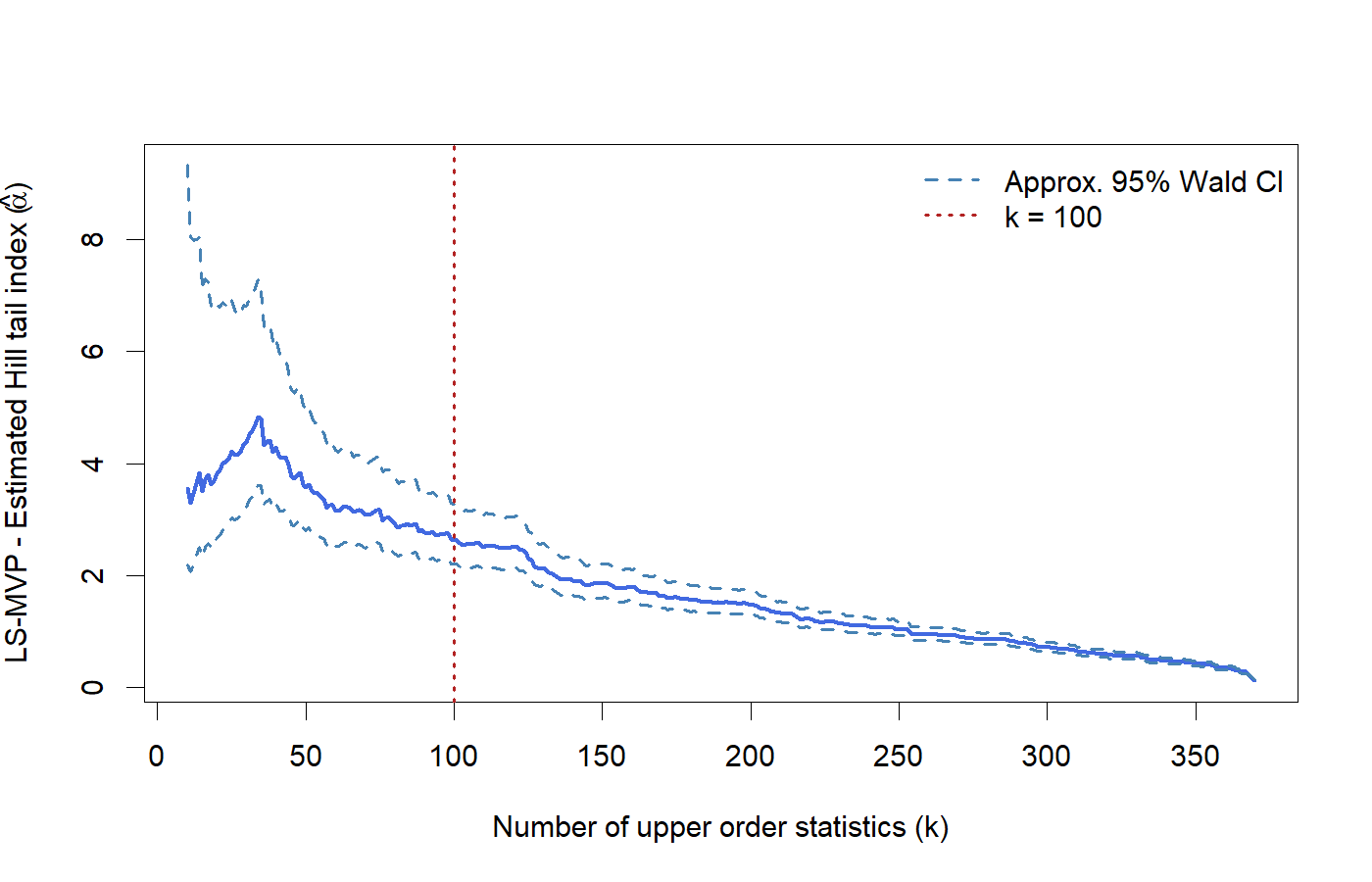}
        \caption{}
        \label{fig:hill_lsmvp}
    \end{subfigure}
    \caption{Estimated Hill tail indices for the right tail of loss magnitudes derived from negative standardized return residuals, with approximate pointwise 95\% Wald confidence intervals: (a) Dow Jones Industrial Average (DJIA), (b) EWP, (c) LS-C95, and (d) LS-MVP portfolios. The vertical reference line indicates $k=100$.}
    \label{fig:hill_left}
\end{figure}

Figure~\ref{fig:hill_left} presents the estimated Hill tail indices for the four series, based on the magnitudes of negative standardized residuals from the fitted AR(1)-GARCH(1,1) models.
For comparability, the Hill tail-index estimates are reported at a common reference threshold of $k=100$, corresponding to the 100 largest negative standardized residual magnitudes obtained from the fitted AR(1)-GARCH(1,1) models. The estimates characterize the tails of the standardized innovations rather than the unconditional portfolio returns. The common threshold provides a consistent reference for comparison, while its sensitivity is examined over a broader range of thresholds. 

Table~\ref{tab:hill_ci} reports the Hill tail-index estimates and their approximate 95\% confidence intervals at $k=100$.
\begin{table}[htb]
\centering
\caption{Hill tail-index estimates and approximate 95\% confidence intervals at $k=100$.}
\label{tab:hill_ci}
\small
\setlength{\arrayrulewidth}{0.5pt}
\renewcommand{\arraystretch}{1.2}
\begin{tabular}{|l|c|c|c|}
\hline
\textbf{Portfolio} &
\textbf{Hill Index ($\hat{\alpha}$)} &
\textbf{95\% CI Lower} &
\textbf{95\% CI Upper} \\
\hline
DJIA   & 2.3409 & 1.9572 & 2.9115 \\
\hline
EWP    & 2.5616 & 2.1418 & 3.1861 \\
\hline
LS-MVP & 2.6349 & 2.2031 & 3.2772 \\
\hline
LS-C95 & 2.6348 & 2.2030 & 3.2771 \\
\hline
\end{tabular}
\end{table}
At $k=100$, LS-MVP and LS-C95 have nearly identical estimated Hill tail indices of 2.6349 and 2.6348, respectively, followed by EWP at 2.5616 and DJIA at 2.3409. Under the assumption of regularly varying tails, a larger Hill tail index corresponds to faster asymptotic tail decay. The point estimates therefore suggest somewhat lighter tails for the standardized innovations of LS-MVP and LS-C95 relative to DJIA and EWP at the selected threshold.

However, the approximate 95\% confidence intervals overlap substantially across the four series. For example, LS-MVP has an estimated interval of [2.2031, 3.2772], compared with [2.1418, 3.1861] for EWP. Consequently, the observed differences should be interpreted descriptively rather than as statistically established differences in extreme-tail behavior.

\begin{figure}[H]
    \centering \includegraphics[width=0.90\textwidth]{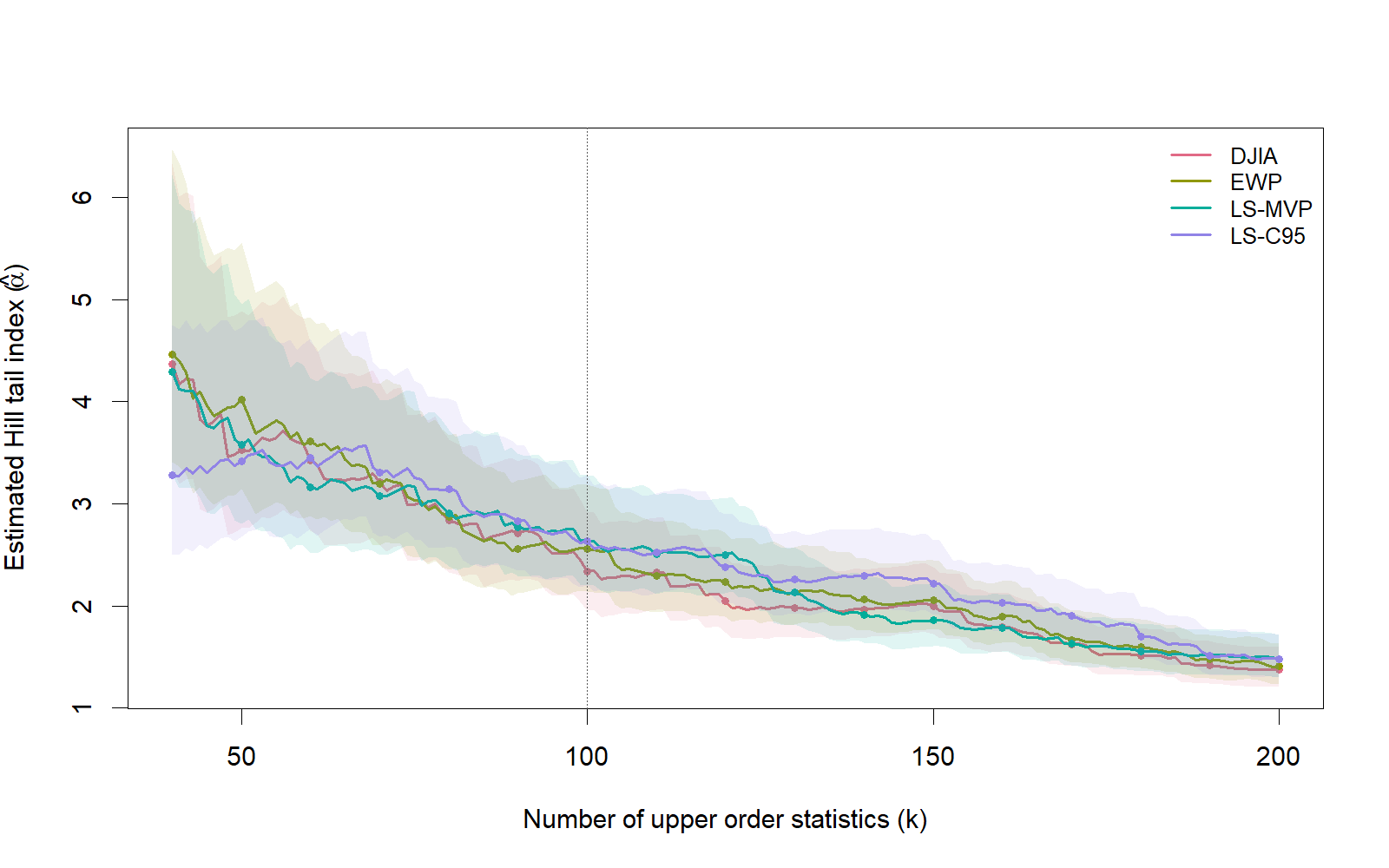}
    \caption{Hill tail-index estimates across thresholds $k=40$--$200$ for DJIA, EWP, LS-MVP, and LS-C95. Shaded regions represent approximate pointwise 95\% confidence intervals. The vertical dotted line indicates the common reference threshold $k=100$.}
    \label{fig:hill_sensitivity}
\end{figure}

To evaluate the sensitivity of the Hill estimates to threshold selection, the analysis is extended to $k=40,50,\ldots,200$. Table~\ref{tab:hill_sensitivity} reports representative thresholds from this expanded range, while Figure~\ref{fig:hill_sensitivity} displays the complete Hill trajectories and their approximate pointwise 95\% confidence intervals.
\begin{table}[htbp]
\centering
\caption{Threshold Sensitivity of Estimated Hill Tail Indices}
\label{tab:hill_sensitivity}
\small
\resizebox{\textwidth}{!}{%
\begin{tabular}{lrrrrrrrrr}
\toprule
& \multicolumn{9}{c}{Estimated Hill Tail Index $\hat{\alpha}(k)$} \\
\cmidrule(lr){2-10}
Portfolio
& $k=40$ & $k=60$ & $k=80$ & $k=100$
& $k=120$ & $k=140$ & $k=160$ & $k=180$ & $k=200$ \\
\midrule
DJIA   & 4.3708 & 3.4280 & 2.8339 & 2.3409 & 2.0513 & 1.9658 & 1.7970 & 1.5138 & 1.3729 \\
EWP    & 4.4629 & 3.6080 & 2.8694 & 2.5616 & 2.2348 & 2.0631 & 1.8926 & 1.5957 & 1.4109 \\
LS-MVP & 4.2945 & 3.1593 & 2.9081 & 2.6349 & 2.5016 & 1.9112 & 1.7819 & 1.5550 & 1.4783 \\
LS-C95 & 3.2788 & 3.4462 & 3.1469 & 2.6348 & 2.3775 & 2.2918 & 2.0309 & 1.6974 & 1.4775 \\
\bottomrule
\end{tabular}%
}
\par\smallskip
\begin{minipage}{\textwidth}
\footnotesize
\textit{Note:} Entries report estimated Hill tail indices $\hat{\alpha}(k)$ computed from the negative standardized AR(1)-GARCH(1,1) residuals. The threshold $k$ represents the number of largest loss magnitudes used in estimation. Selected thresholds from $k=40$ to $k=200$ are reported, with $k=100$ retained as the common reference threshold.
\end{minipage}
\end{table}

The expanded sensitivity analysis reveals a general downward tendency in the estimated Hill tail indices as $k$ increases from 40 to 200. This pattern should be interpreted in light of the trade-off involved in threshold selection. Smaller values of $k$ focus on more extreme observations but may produce estimates with greater sampling variability, whereas larger values include less extreme observations and may introduce bias into the estimation of asymptotic tail behavior. Although some portfolios exhibit relatively stable estimates over shorter intervals, no clearly stable region is observed consistently across all four series.

The relative ordering of the portfolios also varies with the selected threshold. For example, at $k=80$, LS-C95 has the highest estimated tail index, whereas LS-MVP has the highest estimate at $k=120$. At larger thresholds, particularly $k=200$, the estimates converge toward a narrower range of approximately 1.37--1.48. However, increasing $k$ incorporates progressively less extreme observations, which may extend beyond the region where the regularly varying tail approximation is appropriate. Consequently, changes in portfolio rankings at larger thresholds should not necessarily be interpreted as changes in their asymptotic extreme-tail characteristics.

At the common reference threshold $k=100$, LS-MVP and LS-C95 exhibit higher estimated Hill tail indices than EWP and DJIA, suggesting relatively lighter tails for their standardized innovations under the regularly varying tail assumption. Nevertheless, the shaded approximate 95\% confidence intervals in Figure~\ref{fig:hill_sensitivity} overlap considerably across the portfolios, limiting the strength of comparisons based on individual point estimates. The threshold $k=100$ is therefore retained as a common reporting reference rather than identified as an optimal or uniquely stable threshold. The broader sensitivity analysis highlights the uncertainty associated with threshold selection while providing additional context for interpreting the reference estimates.

The Hill tail-index analysis provides an additional perspective on extreme-tail behavior beyond the historical VaR and CVaR measures considered in the previous section. Whereas VaR and CVaR summarize loss thresholds and the average severity of losses beyond those thresholds, the Hill estimator characterizes the rate of asymptotic tail decay under a regularly varying tail assumption.

However, the present estimates characterize the negative standardized GARCH innovations rather than unconditional portfolio losses. Furthermore, the reported confidence intervals are approximate and do not account for uncertainty arising from GARCH parameter estimation, residual dependence, or threshold selection. Accordingly, the Hill results provide descriptive evidence of conditional-innovation tail behavior and its sensitivity to threshold selection, rather than definitive evidence of differences in unconditional portfolio downside risk.

\section{Long Range Dependence in Volatility}\label{sec:long-range}

Understanding the temporal structure of volatility is important for assessing portfolio risk, particularly when volatility shocks persist over time. Financial returns commonly exhibit volatility clustering, with periods of relatively high volatility followed by further periods of high volatility and periods of low volatility followed by low volatility. Long-range dependence provides one possible explanation for persistent volatility dynamics.

In contrast to short-memory processes, whose autocorrelations typically decay exponentially, long-range dependence is associated with hyperbolically decaying autocorrelations. The concept was initially studied in hydrology \citep{hurst1951} and subsequently investigated in financial time series \citep{mandelbrot1968}. Although financial returns often exhibit relatively weak serial correlation \citep{fama1970}, their squared or absolute values may display substantial persistence \citep{ding1993,campbell1997}.

Given the focus of this study on diversified equity--bond ETF portfolios, we examine the volatility dynamics of the equally weighted portfolio (EWP) and two long--short optimized portfolios, LS-MVP and LS-C95.

\begin{figure}[H]
    \centering
    \includegraphics[width=0.90\textwidth]{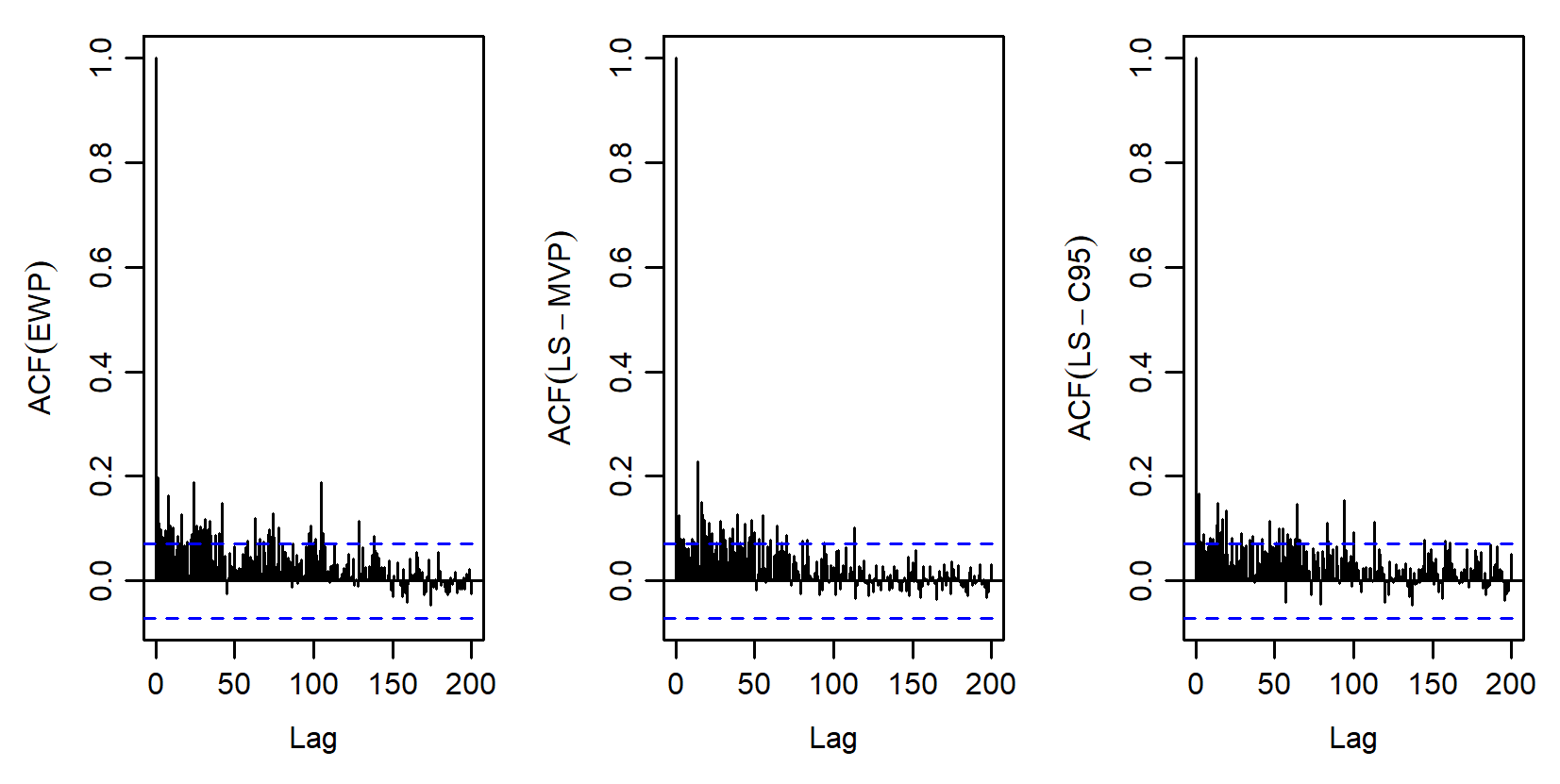}
    \caption{Autocorrelation functions of squared daily simple returns for EWP, LS-MVP, and LS-C95 (from left to right), calculated up to lag 200.}
    \label{fig:acf_lrd}
\end{figure}

Figure~\ref{fig:acf_lrd} presents the autocorrelation functions of squared daily simple returns for the three portfolios, calculated up to lag 200. The plots provide an initial examination of volatility dependence and motivate a comparison of conventional GARCH and fractionally integrated GARCH specifications. However, persistence in squared returns alone does not establish the presence of long-range dependence.

\subsection{Distributional Diagnostics of Portfolio Returns}

\begin{figure}[H]
\centering
\includegraphics[width=0.65\textwidth]{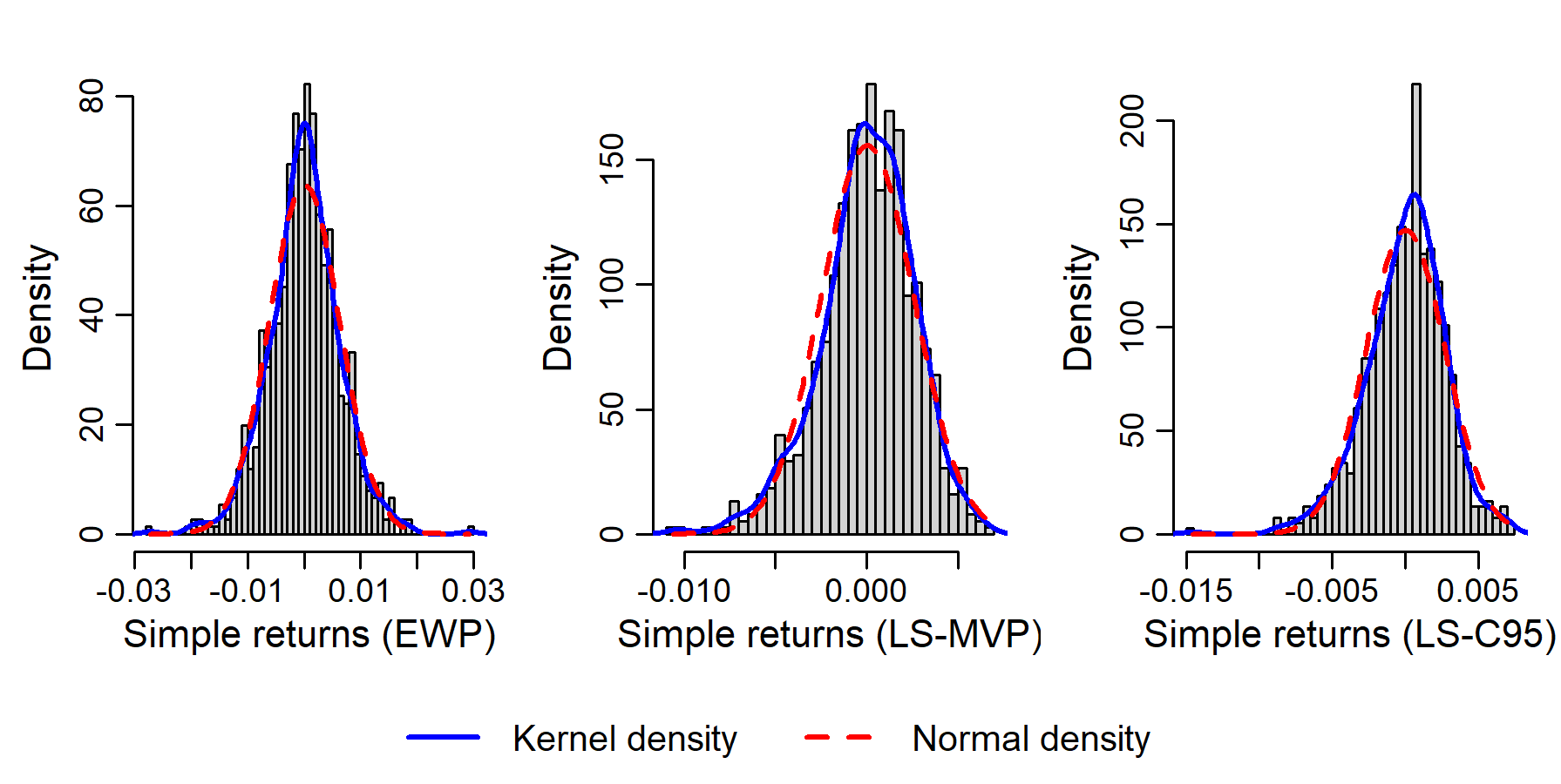}

\vspace{0.5cm}

\includegraphics[width=0.6\textwidth]{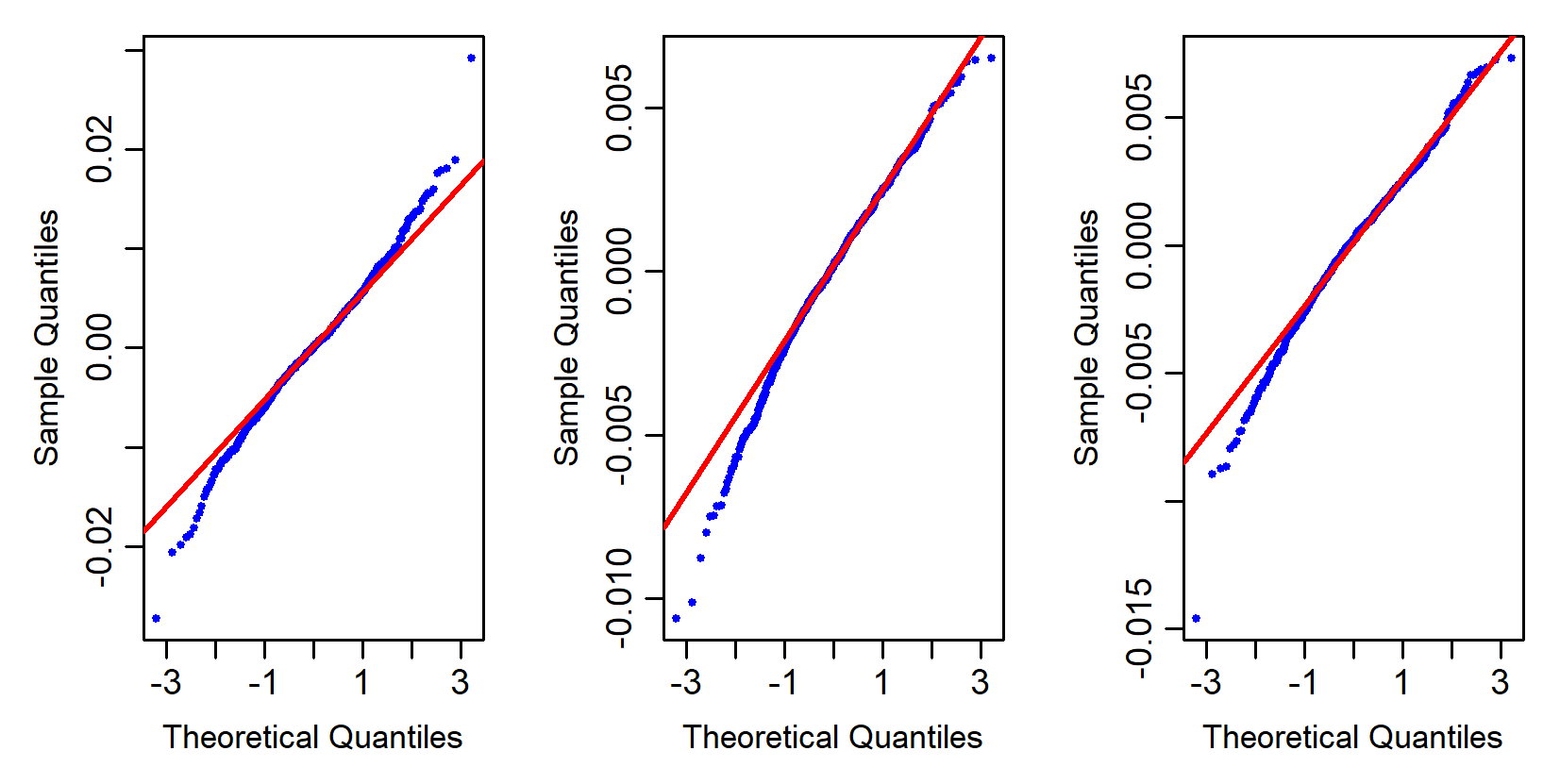}

\caption{Distributional diagnostics of daily simple returns for EWP, LS-MVP, and LS-C95. The upper panels show histograms with kernel density estimates and fitted normal densities, while the lower panels show normal Q--Q plots.}
\label{fig:dist_diag}
\end{figure}

Figure~\ref{fig:dist_diag} presents histograms with kernel and fitted normal densities, together with normal Q--Q plots, for the EWP, LS-MVP, and LS-C95 daily simple returns.

The EWP return distribution has skewness of $-0.0378$ and kurtosis of $4.2892$. Its skewness is close to zero, while its kurtosis exceeds the normal-distribution benchmark of three, indicating heavier tails relative to a normal distribution.

For LS-MVP, the estimated skewness and kurtosis are $-0.5083$ and $3.7481$, respectively. The negative skewness indicates asymmetry toward negative returns, while the kurtosis also exceeds the normal benchmark.

The LS-C95 return distribution has skewness of $-0.4974$ and kurtosis of $4.2875$. These estimates indicate negative skewness and excess kurtosis. Thus, all three return series exhibit kurtosis above three, while the two optimized portfolios display more pronounced negative skewness than EWP.

These diagnostics describe the unconditional distributions of the portfolio returns. The Gaussian innovation assumption used in the volatility models concerns the standardized conditional innovations rather than the unconditional return distributions. The observed skewness and excess kurtosis nevertheless motivate caution when interpreting results from conditionally Gaussian specifications.

\subsection{Model Specification}

We estimate AR(1)--GARCH(1,1) and AR(1)--FIGARCH(1,$d$,1) specifications using the daily simple returns of each portfolio. Both models are estimated by maximum likelihood under conditionally Gaussian standardized innovations.

Let $r_t$ denote the daily simple return of a portfolio. The conditional mean equation is

\begin{equation*}
r_t = \mu + \phi_1 r_{t-1} + \varepsilon_t,
\end{equation*}

where

\begin{equation*}
\varepsilon_t = \sqrt{h_t}\,z_t,
\qquad z_t \overset{\mathrm{i.i.d.}}{\sim} N(0,1),
\end{equation*}
and $h_t$ denotes the conditional variance.

For the GARCH(1,1) specification of \citet{bollerslev1986}, the conditional variance follows

\begin{equation*}
h_t =
\omega +
\alpha_1\varepsilon_{t-1}^{2}
+
\beta_1 h_{t-1},
\end{equation*}

where $\omega>0$, $\alpha_1\geq 0$, and $\beta_1\geq 0$. The coefficients $\alpha_1$ and $\beta_1$ describe the effects of past squared innovations and lagged conditional variance, respectively. The quantity $\alpha_1+\beta_1$ summarizes volatility persistence in the GARCH(1,1) model. When this sum is below but close to one, volatility shocks exhibit high persistence while remaining consistent with the usual covariance-stationarity condition.

To examine whether fractional integration provides an improved description of volatility dynamics, we also estimate the FIGARCH(1,$d$,1) specification introduced by \citet{baillie1996}. Its conditional variance dynamics can be expressed as
\begin{equation*}
(1-\beta_1 L)h_t
=
\omega+
\left[
1-\beta_1 L
-
(1-\phi_1^{(v)}L)(1-L)^d
\right]\varepsilon_t^2,
\end{equation*}

where $L$ denotes the lag operator, $d$ is the fractional-differencing parameter, and $\phi_1^{(v)}$ is a variance-equation parameter distinct from the AR(1) mean coefficient $\phi_1$. The parameter $d$ governs fractional persistence in the FIGARCH specification. Values $0<d<1$ allow for hyperbolically decaying volatility effects, whereas $d=0$ corresponds to a short-memory GARCH-type representation.

\subsection{Volatility Persistence}

Table~\ref{tab:model_comparison} reports the Akaike Information Criterion (AIC) and Bayesian Information Criterion (BIC) for the estimated AR(1)--GARCH(1,1) and AR(1)--FIGARCH(1,$d$,1) models.

The information criteria are reported on a per-observation basis, as calculated by the \texttt{rugarch} package. Both specifications use an AR(1) conditional mean and Gaussian standardized innovations.

\begin{table}[H]
\centering
\caption{Per-observation AIC and BIC for the AR(1)--GARCH(1,1) and AR(1)--FIGARCH(1,$d$,1) models.}
\begin{tabular}{lccc}
\hline
Model & EWP & LS-MVP & LS-C95 \\
\hline
\multicolumn{4}{c}{\textbf{AIC}} \\
AR(1)--FIGARCH(1,$d$,1)
& -7.421631 & -9.176151 & -9.059507 \\

AR(1)--GARCH(1,1)
& \textbf{-7.425491} & \textbf{-9.181173} & \textbf{-9.063849} \\
\hline
\multicolumn{4}{c}{\textbf{BIC}} \\
AR(1)--FIGARCH(1,$d$,1)
& -7.384825 & -9.139344 & -9.022700 \\

AR(1)--GARCH(1,1)
& \textbf{-7.394819} & \textbf{-9.150500} & \textbf{-9.033177} \\
\hline
\end{tabular}
\label{tab:model_comparison}
\end{table}

The AIC and BIC values are close for the two specifications. However, the GARCH model yields slightly lower values of both criteria for each return series. Thus, within the two-model comparison, the additional fractional-integration component does not provide sufficient improvement in fit to offset the additional model complexity.

All six estimated models achieved numerical convergence, with convergence codes equal to zero.

The estimated fractional integration parameters for the FIGARCH models are $d=0.9366$ for EWP, $d=0.9370$ for LS-MVP, and $d=0.9252$ for LS-C95. The corresponding estimated standard errors are $0.0343$, $0.0411$, and $0.0533$, respectively. These large estimates indicate substantial fractional persistence within the fitted FIGARCH specifications. Nevertheless, the slightly lower AIC and BIC values of GARCH suggest that the conventional GARCH specification is preferable for the present analysis. The information-criterion comparisons do not, however, rule out the possibility of long-memory volatility dynamics.

Given these results, Table~\ref{tab:garch_parameters} reports the estimated parameters of the AR(1)--GARCH(1,1) models, together with their estimated standard errors.

\begin{table}[H]
\centering
\caption{Estimated AR(1)--GARCH(1,1) parameters for daily simple returns. Estimated standard errors are reported in parentheses.}
\begin{tabular}{lccc}
\hline
Parameter & EWP & LS-MVP & LS-C95 \\
\hline
$\mu$
& 0.000149 & 0.000099 & 0.000092 \\
& (0.000238) & (0.000074) & (0.000076) \\

$\phi_1$
& 0.153580 & -0.115589 & -0.159084 \\
& (0.036608) & (0.036817) & (0.037399) \\

$\omega$
& $2.6242\times10^{-7}$
& $2.9586\times10^{-8}$
& $2.8855\times10^{-8}$ \\
& $(1.0794\times10^{-6})$
& $(1.0234\times10^{-6})$
& $(6.0290\times10^{-7})$ \\

$\alpha_1$
& 0.033353 & 0.040734 & 0.043821 \\
& (0.015986) & (0.028497) & (0.018703) \\

$\beta_1$
& 0.958917 & 0.954484 & 0.952940 \\
& (0.016359) & (0.027246) & (0.017353) \\

$\alpha_1+\beta_1$
& 0.992269 & 0.995219 & 0.996761 \\
\hline
\end{tabular}
\label{tab:garch_parameters}
\end{table}

The estimated GARCH coefficients indicate high conditional volatility persistence. The sums $\alpha_1+\beta_1$ are approximately $0.9923$, $0.9952$, and $0.9968$ for EWP, LS-MVP, and LS-C95, respectively.

All three persistence estimates are below but close to one. Consequently, the fitted GARCH models indicate that volatility shocks decay slowly, although the estimated persistence parameters alone do not establish fractional long-range dependence.

The estimated standard errors also indicate uncertainty in individual coefficient estimates. In particular, the statistical significance of the ARCH coefficient $\alpha_1$ differs across the portfolios. These results should therefore be interpreted as model-based estimates rather than evidence of identical volatility dynamics across the three return series.

\begin{figure}[H]
    \centering
    \includegraphics[width=0.8\textwidth]{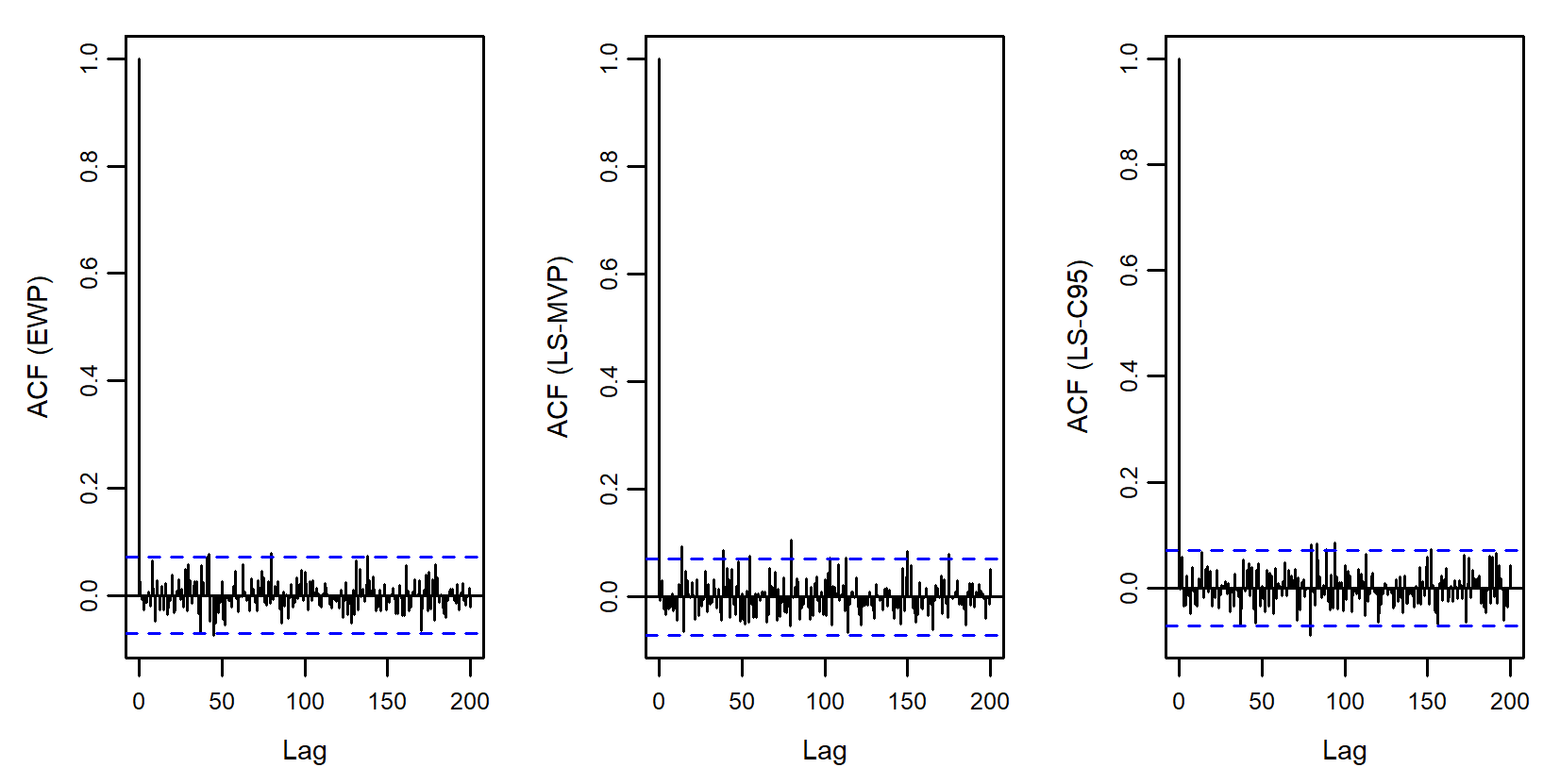}
    \caption{Autocorrelation functions of squared standardized residuals from the AR(1)--GARCH(1,1) models for EWP, LS-MVP, and LS-C95 (from left to right), calculated up to lag 200.}
    \label{fig:acf_residual}
\end{figure}

Figure~\ref{fig:acf_residual} presents the autocorrelation functions of squared standardized residuals from the fitted GARCH models, calculated up to lag 200.

To assess remaining dependence after GARCH filtering, we apply Ljung--Box tests to both the standardized residuals and their squares at lags 10, 20, and 50. Table~\ref{tab:garch_ljungbox} reports the resulting $p$-values.

\begin{table}[H]
\centering
\caption{Ljung--Box test $p$-values for standardized residuals and squared standardized residuals from the fitted GARCH models.}
\begin{tabular}{llccc}
\hline
Residual series & Portfolio & Lag 10 & Lag 20 & Lag 50 \\
\hline
Standardized
& EWP & 0.3062 & 0.3485 & 0.5503 \\
& LS-MVP & 0.9145 & 0.8269 & 0.8616 \\
& LS-C95 & 0.9960 & 0.9421 & 0.9377 \\
\hline
Squared standardized
& EWP & 0.7254 & 0.9194 & 0.5064 \\
& LS-MVP & 0.9460 & 0.5528 & 0.6626 \\
& LS-C95 & 0.3022 & 0.4603 & 0.9113 \\
\hline
\end{tabular}
\label{tab:garch_ljungbox}
\end{table}

For the standardized residuals, the Ljung--Box tests do not reject the null hypothesis of no serial correlation at the examined lags for any of the three portfolios.

Similarly, the tests on squared standardized residuals do not reject the null hypothesis of no remaining autocorrelation at lags 10, 20, or 50. These results are consistent with the residual ACF diagnostics and indicate that the fitted GARCH specifications capture much of the observed conditional volatility dependence. However, failure to reject the Ljung--Box null hypothesis does not establish that all forms of serial dependence or long-range dependence have been eliminated.

Taken together, the information-criterion comparisons, estimated persistence parameters, and residual diagnostics support the use of AR(1)--GARCH(1,1) as a parsimonious conditional volatility specification for these three portfolio return series. Although FIGARCH produces large fractional-integration estimates, the present model comparisons do not demonstrate a clear improvement over conventional GARCH.

\subsection{Implications for Portfolio Modeling}

The empirical results indicate substantial volatility persistence in the EWP, LS-MVP, and LS-C95 daily simple returns. The estimated GARCH persistence coefficients are close to one, suggesting that volatility shocks may influence conditional risk over extended periods.

Within the comparison between GARCH and FIGARCH, the GARCH specification produces slightly lower AIC and BIC values for all three portfolios. Moreover, the standardized residual diagnostics show no statistically significant remaining serial correlation at the examined Ljung--Box lags.

These findings support the use of the AR(1)--GARCH(1,1) model as a relatively parsimonious description of the conditional volatility dynamics in the present sample. They do not establish the absence of long-memory effects, and the conclusions remain conditional on the fitted model specifications and Gaussian innovation assumption.

While this section focuses on conditional volatility persistence, the distributional characteristics of portfolio returns are also relevant for risk modeling. The following section examines the Normal Inverse Gaussian (NIG) L\'evy process and its application to model-implied option pricing.


\section{Option Pricing under the Normal Inverse Gaussian (NIG) L\'evy Process}\label{sec:option-pricing}

The empirical analysis in the preceding sections shows that the EWP return series exhibits non-Gaussian behavior, including asymmetry, excess kurtosis, and heavy tails. These features motivate the use of a more flexible distributional framework for modeling portfolio returns. We employ the Normal Inverse Gaussian (NIG) distribution, which accommodates skewness and heavy-tailed behavior in financial returns \citep{barndorffnielsen1997,barndorffnielsen1998}.

Building on the historical return analysis, we examine the implications of the fitted NIG dynamics for option valuation. The estimated NIG model is incorporated into a martingale-corrected pricing specification to obtain European option prices across different strikes and maturities and construct the corresponding implied-volatility surface. Since the NIG parameters are estimated from historical EWP returns rather than calibrated to observed option prices, the resulting option prices and implied volatilities are model-implied rather than market-calibrated.

\subsection{NIG Model Specification}

Let $X_t$ denote an NIG L\'evy process.\footnote{A L\'evy process is a stochastic process with stationary and independent increments, meaning that changes over non-overlapping time intervals are independent and their probability distribution depends only on the length of the time interval. An NIG L\'evy process has Normal Inverse Gaussian distributed increments \citep{barndorffnielsen1997,barndorffnielsen1998}.} The NIG distribution admits a representation in terms of Brownian motion subordinated by an inverse-Gaussian process \citep{barndorffnielsen1997,barndorffnielsen1998}. In our numerical implementation, the model is evaluated directly using its closed-form cumulant and characteristic functions.

For the parameter vector
\[
\Theta=(\alpha,\beta,\delta,\mu),
\]
the cumulant-generating function of a unit-time increment is
\begin{equation*}
K(w)
=
\mu w
+
\delta
\left[
\sqrt{\alpha^2-\beta^2}
-
\sqrt{\alpha^2-(\beta+w)^2}
\right].
\end{equation*}

Here, $\alpha>0$ controls tail behavior, $\beta$ controls asymmetry, $\delta>0$ is a scale parameter, and $\mu$ is a location parameter. The corresponding characteristic exponent is
\begin{equation*}
\psi(u)=K(iu),
\end{equation*}
so that the characteristic function of an NIG increment over a time interval $t$ is
\begin{equation*}
\phi_X(u;t)=\exp\{t\psi(u)\}.
\end{equation*}
The closed-form characteristic function makes the NIG model particularly convenient for Fourier-based option-pricing methods.

\subsection{Estimation of the NIG Parameters}

The NIG model is fitted to the daily log returns of the same daily-rebalanced EWP used as the benchmark. The EWP simple returns are converted to log returns as $r_{\mathrm{EWP},t}=\ln(1+R_{\mathrm{EWP},t})$. Estimation uses the full sample period (1,257 daily observations) from January 2020 through December 2024.

The NIG parameters are estimated by minimizing an objective function that combines discrepancies between the theoretical and empirical moments with differences between the theoretical and empirical characteristic functions. The estimation objective is given by
\begin{equation*}
J(\Theta)
=
\sum_{q=1}^{4}
\left(
1-\frac{m_q(\Theta)}{\widehat{m}_q}
\right)^2
+
\left[
\sum_{j=1}^{1000}
\left|
1-\frac{\phi_{\Theta}(y_j)}
{\widehat{\phi}(y_j)}
\right|
\right]^2,
\end{equation*}
where $m_q(\Theta)$ and $\widehat{m}_q$ represent the theoretical and empirical mean, variance, skewness, and kurtosis, respectively. The function  $\widehat{\phi}$ denotes the empirical characteristic function.
The characteristic-function component of the objective function is evaluated at 1,000 equally spaced points over the interval $[-1.5,1.5]$. All five components enter the objective with unit coefficients, with the characteristic-function component defined as the squared sum of absolute relative discrepancies.

The optimization is performed using MATLAB's \texttt{fmincon}, subject to the parameter bounds
\[
10^{-6}\leq\alpha\leq500,\qquad
-500\leq\beta\leq500,
\]
\[
10^{-6}\leq\delta\leq1,\qquad
-1\leq\mu\leq1.
\]
Additionally, the nonlinear constraint
\[
\beta^2-\alpha^2+10^{-8}\leq0
\]
is imposed to ensure the NIG admissibility condition $\alpha>|\beta|$. The resulting parameter estimates are reported in Table~\ref{tab:nig-parameters}.

\begin{table}[htb!]
    \centering
    \caption{Estimated NIG parameters for the full-sample EWP return series.}
    \label{tab:nig-parameters}
    \renewcommand{\arraystretch}{1.5}
    \begin{tabular}{|c|c|c|c|c|}
        \hline
        \textbf{Parameters} & $\boldsymbol{\alpha}$ & $\boldsymbol{\beta}$ & $\boldsymbol{\delta}$ & $\boldsymbol{\mu}$ \\
        \hline
        \textbf{Estimated Value} & 69.1337 & -17.4890 & 0.00342448 & 0.00108 \\
        \hline
    \end{tabular}
\end{table}

The fitted characteristic function is evaluated against the empirical characteristic function over the specified 1,000-point frequency grid, with a maximum discrepancy of $1.12\times10^{-7}$, measured by $\left|1-\mathrm{CF}/\mathrm{ECF}\right|$. This indicates close agreement between the characteristic functions on the evaluated grid.

\subsection{Martingale-Corrected NIG Pricing Specification}

For the option-pricing analysis, we employ an exponential L\'evy pricing specification and adjust the drift of the asset-price process using the NIG exponential-moment correction \(K(1)\), consistent with the martingale condition used in exponential L\'evy models \citep{ContTankov2004}. For valuation purposes, we specify the NIG L\'evy dynamics under a pricing probability measure \(\mathbb{Q}\), using the parameters estimated from historical portfolio returns. The drift adjustment ensures that the discounted asset-price process is a martingale under the stipulated pricing law, while retaining the fitted NIG distributional structure.

We begin by expressing the asset price in exponential form as
\begin{equation*}
S_t = S_0 \exp\{ct + X_t\},
\end{equation*}
where \(S_0\) denotes the initial underlying value, \(X_t\) is the NIG L\'evy process introduced above, specified under \(\mathbb{Q}\), and \(c\) is a drift term to be determined. Using the cumulant-generating function \(K(w)\) defined previously, the exponential moment of \(X_t\) satisfies
\begin{equation*}
\mathbb{E}_{\mathbb{Q}}\left[e^{wX_t}\right]
= \exp\{tK(w)\},
\end{equation*}
for values of \(w\) for which the exponential moment exists. Since the asset-price specification contains \(e^{X_t}\), the relevant exponential moment is obtained by setting \(w=1\). Hence,
\begin{equation*}
\mathbb{E}_{\mathbb{Q}}\left[e^{X_t}\right]
= \exp\{tK(1)\},
\end{equation*}
provided that \(K(1)\) is finite.\footnote{For the NIG model, the exponential moment
required for the martingale correction exists when
$|\beta+1|<\alpha$. Using the estimated parameters
$\alpha=69.1337$ and $\beta=-17.4890$, we obtain
$|\beta+1|=16.4890<69.1337$, confirming that the condition is satisfied.}

For the discounted asset-price process to satisfy the martingale condition, we require
\begin{equation*}
\mathbb{E}_{\mathbb{Q}}\left[e^{-rt}S_t\right]
= S_0,
\end{equation*}
where \(r\) denotes the risk-free rate expressed on the same time scale as \(t\). Substituting the exponential price specification gives
\begin{align*}
\mathbb{E}_{\mathbb{Q}}\left[e^{-rt}S_t\right]
&= \mathbb{E}_{\mathbb{Q}}
   \left[e^{-rt}S_0e^{ct+X_t}\right] \\
&= S_0e^{(c-r)t}
   \mathbb{E}_{\mathbb{Q}}\left[e^{X_t}\right] \\
&= S_0e^{(c-r)t}e^{tK(1)}.
\end{align*}

Therefore, the martingale condition requires
\begin{equation*}
c-r+K(1)=0,
\end{equation*}
and hence
\begin{equation*}
c=r-K(1).
\end{equation*}
The resulting martingale-corrected exponential NIG price process is
\begin{equation*}
S_t=S_0\exp\left\{(r-K(1))t+X_t\right\}.
\end{equation*}
Indeed,
\begin{equation*}
\mathbb{E}_{\mathbb{Q}}[S_t]
=S_0e^{(r-K(1))t}e^{tK(1)}
=S_0e^{rt},
\end{equation*}
and consequently,
\begin{equation*}
\mathbb{E}_{\mathbb{Q}}\left[e^{-rt}S_t\right]
=S_0.
\end{equation*}
Furthermore, the independent-increment property of the NIG L\'evy process establishes the conditional martingale property. For \(0\leq s\leq t\),
\begin{align*}
&\mathbb{E}_{\mathbb{Q}}
\left[e^{-rt}S_t\mid\mathcal{F}_s\right] \\
&\quad =
e^{-rs}S_s
\mathbb{E}_{\mathbb{Q}}
\left[
e^{-K(1)(t-s)+X_t-X_s}
\mid\mathcal{F}_s
\right] \\
&\quad =
e^{-rs}S_s
e^{-K(1)(t-s)}
e^{(t-s)K(1)} \\
&\quad =e^{-rs}S_s.
\end{align*}
Thus, the adjustment by \(K(1)\) provides the drift correction required for the discounted asset-price process to be a martingale under the specified NIG pricing law, while retaining the fitted NIG distributional structure.
The drift adjustment \(c=r-K(1)\) follows the mean-correcting martingale approach discussed by \citet{ContTankov2004} and applied to exponential L\'evy option pricing by \citet{lindquist2022}. Under the stipulated NIG pricing law, the correction ensures that the discounted asset-price process is a martingale, provided that \(K(1)\) is finite. However, this drift correction does not, by itself, establish equivalence between the historical probability measure and the pricing measure \(\mathbb{Q}\). Furthermore, we do not establish a unique risk-neutral pricing measure or claim that the specified martingale-corrected NIG dynamics are uniquely determined by the historical return data. Accordingly, the resulting option prices are interpreted as model-implied prices under the stipulated martingale-corrected NIG specification, rather than uniquely determined arbitrage-free market prices.

\subsection{FFT Option Pricing}

European call-option prices are evaluated numerically using the Fast Fourier Transform (FFT) methodology of \citet{CarrMadan1999}. The approach uses the characteristic function associated with the martingale-corrected NIG price process to evaluate option prices across strikes and maturities. Following \citet{CarrMadan1999}, the call price is recovered through Fourier inversion:
\begin{equation*}
C_T(k)
=
\frac{e^{-\alpha_{\mathrm{d}} k}}{\pi}
\operatorname{Re}
\left[
\int_0^\infty e^{-ivk}\psi_T(v)\,dv
\right],
\end{equation*}
where
\[
k=\ln K,
\qquad
\psi_T(v)
=
\frac{
e^{-rT}\phi_T\!\left(v-i(\alpha_{\mathrm{d}}+1)\right)
}{
(\alpha_{\mathrm{d}}+iv)(\alpha_{\mathrm{d}}+1+iv)
}.
\]
Here, $\alpha_{\mathrm{d}}>0$ is the damping parameter, and $\phi_T(u)$ denotes the characteristic function of the martingale-corrected NIG log-price process, given by
\begin{equation*}
\phi_T(u)
=
\exp\left\{
iu\left[\ln S_0+(r-K(1))T\right]
+TK(iu)
\right\}.
\end{equation*}
For the baseline FFT implementation, we use
$N=2^{10}=1024$ integration points and the midpoint integration rule. The numerical integration cutoff is $v_{\max}=575$, yielding
a frequency spacing of $\Delta v=v_{\max}/N=0.56152344$. Since the fitted NIG parameters describe daily log-return increments, maturities $T=1,\ldots,200$ are expressed in trading days, consistent with the continuously compounded daily risk-free rate used in the pricing model.

We set the damping parameter to $\alpha_{\mathrm{d}}=0.45$, for which the required NIG exponential moment is finite.\footnote{The Carr--Madan FFT method requires a positive damping parameter $\alpha_{\mathrm{d}}>0$ and a finite exponential moment at $1+\alpha_{\mathrm{d}}$. For the NIG model, the condition $|\beta+\alpha_{\mathrm{d}}+1|<\alpha$ is satisfied, since $|-17.4890+0.45+1|=16.0390<69.1337$.} The analysis considers maturities from 1 to 200 trading days. European put-option prices are obtained from the corresponding call-option prices through put--call parity, assuming no dividends:
\begin{equation*}
P(K,T)
=
C(K,T)+Ke^{-rT}-S_0.
\end{equation*}
Option prices are reported over the moneyness range $0.8\leq K/S_0\leq1.2$, focusing on strikes around the reference underlying value. To assess the numerical stability of the FFT implementation, we repeat the call-option valuation using progressively finer grids, $N=2^{10},2^{11},2^{12}$, and $2^{13}$. Because changing $N$ produces slightly different strike grids, prices are linearly interpolated onto 100 common moneyness points over $0.81\leq K/S_0\leq1.19$ for comparison. The slightly narrower interval ensures that the common comparison points lie within the reported moneyness range for all four FFT grids. Table~\ref{tab:fft_convergence} reports the resulting convergence diagnostics.

\begin{table}[H]
\centering
\caption{FFT convergence diagnostics under midpoint integration}
\label{tab:fft_convergence}
\resizebox{\textwidth}{!}{
\begin{tabular}{lcccc}
\hline
Grid comparison &
Maximum absolute difference &
Mean absolute difference &
$K/S_0$ at maximum &
Maturity (days) \\
\hline
$2^{10}$ vs.\ $2^{11}$
& $5.5048\times10^{-2}$
& $1.9754\times10^{-3}$
& 1.005758 & 2 \\

$2^{11}$ vs.\ $2^{12}$
& $2.3843\times10^{-2}$
& $4.7740\times10^{-4}$
& 0.998081 & 1 \\

$2^{12}$ vs.\ $2^{13}$
& $7.5322\times10^{-3}$
& $1.2103\times10^{-4}$
& 1.001919 & 1 \\
\hline
\end{tabular}
}
\end{table}

Both the maximum and mean absolute pricing differences decrease as the FFT grid is refined. The maximum absolute difference declines from approximately $5.50\times10^{-2}$ between $N=2^{10}$ and $N=2^{11}$ to $7.53\times10^{-3}$ between $N=2^{12}$ and $N=2^{13}$, while the corresponding mean absolute difference declines from approximately $1.98\times10^{-3}$ to $1.21\times10^{-4}$. The largest differences occur at maturities of one or two trading days and at moneyness levels close to one. These results indicate decreasing pricing differences with finer FFT grids, although numerical discrepancies remain, particularly at short maturities.

As an additional numerical diagnostic, the call-price surface generated using the baseline $N=2^{10}$ grid is checked against the standard no-arbitrage call-price bounds, monotonicity in strike, and convexity in strike over $0.8\leq K/S_0\leq1.2$. A numerical tolerance of $10^{-8}$ is used to identify violations. The diagnostics are performed on the original FFT-generated call prices before any adjustment of negative numerical values. Table~\ref{tab:fft_arbitrage} summarizes the results.

\begin{table}[H]
\centering
\caption{Numerical no-arbitrage diagnostics for the baseline FFT grid}
\label{tab:fft_arbitrage}
\begin{tabular}{lcc}
\hline
Diagnostic & Number of violations & Largest violation \\
\hline
Lower call-price bound & 29 & $3.5264\times10^{-5}$ \\
Upper call-price bound & 0 & --- \\
Monotonicity in strike & 0 & --- \\
Convexity in strike & 1 & $3.1978\times10^{-8}$ \\
\hline
\end{tabular}
\end{table}

At the specified numerical tolerance, the lower-bound and convexity violations are small in magnitude. The largest lower-bound violation is approximately $3.53\times10^{-5}$ and occurs at $K/S_0=0.8042$ for the one-day maturity, while the largest convexity violation is approximately $3.20\times10^{-8}$ and occurs at $K/S_0=0.8131$, also for the one-day maturity. No upper-bound or strike-monotonicity violations are observed. These results indicate that the baseline FFT implementation exhibits small residual numerical violations, particularly at short maturities.

\subsection{Option Pricing Surfaces}

For the option-pricing analysis, the synthetic underlying is based on the daily-rebalanced EWP. We set $S_0=100$ as a normalized reference value for the model-implied option-pricing exercise.

The 3-month U.S. Treasury yield observed on December 31, 2024, is used as a proxy for the risk-free rate. Because the pricing model employs exponential discounting, the annual effective rate is converted to a continuously compounded daily rate according to
\begin{equation*}
r_{\mathrm{daily}}
=
\frac{\ln(1+r_{\mathrm{annual}})}{252}.
\end{equation*}
The resulting continuously compounded daily risk-free rate is \(r_{\mathrm{daily}}=0.0001697305\), corresponding to an annual continuously compounded rate of approximately \(4.2772\%\).

\begin{figure}[H]
\centering

\begin{subfigure}{0.49\textwidth}
\centering
\includegraphics[width=\linewidth]{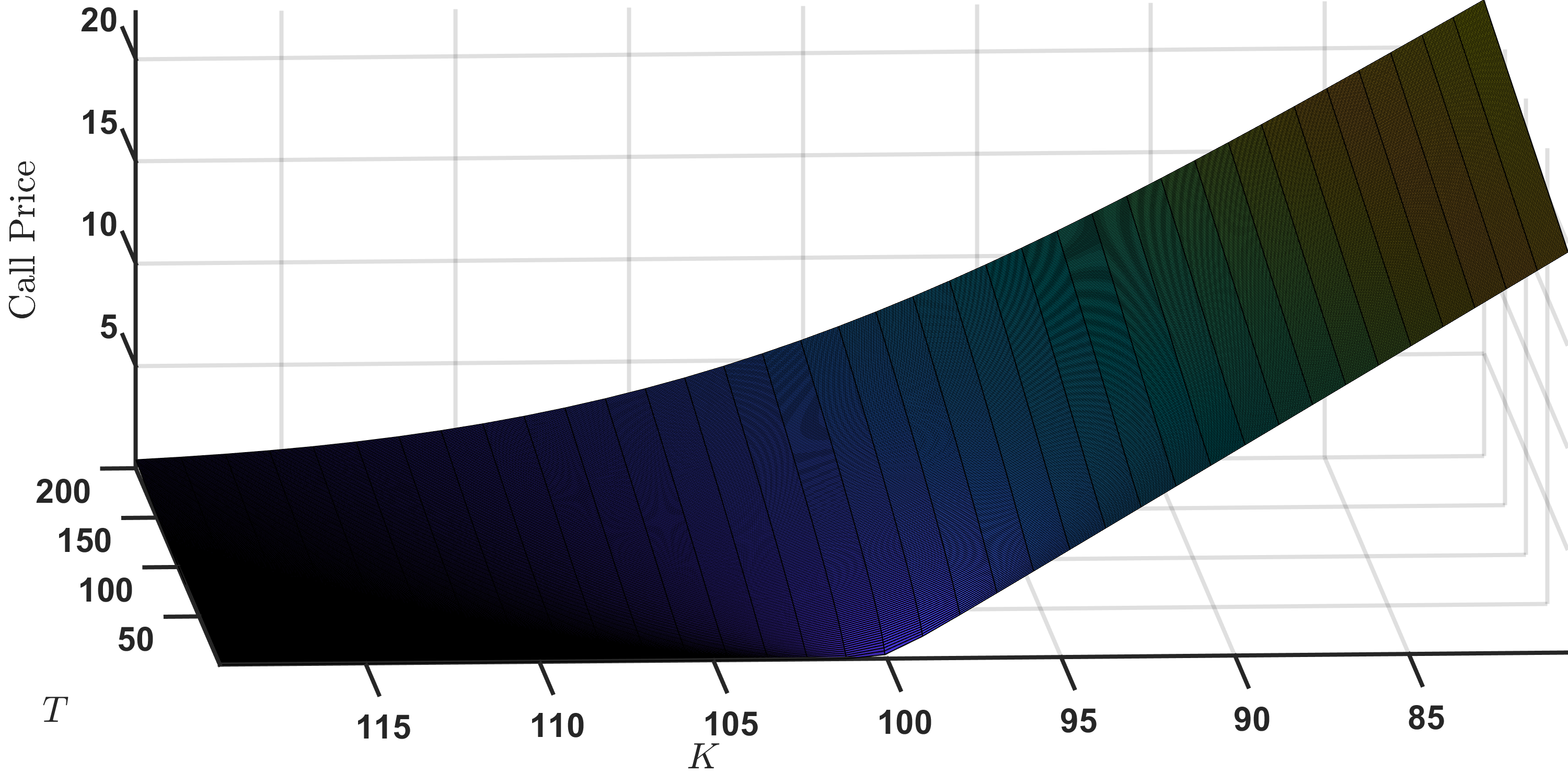}
\caption{European call option price surface}
\label{fig:nig_call_surface}
\end{subfigure}
\hfill
\begin{subfigure}{0.49\textwidth}
\centering
\includegraphics[width=\linewidth]{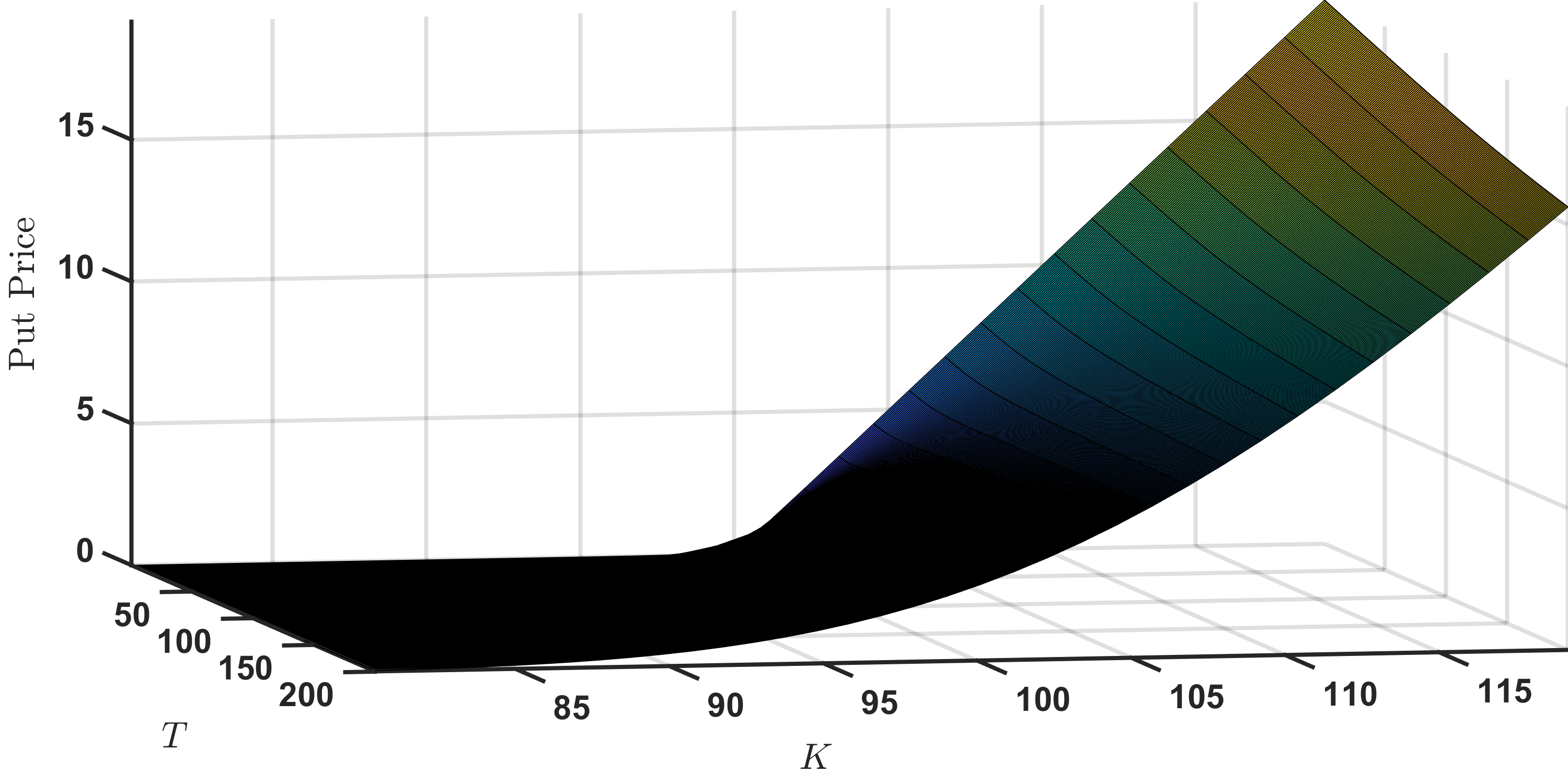}
\caption{European put option price surface}
\label{fig:nig_put_surface}
\end{subfigure}

\caption{Model-generated non-dividend European option-price surfaces. The underlying is based on the daily-rebalanced EWP, with a normalized initial reference value of $S_0=100$ and an annual continuously compounded risk-free rate of $r=4.2772\%$. Maturities are expressed in trading days.}
\label{fig:nig_option_surfaces}
\end{figure}

Figure~\ref{fig:nig_option_surfaces} presents the resulting European call- and put-option price surfaces generated under the fitted NIG model. Call prices generally decrease as the strike price increases, whereas put prices increase with the strike price, consistent with the payoff structure of European call and put options. The surfaces also illustrate how the model-generated option values vary with time to maturity.

The put-option prices are obtained from the corresponding call-option prices through put--call parity. Together, the two surfaces illustrate the variation in model-generated European option values across strike prices and times to maturity under the fitted NIG framework.

\subsection{NIG Model-Implied Volatility}

To express the NIG option prices in the conventional implied-volatility representation, the Black--Scholes pricing equation is inverted numerically.

For each valid NIG call-option price, the implied volatility \(\sigma_{\mathrm{imp}}\) is defined implicitly by
\begin{equation*}
C_{\mathrm{BS}}
\left(
S_0,K,r,T,\sigma_{\mathrm{imp}}
\right)
=
C_{\mathrm{NIG}}(K,T),
\end{equation*}
where \(C_{\mathrm{NIG}}(K,T)\) denotes the call-option price generated under the fitted NIG pricing framework.

Implied volatilities are computed using MATLAB's \texttt{blsimpv} function. A valid NIG call-option price is one for which the Black--Scholes inversion returns a finite implied volatility. Prices that do not admit a valid inversion, including prices outside the Black--Scholes no-arbitrage bounds, are not used in the smoothing procedure. In the baseline calculation, 29 of the 7,400 grid points (0.392\%) produce \texttt{NaN} implied volatilities. These points are excluded from the Gaussian-kernel smoothing, which uses the remaining 7,371 valid observations to construct the displayed surface.

\begin{figure}[H]
\centering
\includegraphics[width=0.75\textwidth]{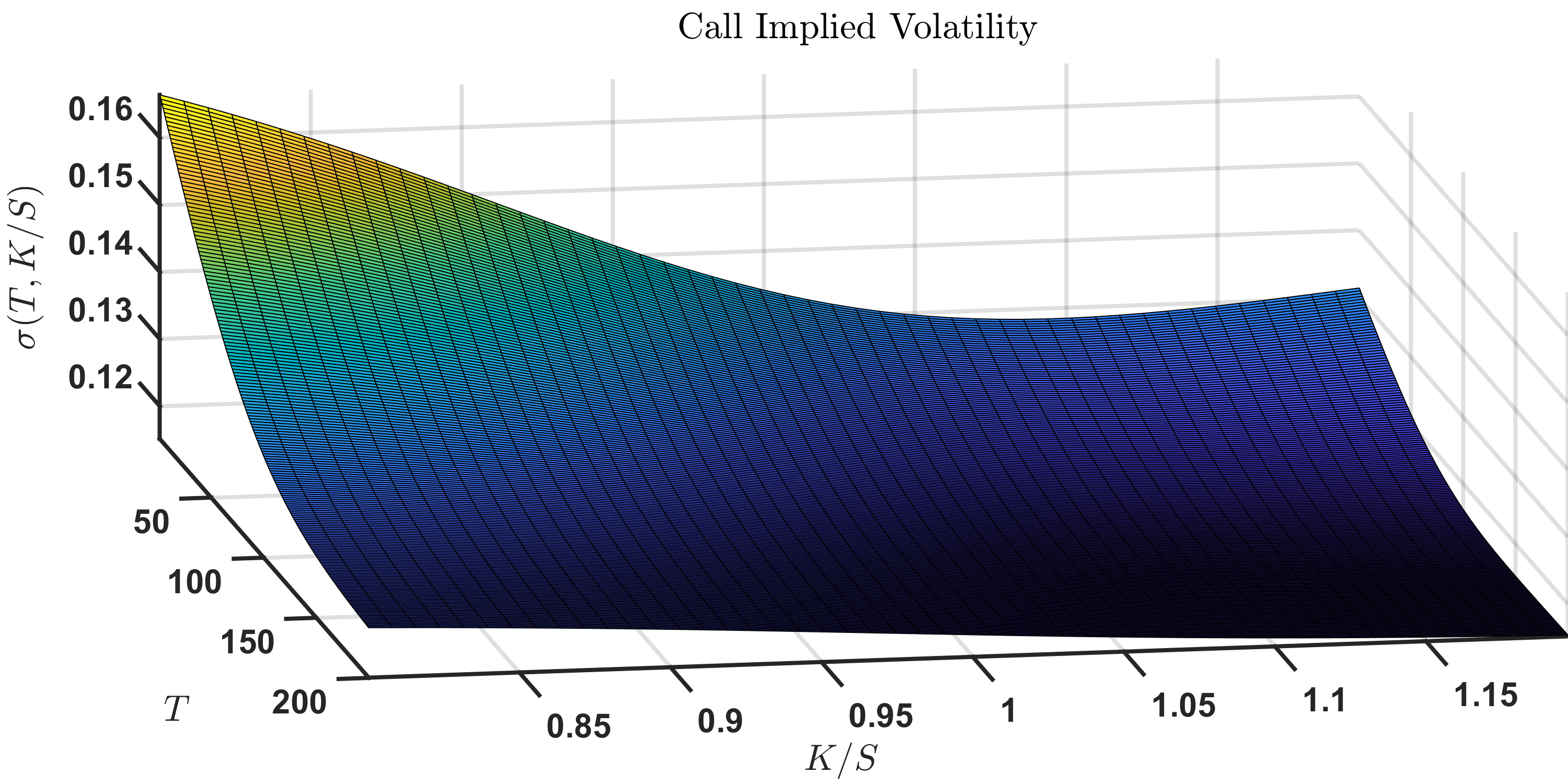}
\caption{Model-implied volatility surface for non-dividend European call options under the fitted NIG framework. The underlying is based on the daily-rebalanced EWP, with a normalized initial reference value of $S_0=100$ and an annual continuously compounded risk-free rate of $r=4.2772\%$. Maturities are expressed in trading days, and implied volatilities are annualized.}
\label{fig:nig_call_iv_surface}
\end{figure}

Figure~\ref{fig:nig_call_iv_surface} presents the resulting annualized model-implied volatility surface across strike prices and maturities, with daily interest rates and maturities converted to annual units using 252 trading days per year.

The implied-volatility surface is non-flat and varies across both strike price and time to maturity. Implied volatility generally decreases as the strike price increases, indicating an asymmetric volatility pattern across strikes. The surface also exhibits variation across maturities, reflecting the non-Gaussian characteristics of the fitted NIG return distribution relative to a constant-volatility specification.

Since the put-option prices are obtained from the call-option prices through put--call parity, a separate put-implied volatility surface does not provide additional independent information. The call-based implied-volatility surface is therefore used to summarize the volatility structure associated with the NIG-generated option prices.

\section{Conclusions}\label{conclusion}
This study examines the performance and risk characteristics of global equity–bond and multi-asset ETFs over 2020–2024, a period marked by the COVID-19 crisis, inflationary pressures, and changing monetary policy conditions. Using a unified empirical framework combining rolling-window portfolio optimization, downside-risk evaluation, extreme-tail estimation, conditional volatility modeling, and L\'evy-based option pricing, the analysis provides evidence on the behavior of alternative portfolio strategies across multiple dimensions of risk.

The empirical results highlight the importance of distinguishing overall return variability from downside and extreme-tail exposure. Although minimum-variance optimization directly targets return dispersion, it does not explicitly distinguish between favorable and unfavorable return deviations. CVaR-based optimization instead focuses on losses in the tail of the return distribution. Over the reported out-of-sample evaluation period, optimized portfolios exhibit lower empirical VaR and CVaR than the equally weighted benchmark, while long–short portfolios generally exhibit lower tail-loss measures than their corresponding long-only strategies. Long–short strategies also record the smallest maximum drawdown and realized volatility. However, the EWP achieves the highest reported Sharpe, Sortino, and Rachev ratios, with all reported Sharpe and Sortino ratios remaining negative. These findings indicate that improvements in downside-risk measures do not necessarily translate into superior risk-adjusted performance across all evaluation criteria.

The extreme-tail analysis further complements the conventional risk measures by examining the behavior of unusually large losses. The Hill tail-index estimates, calculated from standardized GARCH innovations, reveal differences in the estimated tail indices across portfolio strategies. However, these point estimates do not establish statistically significant differences in the unconditional tails of portfolio returns. Together, these results support evaluating portfolio strategies through multiple risk measures rather than relying on a single optimization objective or performance statistic.

The conditional volatility analysis identifies substantial volatility persistence in the portfolio return series. Comparisons between GARCH and FIGARCH specifications indicate that the GARCH model is favored by the reported information criteria, while squared-residual diagnostics show no significant remaining autocorrelation at the examined lags. These findings suggest that the GARCH specification provides an adequate representation of the observed volatility clustering within the models and sample considered, without establishing the absence of long-range dependence more generally.

The NIG L\'evy framework provides a complementary model-based approach to examining the distributional characteristics of portfolio returns and their implications for European option valuation. Using parameters estimated from historical returns and a martingale-corrected pricing specification, the analysis generates model-implied option prices and corresponding implied-volatility surfaces. These results illustrate the pricing implications of the fitted NIG model rather than observed market expectations or option prices calibrated to traded contracts. The option-pricing exercise therefore complements the historical portfolio and risk analysis without establishing a unique risk-neutral pricing measure.

From a practical perspective, the implementation of optimized portfolios must also account for transaction costs, portfolio turnover, short-selling costs, and leverage requirements. The transaction-cost sensitivity analysis provides an additional assessment of how trading expenses affect the reported portfolio performance and the relative attractiveness of alternative allocation strategies. Consequently, the gross performance and risk characteristics of optimized portfolios should not be interpreted as directly attainable net investment outcomes without considering these implementation costs.

Overall, the findings demonstrate that no single portfolio strategy consistently dominates across all dimensions of risk and performance. While optimization can reduce particular measures of portfolio risk, these improvements must be evaluated alongside risk-adjusted returns, extreme-tail behavior, volatility dynamics, and practical implementation costs. Future research could extend the analysis to additional market periods, alternative L\'evy specifications, dynamic allocation methods, and more detailed models of transaction costs and short-selling constraints. Such extensions would help assess the robustness and practical applicability of the portfolio strategies beyond the sample and assumptions considered in this study.


\normalem		

\appendix

\section{Risk Metrics Definitions}\label{risk-metrics}

\subsection{Maximum Drawdown (MDD)}

Let $P_p(t)$ denote the cumulative portfolio wealth of portfolio $p$ at time $t$, and let $M_p(t)$ denote the maximum portfolio wealth observed up to time $t$, defined as
\[
M_p(t)=\max_{0\leq s\leq t}P_p(s).
\]
The maximum drawdown over the interval $[0,T]$, expressed as a percentage, is defined as \footnote{See Section 4.3 in \citet{lindquist2022}.}
\[
\mathrm{MDD}(T)
=
100\max_{0\leq t\leq T}
\left(
\frac{M_p(t)-P_p(t)}{M_p(t)}
\right).
\]

The maximum drawdown measures the largest peak-to-trough decline experienced by the portfolio before a new peak is attained. This metric is particularly important in portfolio management because it directly quantifies realized downside exposure during stressed market conditions \footnote{ We compute the maximum drawdown (MDD) of each portfolio using its cumulative wealth series.
}.

\subsection{Sharpe Ratio}
Let  $r_p(t)$  denote the portfolio return and $r_f(t)$  the risk-free return at time  $t$. The Sharpe ratio over the interval  $[0,T]$  is given by
\[
\mathrm{SR}(T)
=\frac{\mathbb{E}[\,r_p(t) - r_f(t)\,]_{[0,T]}}
     {\sqrt{\mathrm{var}[\,r_p(t)-r_f(t)\,]_{[0,T]}}}
=\frac{\mu_ep[0,T]}{\sigma_ep[0,T]},
\]
where $\mu_ep$   and $\sigma_ep$  represent the mean and standard deviation of excess portfolio returns. 

\subsection{Sortino Ratio}
Unlike the Sharpe ratio, the Sortino ratio focuses exclusively on downside volatility. Define downside semi-deviation as
\[
\mathrm{DSD}
=
\sqrt{\mathbb{E}\!\left[(r_f(t)-r_p(t))^2_{+}\right]}.
\]
The Sortino ratio is then defined as\footnote{See \citet{sortino_classical1994}.}
\[
\mathrm{Sortino}(T)
=
\frac{\mathbb{E}[\,r_p(t)-r_f(t)\,]_{[0,T]}}
     {\mathrm{DSD}_{[0,T]}}.
\]

By concentrating exclusively on downside variation, the Sortino ratio provides a more appropriate measure of performance when return distributions exhibit asymmetry or downside-tail behavior.

\subsection{Rachev Ratio}

For tail confidence levels $\beta$ and $\gamma$, the Rachev ratio is defined as
\[
\mathrm{RR}_{\beta,\gamma}(T)
=
\frac{
\mathbb{E}\left[
r_p(t)-r_f(t)
\mid
r_p(t)-r_f(t) \geq q_{\beta}
\right]
}{
\left|
\mathbb{E}\left[
r_p(t)-r_f(t)
\mid
r_p(t)-r_f(t) \leq q_{1-\gamma}
\right]
\right|
},
\]
where $q_{\alpha}$ denotes the $\alpha$-quantile of the excess portfolio returns.

In our analysis, we set $\beta=\gamma=0.95$, corresponding to the upper and lower 5\% tails of excess portfolio returns. The ratio is calculated empirically as the mean of the highest 5\% of excess returns divided by the absolute value of the mean of the lowest 5\% of excess returns.

The Rachev ratio compares expected upper-tail gains with the magnitude of lower-tail returns and therefore measures tail asymmetry. 


\section{Portfolio Definitions}\label{porfolio-definition}
The appendix definitions below formalize the principal portfolio constructions and risk measures employed throughout the empirical analysis. Several of these definitions are standard within portfolio theory and quantitative risk management; however, they are included here for completeness and notational consistency.

\subsection{Equally Weighted Portfolio (EWP)}
The equally weighted portfolio assigns identical portfolio weights to each asset within the investment universe. Formally,\footnote{EWP is used as a benchmark in \citet{lindquist2022, demiguel2009,Lee2026TaiwanETFs}.}
\[
w_i = \frac{1}{N}, \qquad i = 1,\dots,N,
\]
where $N$ denotes the total number of assets. The corresponding portfolio return is given by 
\[
R_p(t) = \frac{1}{N}\sum_{i=1}^N R_i(t).
\]
The equally weighted portfolio serves as a benchmark specification because of its simplicity, robustness to parameter estimation error, and absence of optimization-induced instability.

\subsection{Value-at-Risk and Conditional Value-at-Risk}
Let $\bar r_p$ denote the daily portfolio return with cumulative distribution 
function $F(x)=\Pr\{\bar r_p \le x\}$. For a confidence level 
$\alpha\in(0,1)$, the Value-at-Risk (VaR) is defined as
\[
\mathrm{VaR}_\alpha(\bar r_p)
=
-F^{-1}(1-\alpha).
\]
Under this sign convention, $\mathrm{VaR}_\alpha$ represents the loss 
threshold associated with the lower $(1-\alpha)$ tail of the portfolio 
return distribution.

The Conditional Value-at-Risk (CVaR) at the same confidence level is 
defined as the average loss over the worst $(1-\alpha)$ portion of the 
return distribution:
\[
\mathrm{CVaR}_\alpha(\bar r_p)
=
-\frac{1}{1-\alpha}
\int_0^{1-\alpha} F^{-1}(u)\,du.
\]
Conditional Value-at-Risk therefore measures the average loss associated 
with the worst $(1-\alpha)$ portion of the return distribution.
\footnote{For more detail, see chapter 7 in \citet{tsay2010} and chapter 3 in \citet{lindquist2022}.}

Unlike VaR, CVaR captures the magnitude of extreme losses beyond the 
quantile threshold and therefore provides a more informative measure of 
downside exposure under heavy-tailed return distributions.

\subsection{Capital Market Line}

Let $(\sigma_M,\mu_M)$ denote the volatility and expected return of the tangent portfolio, respectively. The tangent portfolio is a risky portfolio that maximizes the Sharpe ratio and is obtained as the point of tangency between the efficient frontier and a line drawn from the risk-free rate $(r_f)$. It represents an optimal risky portfolio for mean--variance investors.

The Capital Market Line (CML) describes the relationship between the expected return and volatility of all efficient portfolios formed through combinations of the risk-free asset and the tangent portfolio. It is given by
\begin{equation*}
E[R_p]=r_f+\frac{\mu_M-r_f}{\sigma_M}\sigma_p
\end{equation*}
where $\sigma_p$ denotes the volatility of portfolio $(p)$.


\bibliographystyle{apacite}
\bibliography{references}  

\begin{thebibliography}{}

\bibitem [\protect \citeauthoryear {%
Baillie%
, Bollerslev%
\BCBL {}\ \BBA {} Mikkelsen%
}{%
Baillie%
\ \protect \BOthers {.}}{%
{\protect \APACyear {1996}}%
}]{%
baillie1996}
\APACinsertmetastar {%
baillie1996}%
\begin{APACrefauthors}%
Baillie, R\BPBI T.%
, Bollerslev, T.%
\BCBL {}\ \BBA {} Mikkelsen, H\BPBI O.%
\end{APACrefauthors}%
\unskip\
\newblock
\APACrefYearMonthDay{1996}{}{}.
\newblock
{\BBOQ}\APACrefatitle {Fractionally Integrated Generalized Autoregressive
  Conditional Heteroskedasticity} {Fractionally integrated generalized
  autoregressive conditional heteroskedasticity}.{\BBCQ}
\newblock
\APACjournalVolNumPages{Journal of Econometrics}{74}{1}{3--30}.
\newblock
\begin{APACrefDOI} \doi{10.1016/S0304-4076(95)01749-6} \end{APACrefDOI}
\PrintBackRefs{\CurrentBib}

\bibitem [\protect \citeauthoryear {%
Barndorff-Nielsen%
}{%
Barndorff-Nielsen%
}{%
{\protect \APACyear {1997}}%
}]{%
barndorffnielsen1997}
\APACinsertmetastar {%
barndorffnielsen1997}%
\begin{APACrefauthors}%
Barndorff-Nielsen, O\BPBI E.%
\end{APACrefauthors}%
\unskip\
\newblock
\APACrefYearMonthDay{1997}{}{}.
\newblock
{\BBOQ}\APACrefatitle {{Normal Inverse Gaussian} Distributions and Stochastic
  Volatility Modelling} {{Normal Inverse Gaussian} distributions and stochastic
  volatility modelling}.{\BBCQ}
\newblock
\APACjournalVolNumPages{Scandinavian Journal of Statistics}{24}{1}{1--13}.
\newblock
\begin{APACrefDOI} \doi{10.1111/1467-9469.00045} \end{APACrefDOI}
\PrintBackRefs{\CurrentBib}

\bibitem [\protect \citeauthoryear {%
Barndorff-Nielsen%
}{%
Barndorff-Nielsen%
}{%
{\protect \APACyear {1998}}%
}]{%
barndorffnielsen1998}
\APACinsertmetastar {%
barndorffnielsen1998}%
\begin{APACrefauthors}%
Barndorff-Nielsen, O\BPBI E.%
\end{APACrefauthors}%
\unskip\
\newblock
\APACrefYearMonthDay{1998}{}{}.
\newblock
{\BBOQ}\APACrefatitle {Processes of Normal Inverse Gaussian Type} {Processes of
  normal inverse gaussian type}.{\BBCQ}
\newblock
\APACjournalVolNumPages{Finance and Stochastics}{2}{1}{41--68}.
\newblock
\begin{APACrefDOI} \doi{10.1007/s007800050032} \end{APACrefDOI}
\PrintBackRefs{\CurrentBib}

\bibitem [\protect \citeauthoryear {%
Ben-David%
, Franzoni%
\BCBL {}\ \BBA {} Moussawi%
}{%
Ben-David%
\ \protect \BOthers {.}}{%
{\protect \APACyear {2018}}%
}]{%
bendavid2018etfvolatility}
\APACinsertmetastar {%
bendavid2018etfvolatility}%
\begin{APACrefauthors}%
Ben-David, I.%
, Franzoni, F.%
\BCBL {}\ \BBA {} Moussawi, R.%
\end{APACrefauthors}%
\unskip\
\newblock
\APACrefYearMonthDay{2018}{}{}.
\newblock
{\BBOQ}\APACrefatitle {Do {ETFs} Increase Volatility?} {Do {ETFs} increase
  volatility?}{\BBCQ}
\newblock
\APACjournalVolNumPages{The Journal of Finance}{73}{6}{2471--2535}.
\newblock
\begin{APACrefDOI} \doi{10.1111/jofi.12722} \end{APACrefDOI}
\PrintBackRefs{\CurrentBib}

\bibitem [\protect \citeauthoryear {%
{Board of Governors of the Federal Reserve System (US)}%
}{%
{Board of Governors of the Federal Reserve System (US)}%
}{%
{\protect \APACyear {2026}}%
{\protect \APACexlab {{\protect \BCnt {1}}}}}]{%
FRED_DEXCAUS}
\APACinsertmetastar {%
FRED_DEXCAUS}%
\begin{APACrefauthors}%
{Board of Governors of the Federal Reserve System (US)}.%
\end{APACrefauthors}%
\unskip\
\newblock
\APACrefYearMonthDay{2026{\protect \BCnt {1}}}{}{}.
\newblock
\APACrefbtitle {Canadian Dollars to {U.S.} Dollar Spot Exchange Rate
  [{DEXCAUS}].} {Canadian dollars to {U.S.} dollar spot exchange rate
  [{DEXCAUS}].}
\newblock
\APAChowpublished {Federal Reserve Bank of St. Louis, FRED}.
\newblock
\begin{APACrefURL} \url{https://fred.stlouisfed.org/series/DEXCAUS}
  \end{APACrefURL}
\newblock
\APACrefnote{Daily, not seasonally adjusted; accessed September 28, 2026}
\PrintBackRefs{\CurrentBib}

\bibitem [\protect \citeauthoryear {%
{Board of Governors of the Federal Reserve System (US)}%
}{%
{Board of Governors of the Federal Reserve System (US)}%
}{%
{\protect \APACyear {2026}}%
{\protect \APACexlab {{\protect \BCnt {2}}}}}]{%
FRED_DEXUSAL}
\APACinsertmetastar {%
FRED_DEXUSAL}%
\begin{APACrefauthors}%
{Board of Governors of the Federal Reserve System (US)}.%
\end{APACrefauthors}%
\unskip\
\newblock
\APACrefYearMonthDay{2026{\protect \BCnt {2}}}{}{}.
\newblock
\APACrefbtitle {{U.S.} Dollars to {Australian} Dollar Spot Exchange Rate
  [{DEXUSAL}].} {{U.S.} dollars to {Australian} dollar spot exchange rate
  [{DEXUSAL}].}
\newblock
\APAChowpublished {Federal Reserve Bank of St. Louis, FRED}.
\newblock
\begin{APACrefURL} \url{https://fred.stlouisfed.org/series/DEXUSAL}
  \end{APACrefURL}
\newblock
\APACrefnote{Daily, not seasonally adjusted; accessed September 28, 2026}
\PrintBackRefs{\CurrentBib}

\bibitem [\protect \citeauthoryear {%
Bollerslev%
}{%
Bollerslev%
}{%
{\protect \APACyear {1986}}%
}]{%
bollerslev1986}
\APACinsertmetastar {%
bollerslev1986}%
\begin{APACrefauthors}%
Bollerslev, T.%
\end{APACrefauthors}%
\unskip\
\newblock
\APACrefYearMonthDay{1986}{}{}.
\newblock
{\BBOQ}\APACrefatitle {Generalized Autoregressive Conditional
  Heteroskedasticity} {Generalized autoregressive conditional
  heteroskedasticity}.{\BBCQ}
\newblock
\APACjournalVolNumPages{Journal of Econometrics}{31}{3}{307--327}.
\newblock
\begin{APACrefDOI} \doi{10.1016/0304-4076(86)90063-1} \end{APACrefDOI}
\PrintBackRefs{\CurrentBib}

\bibitem [\protect \citeauthoryear {%
Campbell%
, Lo%
\BCBL {}\ \BBA {} MacKinlay%
}{%
Campbell%
\ \protect \BOthers {.}}{%
{\protect \APACyear {1997}}%
}]{%
campbell1997}
\APACinsertmetastar {%
campbell1997}%
\begin{APACrefauthors}%
Campbell, J\BPBI Y.%
, Lo, A\BPBI W.%
\BCBL {}\ \BBA {} MacKinlay, A\BPBI C.%
\end{APACrefauthors}%
\unskip\
\newblock
\APACrefYear{1997}.
\newblock
\APACrefbtitle {The Econometrics of Financial Markets} {The econometrics of
  financial markets}.
\newblock
\APACaddressPublisher{Princeton, NJ}{Princeton University Press}.
\PrintBackRefs{\CurrentBib}

\bibitem [\protect \citeauthoryear {%
Carr%
\ \BBA {} Madan%
}{%
Carr%
\ \BBA {} Madan%
}{%
{\protect \APACyear {1999}}%
}]{%
CarrMadan1999}
\APACinsertmetastar {%
CarrMadan1999}%
\begin{APACrefauthors}%
Carr, P.%
\BCBT {}\ \BBA {} Madan, D.%
\end{APACrefauthors}%
\unskip\
\newblock
\APACrefYearMonthDay{1999}{}{}.
\newblock
{\BBOQ}\APACrefatitle {Option Valuation Using the Fast {Fourier} Transform}
  {Option valuation using the fast {Fourier} transform}.{\BBCQ}
\newblock
\APACjournalVolNumPages{Journal of Computational Finance}{2}{4}{61--73}.
\newblock
\begin{APACrefDOI} \doi{10.21314/JCF.1999.043} \end{APACrefDOI}
\PrintBackRefs{\CurrentBib}

\bibitem [\protect \citeauthoryear {%
Cont%
}{%
Cont%
}{%
{\protect \APACyear {2001}}%
}]{%
cont2001}
\APACinsertmetastar {%
cont2001}%
\begin{APACrefauthors}%
Cont, R.%
\end{APACrefauthors}%
\unskip\
\newblock
\APACrefYearMonthDay{2001}{}{}.
\newblock
{\BBOQ}\APACrefatitle {Empirical Properties of Asset Returns: {S}tylized Facts
  and Statistical Issues} {Empirical properties of asset returns: {S}tylized
  facts and statistical issues}.{\BBCQ}
\newblock
\APACjournalVolNumPages{Quantitative Finance}{1}{2}{223--236}.
\newblock
\begin{APACrefDOI} \doi{10.1080/713665670} \end{APACrefDOI}
\PrintBackRefs{\CurrentBib}

\bibitem [\protect \citeauthoryear {%
Cont%
\ \BBA {} Tankov%
}{%
Cont%
\ \BBA {} Tankov%
}{%
{\protect \APACyear {2004}}%
}]{%
ContTankov2004}
\APACinsertmetastar {%
ContTankov2004}%
\begin{APACrefauthors}%
Cont, R.%
\BCBT {}\ \BBA {} Tankov, P.%
\end{APACrefauthors}%
\unskip\
\newblock
\APACrefYear{2004}.
\newblock
\APACrefbtitle {Financial Modelling with Jump Processes} {Financial modelling
  with jump processes}.
\newblock
\APACaddressPublisher{Boca Raton, FL}{Chapman \& Hall/CRC}.
\PrintBackRefs{\CurrentBib}

\bibitem [\protect \citeauthoryear {%
DeMiguel%
, Garlappi%
\BCBL {}\ \BBA {} Uppal%
}{%
DeMiguel%
\ \protect \BOthers {.}}{%
{\protect \APACyear {2009}}%
}]{%
demiguel2009}
\APACinsertmetastar {%
demiguel2009}%
\begin{APACrefauthors}%
DeMiguel, V.%
, Garlappi, L.%
\BCBL {}\ \BBA {} Uppal, R.%
\end{APACrefauthors}%
\unskip\
\newblock
\APACrefYearMonthDay{2009}{}{}.
\newblock
{\BBOQ}\APACrefatitle {Optimal versus Naive Diversification: How Inefficient Is
  the 1/N Portfolio Strategy?} {Optimal versus naive diversification: How
  inefficient is the 1/n portfolio strategy?}{\BBCQ}
\newblock
\APACjournalVolNumPages{Review of Financial Studies}{22}{5}{1915--1953}.
\newblock
\begin{APACrefDOI} \doi{10.1093/rfs/hhn098} \end{APACrefDOI}
\PrintBackRefs{\CurrentBib}

\bibitem [\protect \citeauthoryear {%
Ding%
, Granger%
\BCBL {}\ \BBA {} Engle%
}{%
Ding%
\ \protect \BOthers {.}}{%
{\protect \APACyear {1993}}%
}]{%
ding1993}
\APACinsertmetastar {%
ding1993}%
\begin{APACrefauthors}%
Ding, Z.%
, Granger, C\BPBI W\BPBI J.%
\BCBL {}\ \BBA {} Engle, R\BPBI F.%
\end{APACrefauthors}%
\unskip\
\newblock
\APACrefYearMonthDay{1993}{}{}.
\newblock
{\BBOQ}\APACrefatitle {A Long Memory Property of Stock Market Returns and a New
  Model} {A long memory property of stock market returns and a new
  model}.{\BBCQ}
\newblock
\APACjournalVolNumPages{Journal of Empirical Finance}{1}{1}{83--106}.
\newblock
\begin{APACrefDOI} \doi{10.1016/0927-5398(93)90006-D} \end{APACrefDOI}
\PrintBackRefs{\CurrentBib}

\bibitem [\protect \citeauthoryear {%
Elton%
\ \BBA {} Gruber%
}{%
Elton%
\ \BBA {} Gruber%
}{%
{\protect \APACyear {1997}}%
}]{%
elton1997}
\APACinsertmetastar {%
elton1997}%
\begin{APACrefauthors}%
Elton, E\BPBI J.%
\BCBT {}\ \BBA {} Gruber, M\BPBI J.%
\end{APACrefauthors}%
\unskip\
\newblock
\APACrefYearMonthDay{1997}{}{}.
\newblock
{\BBOQ}\APACrefatitle {Modern portfolio theory, 1950 to date} {Modern portfolio
  theory, 1950 to date}.{\BBCQ}
\newblock
\APACjournalVolNumPages{Journal of Banking \& Finance}{21}{11-12}{1743--1759}.
\newblock
\begin{APACrefDOI} \doi{10.1016/S0378-4266(97)00048-4} \end{APACrefDOI}
\PrintBackRefs{\CurrentBib}

\bibitem [\protect \citeauthoryear {%
Fama%
}{%
Fama%
}{%
{\protect \APACyear {1970}}%
}]{%
fama1970}
\APACinsertmetastar {%
fama1970}%
\begin{APACrefauthors}%
Fama, E\BPBI F.%
\end{APACrefauthors}%
\unskip\
\newblock
\APACrefYearMonthDay{1970}{}{}.
\newblock
{\BBOQ}\APACrefatitle {Efficient Capital Markets: A Review of Theory and
  Empirical Work} {Efficient capital markets: A review of theory and empirical
  work}.{\BBCQ}
\newblock
\APACjournalVolNumPages{The Journal of Finance}{25}{2}{383--417}.
\newblock
\begin{APACrefDOI} \doi{10.2307/2325486} \end{APACrefDOI}
\PrintBackRefs{\CurrentBib}

\bibitem [\protect \citeauthoryear {%
B\BPBI M.~Hill%
}{%
B\BPBI M.~Hill%
}{%
{\protect \APACyear {1975}}%
}]{%
hill1975}
\APACinsertmetastar {%
hill1975}%
\begin{APACrefauthors}%
Hill, B\BPBI M.%
\end{APACrefauthors}%
\unskip\
\newblock
\APACrefYearMonthDay{1975}{}{}.
\newblock
{\BBOQ}\APACrefatitle {A simple general approach to inference about the tail of
  a distribution} {A simple general approach to inference about the tail of a
  distribution}.{\BBCQ}
\newblock
\APACjournalVolNumPages{The Annals of Statistics}{3}{5}{1163--1174}.
\newblock
\begin{APACrefDOI} \doi{10.1214/aos/1176343247} \end{APACrefDOI}
\PrintBackRefs{\CurrentBib}

\bibitem [\protect \citeauthoryear {%
J\BPBI B.~Hill%
}{%
J\BPBI B.~Hill%
}{%
{\protect \APACyear {2015}}%
}]{%
hill2015}
\APACinsertmetastar {%
hill2015}%
\begin{APACrefauthors}%
Hill, J\BPBI B.%
\end{APACrefauthors}%
\unskip\
\newblock
\APACrefYearMonthDay{2015}{}{}.
\newblock
{\BBOQ}\APACrefatitle {Tail Index Estimation for a Filtered Dependent Time
  Series} {Tail index estimation for a filtered dependent time series}.{\BBCQ}
\newblock
\APACjournalVolNumPages{Statistica Sinica}{25}{2}{609--629}.
\newblock
\begin{APACrefDOI} \doi{10.5705/ss.2013.200} \end{APACrefDOI}
\PrintBackRefs{\CurrentBib}

\bibitem [\protect \citeauthoryear {%
Hurst%
}{%
Hurst%
}{%
{\protect \APACyear {1951}}%
}]{%
hurst1951}
\APACinsertmetastar {%
hurst1951}%
\begin{APACrefauthors}%
Hurst, H\BPBI E.%
\end{APACrefauthors}%
\unskip\
\newblock
\APACrefYearMonthDay{1951}{}{}.
\newblock
{\BBOQ}\APACrefatitle {Long-Term Storage Capacity of Reservoirs} {Long-term
  storage capacity of reservoirs}.{\BBCQ}
\newblock
\APACjournalVolNumPages{Transactions of the American Society of Civil
  Engineers}{116}{1}{770--799}.
\newblock
\begin{APACrefDOI} \doi{10.1061/TACEAT.0006518} \end{APACrefDOI}
\PrintBackRefs{\CurrentBib}

\bibitem [\protect \citeauthoryear {%
Lee%
, Shirvani%
, Afroz%
, Rachev%
\BCBL {}\ \BBA {} Fabozzi%
}{%
Lee%
\ \protect \BOthers {.}}{%
{\protect \APACyear {2026}}%
}]{%
Lee2026TaiwanETFs}
\APACinsertmetastar {%
Lee2026TaiwanETFs}%
\begin{APACrefauthors}%
Lee, T\BHBI J.%
, Shirvani, A.%
, Afroz, F.%
, Rachev, S\BPBI T.%
\BCBL {}\ \BBA {} Fabozzi, F\BPBI J.%
\end{APACrefauthors}%
\unskip\
\newblock
\APACrefYearMonthDay{2026}{}{}.
\newblock
{\BBOQ}\APACrefatitle {Portfolio Optimization Under Heavy Tails and Asymmetric
  Volatility: Evidence from Taiwan-Exposed {ETFs}} {Portfolio optimization
  under heavy tails and asymmetric volatility: Evidence from taiwan-exposed
  {ETFs}}.{\BBCQ}
\newblock
\APACjournalVolNumPages{Journal of Risk and Financial Management}{19}{10}{769}.
\newblock
\begin{APACrefURL} \url{https://www.mdpi.com/1911-8074/19/10/769}
  \end{APACrefURL}
\PrintBackRefs{\CurrentBib}

\bibitem [\protect \citeauthoryear {%
Lindquist%
, Rachev%
, Hu%
\BCBL {}\ \BBA {} Shirvani%
}{%
Lindquist%
\ \protect \BOthers {.}}{%
{\protect \APACyear {2022}}%
}]{%
lindquist2022}
\APACinsertmetastar {%
lindquist2022}%
\begin{APACrefauthors}%
Lindquist, W\BPBI B.%
, Rachev, S\BPBI T.%
, Hu, Y.%
\BCBL {}\ \BBA {} Shirvani, A.%
\end{APACrefauthors}%
\unskip\
\newblock
\APACrefYear{2022}.
\newblock
\APACrefbtitle {Advanced {REIT} Portfolio Optimization: Innovative Tools for
  Risk Management} {Advanced {REIT} portfolio optimization: Innovative tools
  for risk management}.
\newblock
\APACaddressPublisher{Cham}{Springer}.
\newblock
\begin{APACrefDOI} \doi{10.1007/978-3-031-15286-3} \end{APACrefDOI}
\PrintBackRefs{\CurrentBib}

\bibitem [\protect \citeauthoryear {%
Madhavan%
}{%
Madhavan%
}{%
{\protect \APACyear {2012}}%
}]{%
madhavan2012}
\APACinsertmetastar {%
madhavan2012}%
\begin{APACrefauthors}%
Madhavan, A.%
\end{APACrefauthors}%
\unskip\
\newblock
\APACrefYearMonthDay{2012}{}{}.
\newblock
{\BBOQ}\APACrefatitle {Exchange-Traded Funds, Market Structure, and the Flash
  Crash} {Exchange-traded funds, market structure, and the flash crash}.{\BBCQ}
\newblock
\APACjournalVolNumPages{Financial Analysts Journal}{68}{4}{20--35}.
\newblock
\begin{APACrefDOI} \doi{10.2469/faj.v68.n4.6} \end{APACrefDOI}
\PrintBackRefs{\CurrentBib}

\bibitem [\protect \citeauthoryear {%
Mandelbrot%
}{%
Mandelbrot%
}{%
{\protect \APACyear {1963}}%
}]{%
mandelbrot1963}
\APACinsertmetastar {%
mandelbrot1963}%
\begin{APACrefauthors}%
Mandelbrot, B\BPBI B.%
\end{APACrefauthors}%
\unskip\
\newblock
\APACrefYearMonthDay{1963}{}{}.
\newblock
{\BBOQ}\APACrefatitle {The Variation of Certain Speculative Prices} {The
  variation of certain speculative prices}.{\BBCQ}
\newblock
\APACjournalVolNumPages{The Journal of Business}{36}{4}{394--419}.
\newblock
\begin{APACrefDOI} \doi{10.1086/294632} \end{APACrefDOI}
\PrintBackRefs{\CurrentBib}

\bibitem [\protect \citeauthoryear {%
Mandelbrot%
\ \BBA {} Wallis%
}{%
Mandelbrot%
\ \BBA {} Wallis%
}{%
{\protect \APACyear {1968}}%
}]{%
mandelbrot1968}
\APACinsertmetastar {%
mandelbrot1968}%
\begin{APACrefauthors}%
Mandelbrot, B\BPBI B.%
\BCBT {}\ \BBA {} Wallis, J\BPBI R.%
\end{APACrefauthors}%
\unskip\
\newblock
\APACrefYearMonthDay{1968}{}{}.
\newblock
{\BBOQ}\APACrefatitle {{{Noah, Joseph, and Operational Hydrology}}} {{{Noah,
  Joseph, and Operational Hydrology}}}.{\BBCQ}
\newblock
\APACjournalVolNumPages{Water Resources Research}{4}{5}{909--918}.
\newblock
\begin{APACrefDOI} \doi{10.1029/WR004i005p00909} \end{APACrefDOI}
\PrintBackRefs{\CurrentBib}

\bibitem [\protect \citeauthoryear {%
Markowitz%
}{%
Markowitz%
}{%
{\protect \APACyear {1952}}%
}]{%
markowitz1952}
\APACinsertmetastar {%
markowitz1952}%
\begin{APACrefauthors}%
Markowitz, H.%
\end{APACrefauthors}%
\unskip\
\newblock
\APACrefYearMonthDay{1952}{}{}.
\newblock
{\BBOQ}\APACrefatitle {Portfolio Selection} {Portfolio selection}.{\BBCQ}
\newblock
\APACjournalVolNumPages{The Journal of Finance}{7}{1}{77--91}.
\newblock
\begin{APACrefDOI} \doi{10.1111/j.1540-6261.1952.tb01525.x} \end{APACrefDOI}
\PrintBackRefs{\CurrentBib}

\bibitem [\protect \citeauthoryear {%
McNeil%
\ \BBA {} Frey%
}{%
McNeil%
\ \BBA {} Frey%
}{%
{\protect \APACyear {2000}}%
}]{%
mcneil2000}
\APACinsertmetastar {%
mcneil2000}%
\begin{APACrefauthors}%
McNeil, A\BPBI J.%
\BCBT {}\ \BBA {} Frey, R.%
\end{APACrefauthors}%
\unskip\
\newblock
\APACrefYearMonthDay{2000}{}{}.
\newblock
{\BBOQ}\APACrefatitle {Estimation of Tail-Related Risk Measures for
  Heteroscedastic Financial Time Series: An Extreme Value Approach} {Estimation
  of tail-related risk measures for heteroscedastic financial time series: An
  extreme value approach}.{\BBCQ}
\newblock
\APACjournalVolNumPages{Journal of Empirical Finance}{7}{3--4}{271--300}.
\newblock
\begin{APACrefDOI} \doi{10.1016/S0927-5398(00)00012-8} \end{APACrefDOI}
\PrintBackRefs{\CurrentBib}

\bibitem [\protect \citeauthoryear {%
Pflug%
}{%
Pflug%
}{%
{\protect \APACyear {2000}}%
}]{%
pflug2000}
\APACinsertmetastar {%
pflug2000}%
\begin{APACrefauthors}%
Pflug, G\BPBI C.%
\end{APACrefauthors}%
\unskip\
\newblock
\APACrefYearMonthDay{2000}{}{}.
\newblock
{\BBOQ}\APACrefatitle {Some Remarks on the {V}alue-at-{R}isk and the
  {C}onditional {V}alue-at-{R}isk} {Some remarks on the {V}alue-at-{R}isk and
  the {C}onditional {V}alue-at-{R}isk}.{\BBCQ}
\newblock
\BIn{} S\BPBI P.~Uryasev\ (\BED), \APACrefbtitle {Probabilistic Constrained
  Optimization: Methodology and Applications} {Probabilistic constrained
  optimization: Methodology and applications}\ (\BPGS\ 272--281).
\newblock
\APACaddressPublisher{Dordrecht}{Springer}.
\newblock
\begin{APACrefDOI} \doi{10.1007/978-1-4757-3150-7_15} \end{APACrefDOI}
\PrintBackRefs{\CurrentBib}

\bibitem [\protect \citeauthoryear {%
Rachev%
, Menn%
\BCBL {}\ \BBA {} Fabozzi%
}{%
Rachev%
\ \protect \BOthers {.}}{%
{\protect \APACyear {2005}}%
}]{%
rachev2005}
\APACinsertmetastar {%
rachev2005}%
\begin{APACrefauthors}%
Rachev, S\BPBI T.%
, Menn, C.%
\BCBL {}\ \BBA {} Fabozzi, F\BPBI J.%
\end{APACrefauthors}%
\unskip\
\newblock
\APACrefYear{2005}.
\newblock
\APACrefbtitle {Fat-Tailed and Skewed Asset Return Distributions: Implications
  for Risk Management, Portfolio Selection, and Option Pricing} {Fat-tailed and
  skewed asset return distributions: Implications for risk management,
  portfolio selection, and option pricing}.
\newblock
\APACaddressPublisher{Hoboken, NJ}{John Wiley \& Sons}.
\PrintBackRefs{\CurrentBib}

\bibitem [\protect \citeauthoryear {%
Resnick%
}{%
Resnick%
}{%
{\protect \APACyear {2007}}%
}]{%
resnick2007}
\APACinsertmetastar {%
resnick2007}%
\begin{APACrefauthors}%
Resnick, S\BPBI I.%
\end{APACrefauthors}%
\unskip\
\newblock
\APACrefYear{2007}.
\newblock
\APACrefbtitle {Heavy-Tail Phenomena: Probabilistic and Statistical Modeling}
  {Heavy-tail phenomena: Probabilistic and statistical modeling}.
\newblock
\APACaddressPublisher{New York}{Springer}.
\newblock
\begin{APACrefDOI} \doi{10.1007/978-0-387-45024-7} \end{APACrefDOI}
\PrintBackRefs{\CurrentBib}

\bibitem [\protect \citeauthoryear {%
Rockafellar%
\ \BBA {} Uryasev%
}{%
Rockafellar%
\ \BBA {} Uryasev%
}{%
{\protect \APACyear {2000}}%
}]{%
rockafellar_uryasev_2000}
\APACinsertmetastar {%
rockafellar_uryasev_2000}%
\begin{APACrefauthors}%
Rockafellar, R\BPBI T.%
\BCBT {}\ \BBA {} Uryasev, S.%
\end{APACrefauthors}%
\unskip\
\newblock
\APACrefYearMonthDay{2000}{}{}.
\newblock
{\BBOQ}\APACrefatitle {Optimization of Conditional Value-at-Risk} {Optimization
  of conditional value-at-risk}.{\BBCQ}
\newblock
\APACjournalVolNumPages{The Journal of Risk}{2}{3}{21--41}.
\newblock
\begin{APACrefDOI} \doi{10.21314/JOR.2000.038} \end{APACrefDOI}
\PrintBackRefs{\CurrentBib}

\bibitem [\protect \citeauthoryear {%
Sharpe%
}{%
Sharpe%
}{%
{\protect \APACyear {1966}}%
}]{%
sharpe1966}
\APACinsertmetastar {%
sharpe1966}%
\begin{APACrefauthors}%
Sharpe, W\BPBI F.%
\end{APACrefauthors}%
\unskip\
\newblock
\APACrefYearMonthDay{1966}{}{}.
\newblock
{\BBOQ}\APACrefatitle {Mutual Fund Performance} {Mutual fund
  performance}.{\BBCQ}
\newblock
\APACjournalVolNumPages{The Journal of Business}{39}{1}{119--138}.
\newblock
\begin{APACrefDOI} \doi{10.1086/294846} \end{APACrefDOI}
\PrintBackRefs{\CurrentBib}

\bibitem [\protect \citeauthoryear {%
Sortino%
\ \BBA {} Price%
}{%
Sortino%
\ \BBA {} Price%
}{%
{\protect \APACyear {1994}}%
}]{%
sortino_classical1994}
\APACinsertmetastar {%
sortino_classical1994}%
\begin{APACrefauthors}%
Sortino, F\BPBI A.%
\BCBT {}\ \BBA {} Price, L\BPBI N.%
\end{APACrefauthors}%
\unskip\
\newblock
\APACrefYearMonthDay{1994}{}{}.
\newblock
{\BBOQ}\APACrefatitle {Performance Measurement in a Downside Risk Framework}
  {Performance measurement in a downside risk framework}.{\BBCQ}
\newblock
\APACjournalVolNumPages{Journal of Investing}{3}{3}{59--65}.
\newblock
\begin{APACrefDOI} \doi{10.3905/joi.3.3.59} \end{APACrefDOI}
\PrintBackRefs{\CurrentBib}

\bibitem [\protect \citeauthoryear {%
Tsay%
}{%
Tsay%
}{%
{\protect \APACyear {2010}}%
}]{%
tsay2010}
\APACinsertmetastar {%
tsay2010}%
\begin{APACrefauthors}%
Tsay, R\BPBI S.%
\end{APACrefauthors}%
\unskip\
\newblock
\APACrefYear{2010}.
\newblock
\APACrefbtitle {Analysis of Financial Time Series} {Analysis of financial time
  series}\ (\PrintOrdinal{3rd}\ \BEd).
\newblock
\APACaddressPublisher{Hoboken, NJ}{John Wiley \& Sons}.
\newblock
\begin{APACrefDOI} \doi{10.1002/9780470644553} \end{APACrefDOI}
\PrintBackRefs{\CurrentBib}

\end{thebibliography}
\end{document}